# Total charge changing cross sections measurements for $^{12}$C ion interactions with Al using CR-39 nuclear track detector

By

**Iftekhar Ahmed**

A Dissertation Submitted to the Department of Physics, University of Chittagong, Bangladesh, in Partial Fulfillment for the Master Degree in Physics.

Thesis Advisors

**Dr. Nakahiro Yasuda**

Research Institute of Nuclear Engineering, University of Fukui, Tsuruga, Fukui, Japan

**Dr. Quazi Muhammad Rashed Nizam**

Department of Physics, University of Chittagong, Chittagong, 4331, Bangladesh

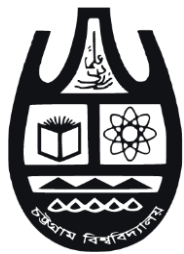

**Department of Physics, University of Chittagong, Chittagong-4331, Bangladesh.**

**December 2021**

*DEDICATED*
*TO*
*MY BELOVED PARENTS AND*
*TEACHERS*

# Authors Declaration

I do hereby declare that the entire research work presented as a thesis entitle "**Total charge changing cross sections measurements for $^{12}$C ion interactions with Al using CR-39 nuclear track detector**" has been submitted as partial fulfillment of the requirement for the degree of Masters of Science in Physics from University of Chittagong. It is hereby declared that this thesis or any part of it has not been submitted elsewhere for the award of any degree or diploma.

# Abstract

The study of nucleus-nucleus interaction is significant for designing space radiation shielding and predicting effects on the human body during heavy ion cancer therapy. Two main processes follow the interaction of heavy ions, called the electromagnetic and nucleus-nucleus interactions while propagating through the material. The theory of electromagnetic interaction is well developed, but no concrete theory for the nucleus-nucleus collision yet. Models for nucleus-nucleus interaction have still been under progress through new experimental data. Since ion with an energy range of 100-400 MeV/n is useful for cancer therapy as well as the flux of Galactic Cosmic Ray (GCR) and Solar Particle Events (SPE) is high in this energy range, so we are trying to measure the total charge changing cross-section for $^{12}$C on Al in the energy around 100 MeV/n in this study.

We are investigating the total charge changing cross-section of $^{12}$C by using the CR-39 nuclear track detector. Target sandwiched with CR-39 of 5 cm x 5 cm area was irradiated with $^{12}$C ion at 55 MeV/n and 135 MeV/n at the Wakasa wan Energy Research Center (WERC) and Heavy Ion Medical Accelerator in Chiba (HIMAC), Japan, respectively. Exposed CR-39s were etched in 7N NaOH solution at 70°C to visualize the tracks produced by the primary ion beam and its fragmentations. The etching time was chosen between 15 and 25 hours according to exposed densities of $^{12}$C ions. The SEIKO FSP-1000 imaging microscope was used to simultaneously get the front and back surface etch pit images. Then by counting the projectile track before and after the target, the total charge changing cross-section has been measured. The measured data for the energies 14.5±1.0, 24.1±0.9, 31.5±0.5, 89.5±0.2, 92.6±0.2, 95.7±0.2, 98.6±0.2, 101.5±0.2, 104.4±0.2, 107.2±0.2, and 110.0±0.2 MeV/n were 2990.0±472.0, 2748.4±451.0, 2437.2±424.0, 1008.5±304.0, 1187.2±329.0, 545.6±222.0, 545.0±222.0, 1445.7±361.2, 1711.2±391.0, 1705.1±391.2, 1966.5±419.0 mb. Since there are no previous experimental data to compare, PHITS and Glauber model calculated theoretical values were compared with the previous and this experimental result. From that, the measured data were found in good agreement with the previous and calculated value except for two points at energy 95.7 and 98.6 MeV/n. These two points need more analysis which will be done in the next phase. Measured data in this study will be the first experimental TCC data for $^{12}$C+Al interactions in the energy ~100 MeV/n. These measured data will be useful to verify the models, optimize shielding in space radiation circumstances, and improve treatment planning on therapy.

# Table of Contents

# Chapter 4
# Experimental Method

# Chapter 5
# Results and Discussions

# Chapter 6
# Conclusions and Future Works

# List of Figures

# List of Tables

# Chapter 1
# Introduction

## 1.1 Introduction

The discovery of the X-ray by German physicist Wilhelm Conrad Rontgen in December 1895 explored new physics and opened the gate for new treatment. The use of X-rays for cancer treatment was started a few days after that discovery without knowing the adverse effect of this ray which is still used in modern days as radiotherapy. Since then, scientists have been trying to understand the interaction mechanism, proper use and disadvantages of this new treatment system which led us to the branch of radiation physics. In this branch, researchers are trying to understand the interactions of radiation with matters for proper use and understand the characteristics of the radiation. Also, the source of nuclear radiation, the nucleus, from where the different types of radiation can emerge is a matter of investigation regarding its shape, size, and formation, which are also under research till now. One of the ways to understand the nucleus is through heavy ion interaction, where we can understand the different properties of the nucleus. During these kinds of heavy ion interactions, it is found that heavy ion (Z>2) [1] is more convenient than conventional radiotherapy. Also, this study is useful for the analysis of cosmic rays. The scope of the heavy ion interactions in the different fields is briefly described below.

## 1.2 Significance of heavy ion interaction

Heavy ion study is involved in explaining several important sectors. Heavy ion study allows studying one of the fundamental quantum field theories (named QCD) at non-zero temperature and density. It is also important for condensed matter physics and cosmology. It is an active field of research. Ongoing large experimental programs at the LHC (CERN) by the collaborations ALICE, ATLAS, CMS, LHCb, and at RHIC (BNL) by the collaborations Phenix and STAR. Quark gluon plasma has filled the universe from about $10^{12}$ s to $10^6$ s after the big bang. Heavy ion collisions allow learning something about this state from laboratory experiments. Understanding of heavy ion interaction mechanism is also very important for the proper use of heavy ions in cancer therapy and shielding of space radiation. These are the most significant aspect of this study.

## *1.2.1 Heavy ion cancer therapy*

We know that an ion is a charged atom or molecule where the proton and electron number are not the same due to the gain or loss of electrons. And ions heavier than Z>2 are called heavy ions. Scientists are conducting research on heavy ions such as carbon ions to make improvements in clinical practice, education, and research. Because heavy ion therapy, like carbon, achieved substantial interest due to an excellent combination of physical and radiobiological advantages compared to conventional photon therapy [2]. Carbon ion therapy has high dose distribution that may allow dose escalation to the target region (tumor) while reducing radiation dose to the entrance channel (normal tissue) [3]. So, the study of $^{12}$C ion interactions is very important for the use of heavy-ion in cancer therapy. Carbon ion therapy started in 1994 at the National Institute of Radiological Science, Heavy Ion Medical Accelerator (HIMAC), Japan [4], [5].

Radiation therapy (called radiotherapy) is a type of cancer treatment that uses a beam of intense energy to destroy cancer cells or slow their growth by damaging their DNA and shrinking tumors. Radiotherapy was applied just three days after x-rays were discovered in December 1895 by W.C. Rontgen. Emil Grubbe applied the x-rays for the first time to treat a woman who was suffering from breast cancer. Then, C0-60 beam therapy emitting gamma rays took place in x-ray therapy many years later and became the common radiotherapy in cancer treatment [6]. In order to treat a deep-seated tumor by conventional radiotherapy, scientists found out that radiation can also damage healthy cells before the tumor region. Due to several side effects, scientists were looking for advanced techniques that can minimize the damage to normal tissue surrounding the tumor to a level as low as possible [3], [7]. Studies on heavy ions were initially suggested for clinical trial due to good relative biological effectiveness owing to the high LET (linear energy transfer) of heavy ions [7], [8].

The application of heavy ions offers significant advantages in achieving accurate dose localization in the target region in comparison to megavolt photon therapy. In 1952, the first human patients were treated with neon and helium ion beams [8]. 433 patients were treated with neon ions and 2,054 patients with helium ions until it was closed. In the meantime, interest in heavy ion beams expanded, with proton facilities arising around the world. Lawrence and Tobias first initiated the clinical application of proton beams for the first time in the world between 1954 and 1957 at the Lawrence Berkeley National Laboratory (LBNL), the University of California, which marked the beginning of modern radiotherapy [9]. Since then, over 170,000 patients have been treated with proton therapy. Currently, there are a total of 89 proton therapy centers worldwide, with still more being planned in the world. Nowadays, scientists are working on carbon ion therapy that was first started clinically by Heavy Ion Medical Accelerator (HIMAC) in Chiba, Japan, in 1994. Among 12 carbon ion therapy facilities all over the world, Japan has six carbon ion therapy centers, and they have treated more than 19,000 cancer patients successfully from 1994 to 2019 [5].

So, therapeutic radiation is classified into two basic types: photon beam (X-rays and gamma rays) and particle beam (protons and heavy ions, etc.), as shown in Fig. 1.1. This figure shows the type of radiation for cancer treatment and the way how proton ion and $^{12}$C ion can be produced by removal of orbital electron form the atom. A photon beam is a kind of electromagnetic wave, whereas a particle beam can move with a velocity close to the speed of light that is made up of high-energy ions.

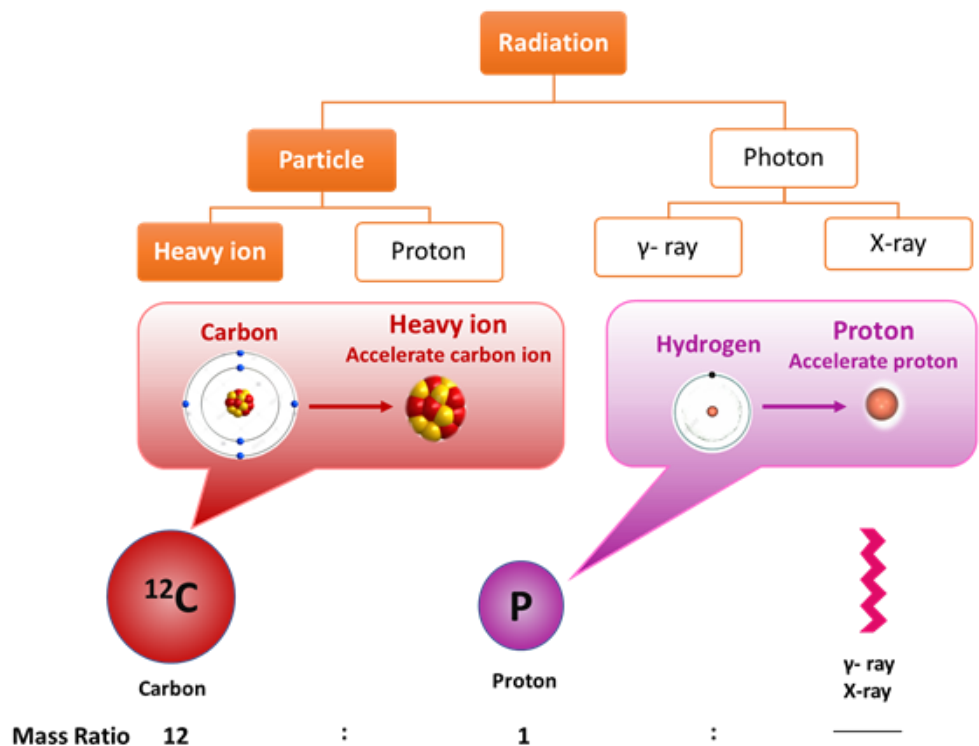

*Figure 1.1 Therapeutic radiation treatment. This figure shows that how an ion ($^{12}$C or proton) has been produced by removal of electron. Only the nucleus remain that interact with the cancer cell in case of $^{12}$C ion therapy and proton therapy respectively.*

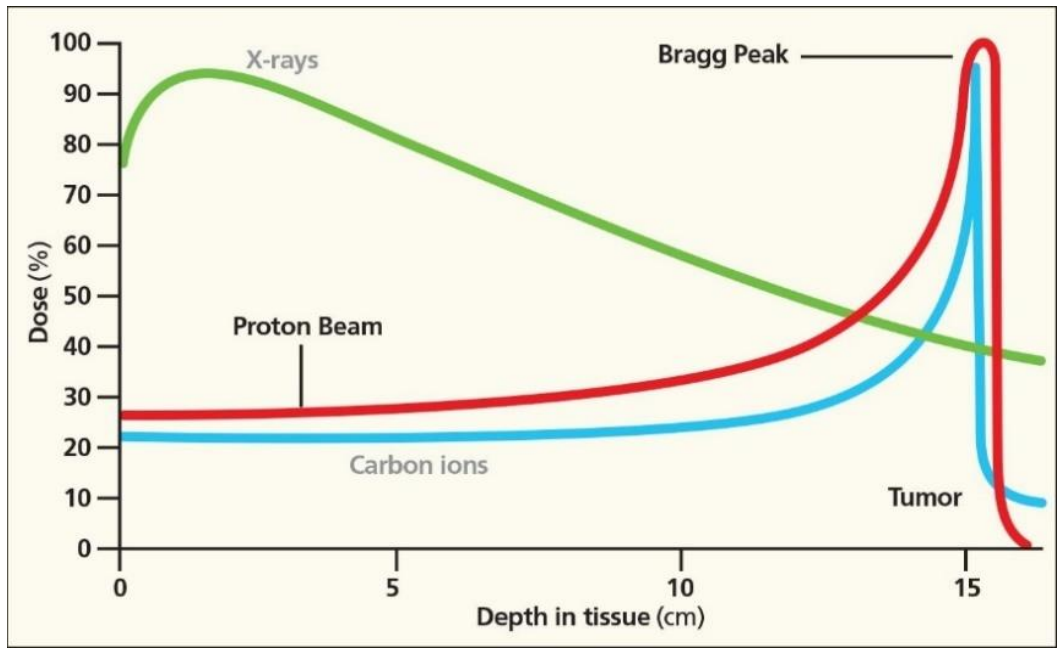

*Figure 1.2 Relationship of depth and relative dose in photon, proton and carbon ion beams. Proton and carbon can produce Bragg peak [8] at the desire depth in case of deep seated tumor whereas photon loss all of its energy at the entry point and dose become lower with depth.*

Carbon ion irradiation can pinpoint the location of the tumor with precise irradiation based on their Bragg peak (showed in Fig. 1.2 [15]). Therefore, provide sufficient damage to the target tumor while causing the least amount of damage to healthy tissue. Carbon ion beams can deliver a highly concentrated dose in the cancer lesion conformably compared to X-ray beams showed in Fig. 1.3 [4], [15].

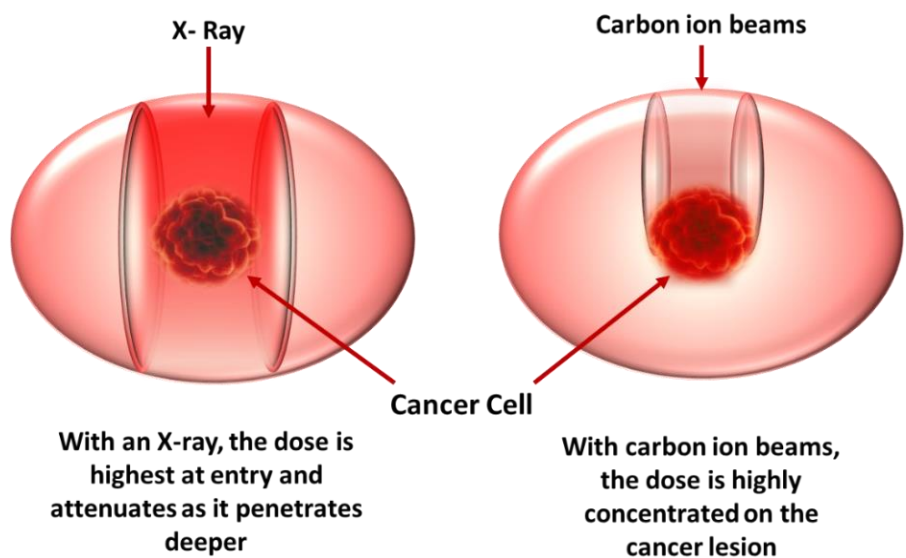

*Figure 1.3 Comparison of dose-distribution between an X-ray and carbon ion beams which shows the maximum deposition of dose at the center of desired area by the $^{12}C$ ion.*

## *1.2.2 Space radiation shielding*

Another significance of the heavy ion interaction study implies the shielding of radiation in the international space station (ISS). Space radiation is a major concern in planning long-term human exploratory missions in outer space. In Low-Earth orbit, our magnetosphere shield protects us from high-energy charged particle radiation, which deflects most of these charged particles. However, outside of the Earth's magnetosphere, astronauts are expected to encounter by cosmic rays originating from the sun and galaxies [10]. And, Astronauts will be exposed to ionizing radiation that is different in dose than electromagnetic radiation emitted by the Earth's surface while they explore the solar system. As a consequence, space radiation can lead to long-term effects on astronauts, such as cancer, sterility, radiation sickness, and degenerative diseases [11]. So, radiation shielding in space is a crucial issue to protect astronauts by reducing the space radiation dose.

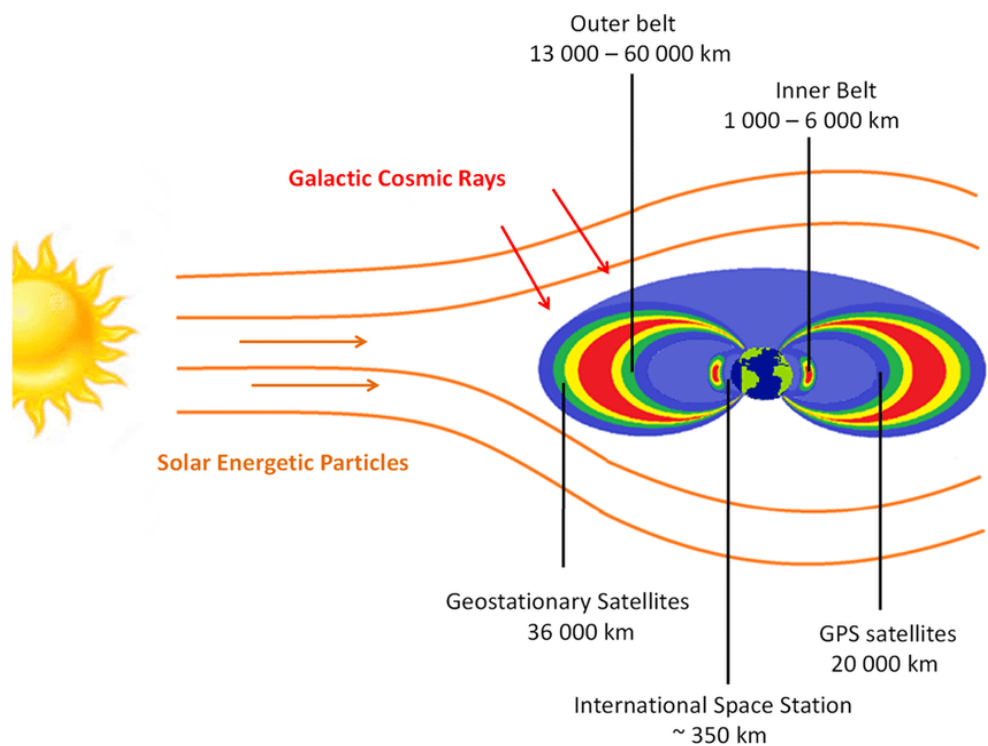

*Figure 1.4 The interplanetary space environment showing the toxic combination of galactic cosmic radiation (GCR) and (largely) proton radiation due to solar particle events (SPEs). Figure courtesy of NASA/JPL-Caltech.*

Several forms of ionizing radiation exist in the space environment beyond low Earth orbit (LEO). The solar wind, which produces a steady flux of low linear energy transfer (LET) radiation, is responsible for the majority of the energetic particles observed in interplanetary space. Galactic cosmic radiation (GCR) will contribute a considerable amount of the radiation dose accumulated by astronaut crew members during missions outside of LEO. GCR ions come far beyond our solar system and are mostly made up of highly energetic protons and alpha particles, with a minor amount of HZE nuclei moving at relativistic speeds and energies [12]. In addition to GCR, solar particle events (SPEs) are unpredictable and intermittent phenomena that can produce enormous plasma clouds containing highly energetic protons and some heavy ions, which can cause a rapid surge of radiation both outside and inside a spacecraft (Fig. 1.4).

Moon bases, contact with a near-Earth object, such as an asteroid, and, potentially, settlements on the surface of Mars are all possibilities for future human spaceflight missions. The shielding given by the Earth's magnetic field attenuates the significant biological effects of space radiation exposures for current space missions in LEO. Future spaceflight missions to a NEO or Mars, however, will need extended transit beyond the protection of the Earth's magnetosphere, increasing the risks of space radiation.

## *1.2.3 Galactic cosmic radiation*

GCR nuclei are high-LET relativistic particles that come from outside our solar system and have enough energy to penetrate any shielding equipment utilized on current mission spacecraft. The GCR spectrum is made up of about 87% hydrogen ions (protons) and 12% helium ions (alpha particles), with the remaining 1%–2% of particles being HZE nuclei with charges ranging from $Z = 3$ (lithium) to $Z = 28$ (helium) (nickel). Iron ($Z = 26$) and other ionized transition metals are biologically hazardous, and no reasonable amount of spaceship material can protect them (Fig. 1.2.2.1.1). The fluence of ionized nuclei in GCR is inversely related to the solar cycle, decreasing by a factor of two during solar maximum. The GCR fluence rate and spectrum outside of LEO have been broadly defined by observations taken by unmanned spacecraft, including the Mars Science Laboratory (MSL) spacecraft, which transported the Mars Curiosity rover to the red planet from December 2011 to July 2012 [13]. As a result, recent evidence suggests that the absorbed dose and dose equivalent from incident particles can be accurately predicted ahead of future exploration-class space missions.

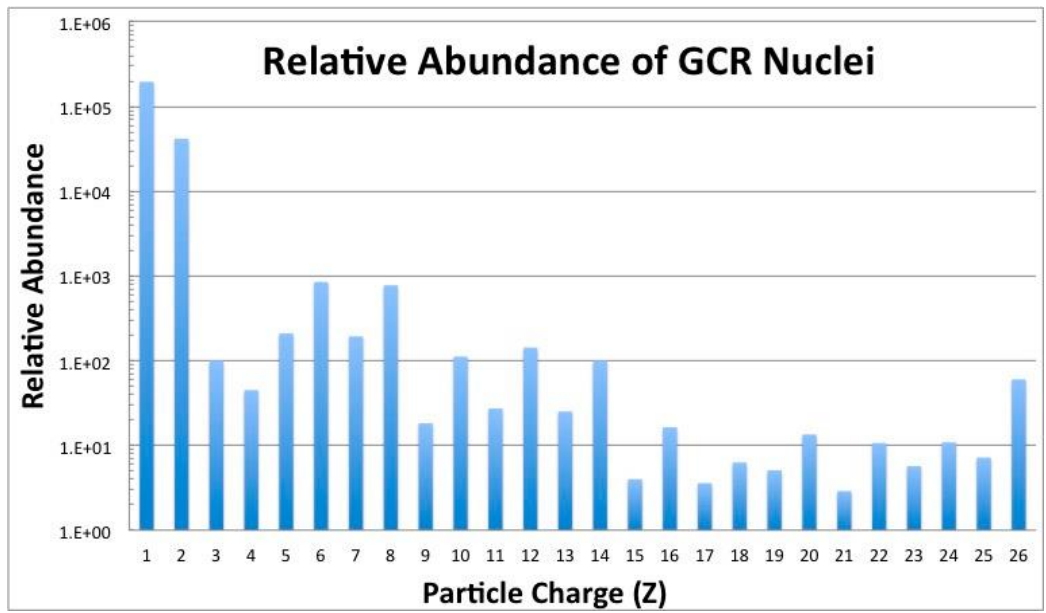

*Figure 1.5 The GCR spectrum consists of approximately 87% hydrogen ions (protons) and 12% helium ions (alpha particles), with the remaining 1%–2% of particles being HZE nuclei with charges ranging from Z = 3 (lithium) to approximately Z = 28 (nickel). Ionized transition metals, such as iron (Z = 26), are biologically harmful, as no reasonable amount of spacecraft material can shield them. .Relative abundance of GCR nuclei from hydrogen (Z = 1) to iron (Z = 26) [14].*

Because of their high ionization power, GCR ions present a significant health risk to humans and are one of the most important barriers to crewed spacecraft interplanetary travel. The energies of GCR particles are high enough to penetrate several centimeters of biological tissue as well as other organic and inorganic materials. Given the penetrating ability of HZE ions, shielding only partially lessens the doses experienced within a spaceship. While thicker shielding could theoretically provide more protection, the practical limitations of current spacecraft launch technologies limit the amount of shielding that can be sent into orbit. During transit outside of LEO, every cell nucleus within an astronaut would be traversed, on average, by a hydrogen ion every few days and by heavier HZE nuclei (e.g., $^{12}$C, $^{16}$O, $^{28}$Si, $^{56}$Fe) every few months. Consequently, despite their low flux, HZE ions constitute a biological concern and comprise a considerable amount of the total GCR dose that astronauts will get outside of LEO [14].

## *1.2.4 SPE radiation*

SPEs can produce massive amounts of energetic protons with fluences exceeding $10^9$ protons/cm$^2$, and they are dangerous and unpredictable. Low-LET protons with energy up to 1 GeV/n make up the majority of SPEs, which can be easily shielded by spaceship hulls.

Even yet, astronauts in thinly shielded spacecraft and habitats may be concerned about very high density fluxes of protons with energy of more than 30 MeV. SPE dose rates vary during an event, ranging from zero to 100 mGy/h for an astronaut within a spacecraft and zero to 500 mGy/h for an astronaut exposed during an extravehicular activity (EVA) on missions outside of LEO. SPEs are related to sunspot activity, and their frequency varies with the phase of the 11-year solar cycle, peaking when equatorial sunspot activity is at its maximum. The intensity of SPEs is not determined by the phase of the solar cycle, and some of the largest observed solar events have happened during off-peak periods, when there is a major reduction in recorded sunspots (Fig. 1.6). Interplanetary transits will be a part of exploration missions outside of LEO, and the spacecraft's shielding may not be enough to protect astronaut crews from the effects of an SPE. Moreover, crewmembers are likely to be exposed to multiple SPEs during such missions [14], [15]. The main consequences are skin lesions, hematological, and immunological dysfunctions. Shielding is an efficient counter-measure for the same reason. SPE is not likely to produce major direct changes in CNS function, despite the fact that it may contribute to cancer risk and tissue degeneration.

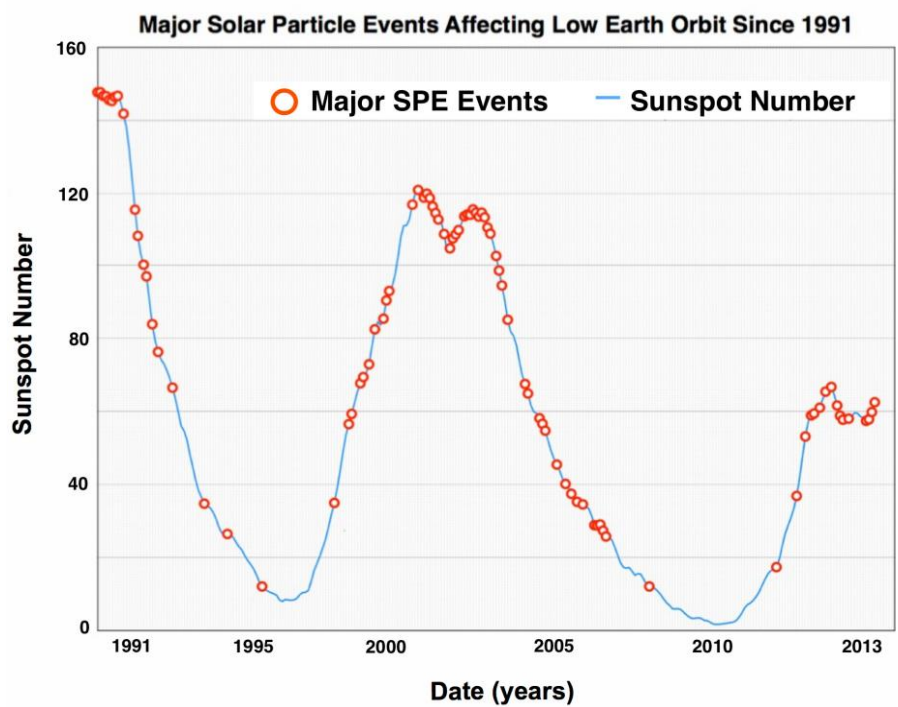

*Figure 1.6 Energetic SPEs affecting low Earth orbit (LEO) space missions since 1991 are plotted as a function of the solar cycle. Shown here are events (red circles) that have been measured since 1991 to 2013 and include Solar Cycle 22 (partially), 23 and 24 (partially). Energetic solar events contain a higher fluence of >100 MeV protons that can penetrate typical spacecraft shielding and significantly impact the health of astronauts [14].*

## *1.2.5 Intravehicular Radiation*

Interaction of energetic SPE protons and heavy-charged GCR particles with the spacecraft body can result in a secondary intravehicular radiation risk. Protons, alpha particles, beta particles, gamma rays, x-rays, neutrons, and heavy-charged particles are secondary particles produced in nuclear fission reactions. These fission products, which are created by traveling through spacecraft shielding, can give a major fraction of the total mission dose and can harm important cellular components while passing through the tissues of the body. An efficacious shield against galactic cosmic rays is one that effectively splits up heavy ions into lighter ions. Put in other words, the largest cross-sections per unit mass of shielding materials are called to be the best shield for radiation protection [16]. Therefore, secondary neutrons produced by heavy ions must be considered, particularly for future long-term and deep space missions.

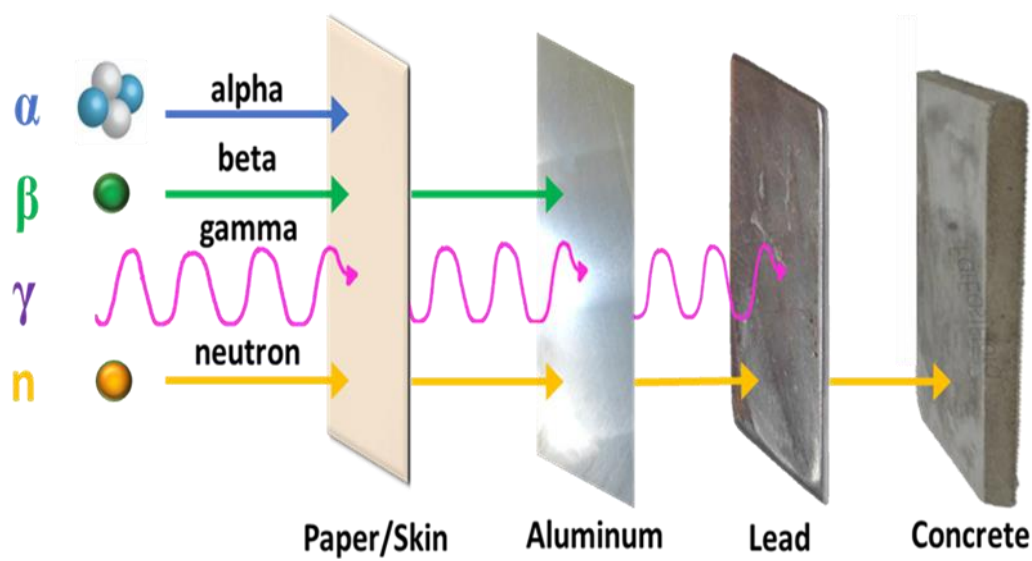

*Figure 1.7 The illustration shows the penetrating power of different types of ionizing radiation, ranging from the least penetrating alpha particles to the most secondary particle (neutron) through skin, aluminum, lead and concrete.*

## *1.2.6 Biomedical Consequences of Exposure to Space Radiation*

It's still uncertain how SPEs of moderate to high amplitude, along with continuous GCR exposure, will affect astronaut crews' health and performance during interplanetary transits. When it comes to whole-body irradiation, SPEs have a particular dose distribution. Because of its superficial position and tendency to absorb in the low energy spectra of protons and nuclei, skin doses are 5–10 times higher than those experienced by internal organs.

Prodromal symptoms (e.g., nausea and vomiting), skin injury, hematological disorders, and immune system malfunction can all occur as a result of SPE radiation and the synergistic effects of spaceflight. According to current estimates, the probability of fatality from a massive solar event or the combined effect of multiple SPEs is negligible [14]. GCR exposure accumulated over the course of exploration class missions to Mars, the Moon, or an asteroid is claimed in a similar way. Despite the existence of a considerable amount of research addressing the effects of projected absorbed dose ranges, this insufficient understanding remains. Because of the complexity of the radiation environment, the shielding effects of the vehicle or space suit, and human anatomy and physiology, calculating astronaut radiation exposures in a precise and realistic manner is difficult. The requirement for models that can identify particle energy and species on an event-by-event basis is highlighted by uncertainties in dose toxicity and the complex variation in SPE and GCR spectra likely to be encountered in future exploration missions. Integrating micro dosimetry measurements with radiobiological investigations is also critical for minimizing uncertainty in dose projections during mission planning and spaceflights, as well as informing post-flight study on astronaut health. Multiple organs and physiological systems are affected in complex ways by space radiation (Fig. 1.8). The biomedical consequences are divided into four risk categories by NASA. For a given absorbed dose, alpha particles cause more damage than beta particles, gamma rays, and x rays because they deposit their energy thousands of times more effectively. While lower-energy electrons can pass through the gaps between DNA strands without interacting, some high-energy heavy ions leave an ionization trail that kills almost every cell they pass through. So, space suit has as many as 20 layers to protect against heat, cold, pressure, and mostly space radiation. Modern materials such as aluminum, Mylar, Dacron, nylon, and Teflon constituted the many layers [14], [17].

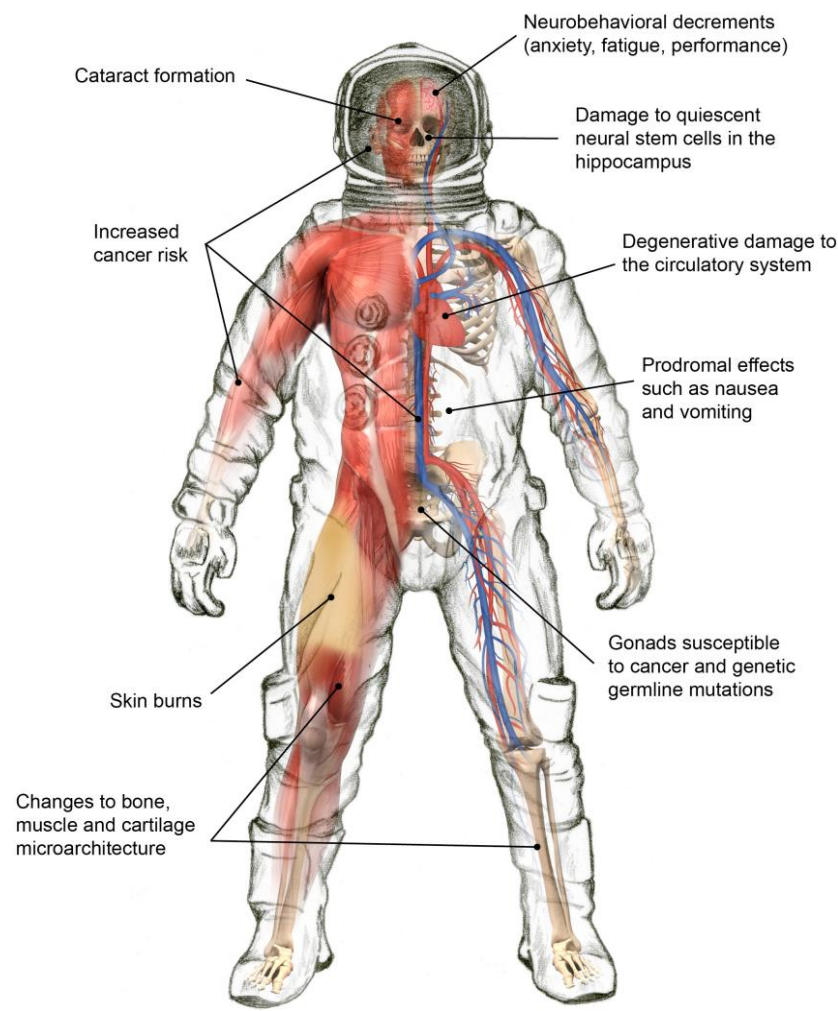

*Figure 1.8 Exposure to space radiation affects multiple organs and physiological systems in complex ways. NASA categorizes the biomedical consequences into four risk areas. Select health effects due to space radiation exposures [14].*

## *1.2.7 Shielding of the accelerator infrastructure*

The nuclear reaction of heavy ions with the accelerator components or target produces a variety of secondary particles in a heavy ion accelerator. Among all secondary particles, neutrons are the most abundant products. Furthermore, because neutrons cannot directly cause ionization, they may have the potential to affect a vast area. In cancer therapy, this means that the neutrons would influence both healthy tissue and tumor in a patient body. As a result, neutrons are the most significant factor to consider in both the safety assessment of heavy ion therapy and environmental protection and shielding of the accelerator infrastructure. A thorough understanding of the neutron yield could be particularly beneficial in determining the failure rate of the front-end electronics inside the accelerator hall [18]. The neutron energy spectra are a basic measurable quantity that is required for the protection of the accelerator from radiation. The barriers of the treatment rooms should be considered with the patient as a secondary radiation source. Despite the fact that the patient's body is mainly made up of light elements that produce secondary particles with a low yield per primary, it can be a significant source of secondary radiation since it totally absorbs the treatment beam [19]. So, it is essential to minimize the unwanted secondary neutron exposure in passive radiotherapy without affecting the clinical beam in order to lower the risk of secondary cancer [20].

From the above discussion on the significance of heavy ions, we now know that the propagation and interaction of intermediate and high energy heavy ions in the material is a topic of interest in the fields of cosmology, radio-biology, and radiation protection.

In the case of shielding space radiation, among different heavy ions, $^{12}$C has a significant number of fluxes as space radiation components which need to be counted from the radiation protection point of view for the astronauts.

In both cases, a better understanding of the mechanism of carbon ion interaction with different target nuclei is very important. The scattering angle of $^{12}$C due to multiple Coulomb interactions inside the traveling medium is the key interest to understanding the interaction mechanism. Furthermore, there are some calculation codes to assess the interaction scenario by computer simulation. In these codes, some theoretical model has been used to imply the physical process during heavy-ion interactions [21]. From using these models, we can assess the physical parameters that can help to understand the interaction mechanism, like scattering angle, total charge changing cross-sections, etc. For assessing by simulation, we need an accurate theoretical model. At present, there are many calculation codes, but the result differs from each other. Also, there are discrepancies between calculation results and experimental results [22]. So, the researchers have been trying to find out the experimental results for angular measurement for the nucleus-nucleus interaction for several decades. But still, there is lacking experimental data for angular measurement of the projectile and fragments produced during heavy ion interactions. A recent study shows that the higher precision of angular measurement of heavy-ion like $^{12}$C interactions with different target materials can be achieved by using the plastic nuclear track detector, CR-39 [23].

## 1.3 Objective of the thesis

Since carbon ion has various aspects of use, it is very important for us to understand the interaction mechanism of $^{12}$C with different target materials. The interaction of heavy ions is mainly governed by the total reaction cross-section, which is one of the most fundamental quantities for understanding the nucleus-nucleus interaction mechanism but is difficult to measure experimentally due to various reaction channels [24]. Instead, since most reactions undergo a proton change, so the total charge changing cross-section ($\sigma_{TCC}$), which measures the probability of changing the charge (proton) of a projectile, is the alternative way to understand the reaction cross section. So, in this study, we focus on the measurement of $\sigma_{TCC}$ for the $^{12}$C+Al interaction in the energy range below 100 MeV/n which is very crucial. Al is used for the hull of the spaceship. A precise and reliable database of total charge changing cross sections of carbon ion is a vital tool for both radiotherapy and space radiation protection. Nuclear track detectors have been utilized successfully in a variety of fields for a long time. CR-39 $(C_{12}H_{18}O_7)_n$, a track etch detector, is used in a variety of experiments, including searches for heavy ion fragmentation processes, magnetic monopoles, and heavy ionizing particle production. CR-39 has several advantages, like the detection of multi-fragments from a single interaction, detection of um range fragments over the conventional active detectors Si, NaI, PSD, etc. So, we used the CR-39 track detector for the measurement of the total charge changing cross-section in this study.

# Chapter 2
# Theoretical Background & Literature Review

## 2.1 Introduction

The importance of fragmentation of heavy ions at intermediate and high energy is particularly in the fields of astrophysics, nuclear physics, radiobiology, and radiation protection. Fragmentation research make to understand not only the propagation process of galactic cosmic rays but also the fundamental nuclear physics processes involved in nuclear collisions and fragmentation. Nucleus–nucleus interactions at lab energies ≤10 A GeV, which are critical for radiotherapy and radioprotection purposes, have recently received a lot of attention [25]. The nuclear reaction of heavy ions with the accelerator components or target produces a variety of secondary particles in a heavy ion accelerator by the effects of fragmentation. Total fragmentation cross-section is an essential physical quantity for understanding the nuclear structure and its behavior during high-energy collisions [26], [27]. Therefore, it is important to have precise values of cross sections explaining the production of fragments by heavier ions interacting with the target. The source abundances are critically dependent on the accuracy of the estimated cross-section. Measuring all the required cross sections is an enormous task that never be completed. The cosmic radiation arriving at earth's orbit has its highest intensity in the energy range of $10^2$-$10^4$ MeV/nucleon, and the nuclear component consists of ~ 87% H, ~12% He, and the remaining 1% heavier nuclei ranging from carbon to actinium [28]. Even though the heavy charged particles have a lower abundance than protons and helium, these are considered to be crucial because of their higher relative biological effectiveness. Carbon ions are one of the most abundant components among the heavy component of GCR, and they play a significant role in radiation therapy and radiation shielding [29]. The partial cross sections of projectile fragment production are not well in agreement with each other for the same kind of fragment at the near same beam energies. A brief description of some important theoretical background topics related to this research is described in this chapter.

## 2.2 Electromagnetic and Atomic interactions

Ionization and excitation effects, as well as Coulomb scattering, are generally the most important atomic interactions of charged particles with matter. Depending on the proximity of the encounter, secondary particles such as neutrons interacting with matter first produce charged particles which transfer energy mostly through excitation and ionization [30]. Excitation occurs when the charged particle can transfer energy to the atom, raising electrons to a higher energy level. Where ionization can occur when the charged particle has enough energy to remove an electron (Fig. 2.1). This results in a creation of ion pairs in surrounding matter. Interaction of heavy ion is followed by two main process, called the electromagnetic and nucleus-nucleus interactions while propagating through the matter. The theory of electromagnetic interaction is well developed but no concreter theory for the nucleus-nucleus collision yet. Models for nucleus-nucleus interaction have been still under progress through new experimental data. Some of the fundamental features that are necessary to understand energy deposition and dissipation due to energy losses of particles moving through matter will be discussed.

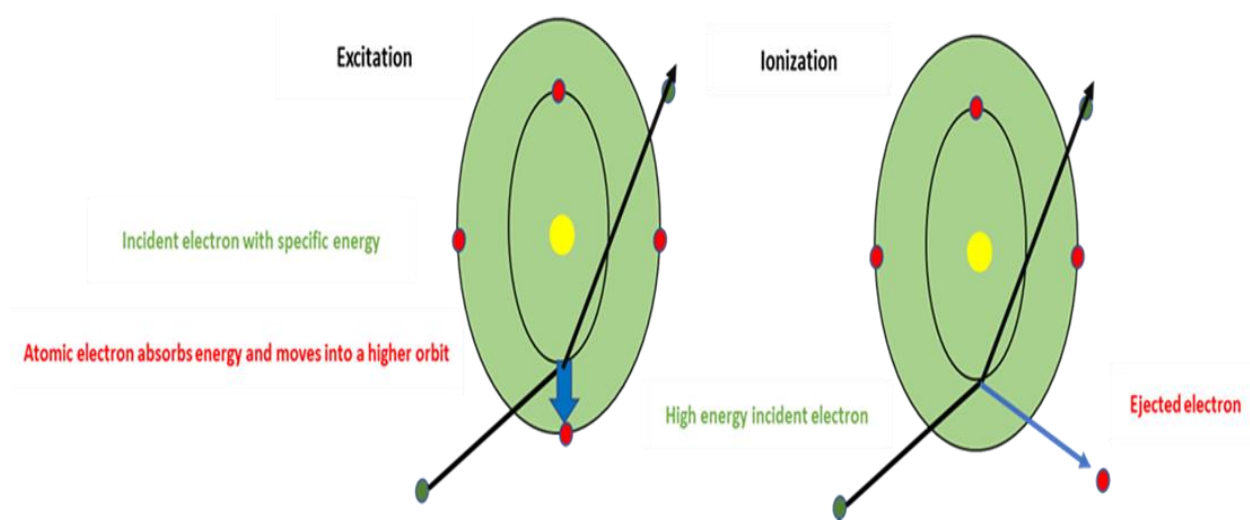

*Figure 2.1 Excitation occurs when the charged particle can transfer energy to the atom, raising electrons to a higher energy level. Where ionization can occur when the charged particle has enough energy to remove an electron. Excitation and Ionization process.*

### 2.2.1 Nature of the interaction

The coulomb forces between the positive charge of the heavy charged particles and the negative charge of the orbital electrons within the absorber atoms are the principal mode by which they interact with matter. Although interactions between the particle and nuclei (as in Rutherford scattering) are possible, such contacts are rare and have little impact on the response of radiation detectors. Charged particle detectors, on the other hand, must rely on the results of electron interactions for their response.

When a charged particle enters an absorbing medium, it interacts with a large number of electrons simultaneously. As the particle passes in its vicinity, the electron experiences an impulse from the attractive coulomb force. This impulse may be sufficient to either raise the electron to a higher-lying shell within the absorber atom (excitation) or to eliminate the electron completely from the atom, depending on the proximity of the encounter (ionization). The velocity of charged particle is decreased as a result of the interaction since the energy transferred to the electron must come at the expense of the charged particle. In a single collision, the maximum energy transfer from a charged particle of mass m with kinetic energy E to an electron of mass $m_o$ is $4Em_0/m$, or roughly 1/500 of the particle energy per nucleon. Because this is such a small fraction of the overall energy, the initial particle must lose energy in a series of interactions when it passes through an absorber. Because the particle is constantly interacting with a large number of electrons, the net effect is to reduce the particle's velocity until it stops.

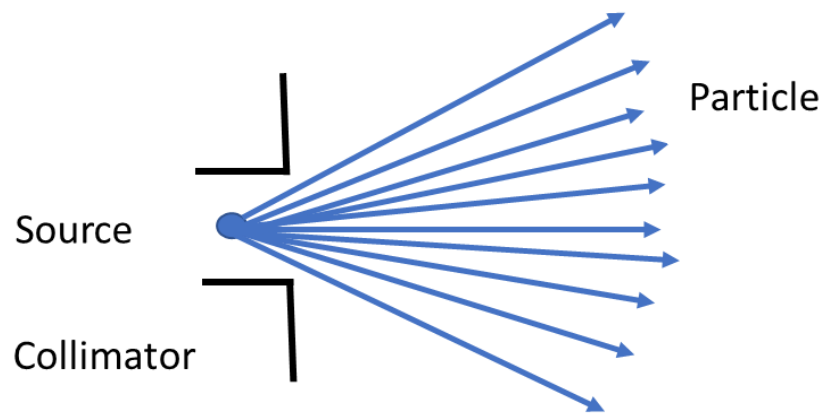

*Figure 2.2 The illustration shows heavy charged particles in their slowing down process. The particle is not much deflected by any one encounter, and interactions occur in all directions at the same time, the tracks tend to be relatively straight, except at the very end.*

In the diagram above, the paths taken by heavy charged particles in their slowing down process are schematically illustrated. Because the particle is not much deflected by any one encounter, and interactions occur in all directions at the same time, the tracks tend to be relatively straight, except at the very end. In a given absorber material, charged particles are therefore characterized by a definite range. The range, which will be explained in further detail below, is the maximum distance beyond which no particles will penetrate. Excited atoms or ion pairs are the products of these encounters in the absorber. A free electron and the corresponding positive ion of an absorber atom from which an electron has been completely removed constitute each ion pair. Ion pairs have a natural tendency to recombine to form neutral atoms, but this recombination is prevented in some types of detectors, allowing the ion pairs to be employed as the base of the detector response.

In particularly close encounters, an electron may experience a significant enough impulse that it retains enough kinetic energy to produce more ions after leaving its parent atom. These energetic electrons, also known as delta rays, are an indirect method of transferring the charged particle energy to the absorbing medium. The majority of the energy loss of the charged particle occurs via these delta rays under standard conditions. Because the range of delta rays is always small in comparison to the range of the incident energetic particle, ionization occurs along the primary track. On a microscopic scale, one impact of this process is that the ion pairs do not generally appear as randomly spaced single ionizations, but rather as many "clusters" of multiple ion pairs spread along the particle's track [27], [31].

## *2.2.2 Elastic scattering*

The numerous collisions with atomic electrons go through by a heavy charged particle as it traverses material affect its direction negligibly because of their large mass difference. The atomic nuclei of the target, however, may scatter a heavy charged particle. The deflection of the projectile in the repulsive Coulomb field of target nuclei is responsible for the majority of these elastic scatter processes.

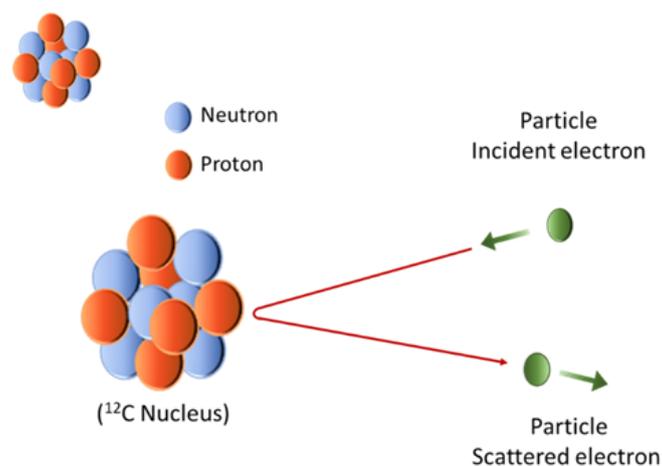

*Figure 2.3 The illustration shows elastic scattering of electron after colliding with a carbon nucleus. This scattering occurs when there is no loss of energy of the incident primary electron. Elastically scattered electrons can change direction but do not change their wavelength.*

As heavy ion beams pass through materials, elastic scattering leads them to broaden. Multiple Coulomb scattering models can well represent the effect of many single scatter events on a particle beam [36]. The projectile can be elastically scattered by the short-range attractive nuclear force if it moves very close to a target nucleus (as close as a few fm).

### 2.2.2.1 Coulomb/Rutherford scattering

The interaction is driven by Coulomb forces and characterized by very large η (η>>1) when the projectile's incident energy is substantially below the barrier energy of the two interacting nuclei. As seen in Fig. 3, the scattering angle is dependent on the impact parameter. Large scattering angles are produced by small impact parameters, and vice versa.

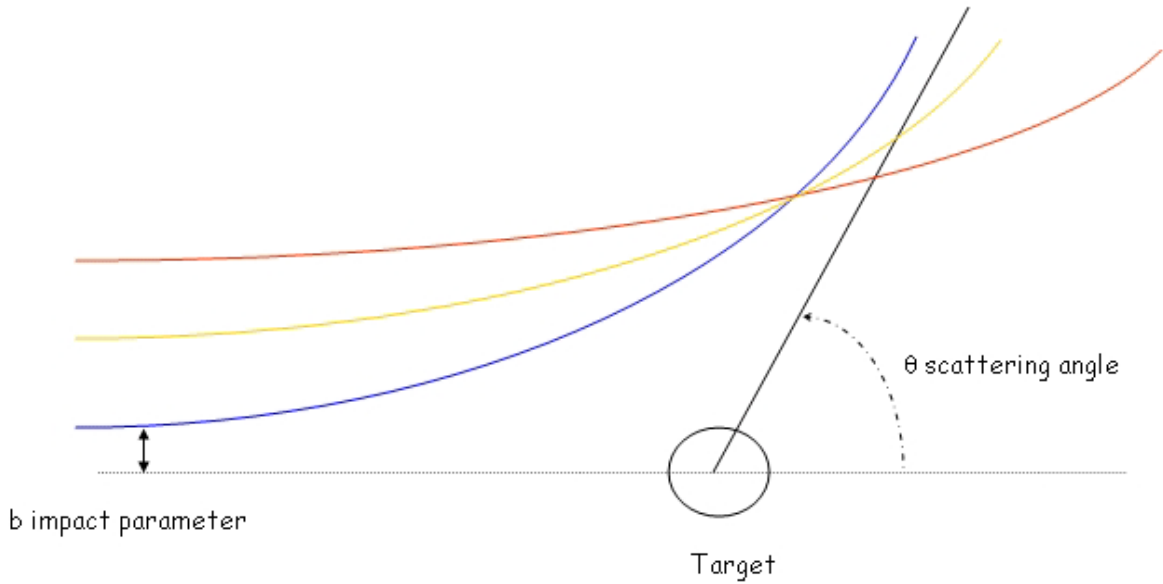

*Figure 2.4 Illustration shows the relation between impact parameter and scattering angle [32].*

The Coulomb force prevents the projectile and target getting close enough for nuclear interactions to take place. Rutherford scattering is the limit of pure Coulomb scattering and is given by:

$$\frac{d\sigma_{Ruth}}{d\Omega} = \left(\frac{Z_p Z_t e^2}{4E}\right)^2 \frac{1}{\sin^4 \frac{\theta}{2}} \qquad (2.1)$$

where θ and E are both in the center of mass frame. As the incident energy climbs, some part of the incident flux interacts through the nuclear potential. Nuclear reactions and significant absorption from the elastic channel are also caused by the presence of the nuclear potential [33]. This phenomenon is further divided into two types: Fresnel diffraction and Fraunhofer diffraction.

- *Fresnel diffraction*

Fresnel diffraction is characterized by n>>1 and occurs at incident energy slightly over the barrier. In a classical sense, distant trajectories far from the nuclear interaction region are solely scattered by the Coulomb potential, and thus follow Rutherford scattering. Grazing collisions can cause comparatively closer trajectories to be scattered to the same scattering angle as certain distant trajectories. The elastic scattering angular distribution near the grazing angle at energies slightly above the barrier shows a noticeable peak as a result of interference between these two trajectories. Fresnel scattering is characterized by this peak.

- *Fraunhofer diffraction*

When the incident energy exceeds the Coulomb barrier such that n<1, the attractive nuclear potential begins to dominate the interaction between the two interacting nuclei. As the nuclear force is attractive, one can envisage scattering to an angle (θ, φ) through a pure Coulomb interaction (near-side scattering) or from a strong nuclear interaction for a trajectory originating from the opposite side of the target (far-side scattering). In the elastic scattering angular distribution, the interference between these two produces an oscillatory pattern with constructive and destructive interferences causing maxima and minima. The nuclear potential can absorb a fraction of the incoming trajectories. As a result, non-elastic components can be found in both Fresnel and Fraunhofer diffractions.

## 2.2.2.2 Multiple Coulomb scattering

An accumulation effect will occur as a result of many small-angle events, affecting the emittance of the beam. The cumulating effect of several slight deviations will result in an increase in emittance, which will affect the beam's lifetime. A simple formula can be used to approximate the r.m.s. value of scattering angle [32]:

$$\theta_{r.m.s.} \propto \frac{1}{p}\sqrt{\frac{L}{X_0}} \qquad (2.2)$$

where $L/X_0$ denotes the amount of material detected by the particle in terms of radiation lengths, and p denotes the momentum of incoming particle.

## 2.2.3 Bremsstrahlung

High-energy charged particles lose energy in interactions with the Coulomb field of the nuclei of the medium in addition to ionization loss. When charged particles are de-accelerated in the Coulomb field, bremsstrahlung occurs, in which a part of the particle's kinetic energy is emitted as photons. The energy loss due to bremsstrahlung is determined:

$$-\frac{dE}{dx} = 4.\alpha.N_A.\frac{Z^2}{A}.z^2.(\frac{1}{4\pi\varepsilon_0}\frac{e^2}{mc^2})^2.E.\ln(\frac{183}{Z^{1/3}}) \qquad (2.3)$$

where A and Z are atomic mass and atomic number of the medium, and m, E, and z are the mass, energy, and atomic number of the incident particle, respectively.

For electrons, the energy loss due to bremsstrahlung is expressed as

$$-\frac{dE}{dx} = 4.\alpha.N_A.\frac{Z^2}{A}.r_e^2.E.\ln(\frac{183}{Z^{1/3}})(\frac{MeV}{g/cm^2}) \qquad (2.4)$$

if $E \gg m_e . c^2 / \alpha . Z^{1/3}$.

The ionization loss beyond the minimum ionization is proportional to the logarithm of the energy, whereas the bremsstrahlung loss is proportional to the energy and inversely proportional to the squared mass of the incident particle [31]. The average energy loss by bremsstrahlung and ionization is the same at the so-called critical energy $E_c$:

$$-\frac{dE}{dx}(E_c)|ionization| = -\frac{dE}{dx}(E_c)|bremsstrahlung| \qquad (2.5)$$

For medium with $Z \geq 13$, the critical energy $E_c$ is approximately given as

$$E_c = \frac{(500)MeV}{Z} \qquad (2.6)$$

with Z the atomic number of the target material.

## 2.2.4 Nuclear fragmentation

In heavy nuclei collisions, the reaction cross section is particularly essential since it provides information about the size of the collision partners as well as the open channels [34]. A projectile nucleus is shot against a target matter during cross-section measurements and elastic or inelastic processes may occur between projectile and target nuclei. When elastic interactions occur, there is no change in the composition of the nuclei and the system's total kinetic energy is conserved. Inelastic interactions, on the other hand, have two outcomes. The first is that the projectile and target nuclei remain intact and the total kinetic energy is not conserved due to excitation processes occurring. The second outcome happens when the projectile and target collide, and one or both of them break apart, producing secondary nuclei. This nuclear reaction process is known as fragmentation and the secondary nuclei are called either projectile or target fragments, depending on which nucleus they originate from [35]. So, the reaction cross-section is the cross-section in which a nuclear process takes place.

### *2.2.4.1 Activation*

The majority of nuclear reaction fragments are unstable. After proton and heavy ion irradiation, the material becomes activated. Many of the radioactive nuclei produced are neutron-deficient isotopes, such as $^{11}$C, and thus $\beta^+$-emitters that can be used to verify in vivo range. Because the half-lives of the produced nuclides are so short, light materials (like biological tissue) often disintegrate within minutes or hours of irradiation. Longer-lived nuclides are produced in heavy materials like brass, steel, and lead, that could require the storage of irradiation metal components permanently. Protons do not fragment themselves; instead, they produce target fragments, however radioactive projectile fragments can be produced in nucleus-nucleus reactions. These projectile fragments contribute to the activation of the irradiated target if they stop inside it [36].

### *2.2.4.2 Influence on dose profiles*

Nuclear fragmentation reactions have a significant influence on protons and heavy ions dose profiles, particularly at high penetration depths. As a result, for radiation transport codes to be utilized for dose calculation, precise models to represent these reactions are an essential requirement. Fig. 2.5 depicts a spread out Bragg peak (SOBP) produced by $^4$He ions when different nuclear reaction cross sections $\sigma_R$ are used.

Following irradiation with different energies to cover the target volume with a homogeneous dose in depth, such SOBPs can be produced. The SOBP dose reduces as the nuclear reaction cross section increases because more ions fragment before reaching the Bragg peak depth, increasing the dose in the fragment tail, and vice versa. During real irradiation, a flat SOBP that has been designed with incorrect nuclear models can appear with a gradient [31].

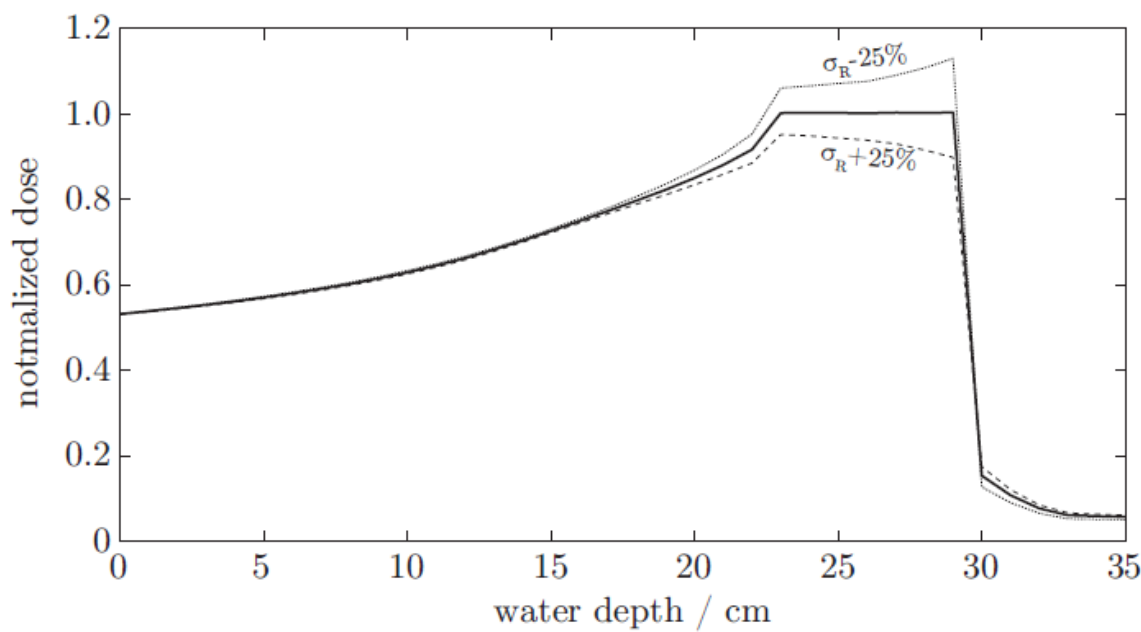

*Figure 2.5 The impact of a nuclear reaction model on a $^4$He SOBP in water was investigated using a simple transport model based on a one-dimensional forward computation [36].*

## 2.3  Concept of cross sections

In nuclear or particle physics, cross-section refers to the interaction of elementary particles, nuclei, or atoms with one another. The probability that two particles will interact when they approach each other is the cross-section for this process. This is the most general definition of cross-section. The cross-section can be thought of as the area within which a reaction will occur in a simple geometrical interpretation. As a result, the cross-sectional units are the same as the area units. The following definition was introduced in the early days of nuclear physics:

$$1 \text{ barn} = 10^{-24} \text{ cm}^2$$

During the early experiments the physicists discovered that the interactions were far more probable than expected. The nucleus was 'as big as a barn'. So, the cross-section has always been measured in barns. Partial or double differential cross sections are used to express the angular and energy distribution of particles produced in nuclear reactions with materials. The probability of producing a particle of energy E in the energy range E + dE during an interaction of particles with matter in the solid angle region between $\Omega$ and $\Omega + d\Omega$ is denoted by [30].

$$\frac{d^2\sigma}{dE} d\Omega$$

## *2.3.1 Total reaction cross sections in nucleus-nucleus reactions*

The target nucleus, as well as the projectile nucleus, can fragment in nucleus-nucleus reactions. The velocity of the target fragments is nearly zero and the velocity of the projectile is pretty much equal to that of the projectile ion in a simplified model. The abrasion-ablation model is a commonly used model to describe nucleus-nucleus collisions [37]. The fragmentation of the projectile into one $^4$He nuclei, a proton and a neutron is shown in the example reaction for the $^{12}$C nucleus + target nucleus collision, which has a higher probability due to the high stability of the $^4$He nucleus. Figure shows examples of nuclear reactions that are relevant to particle therapy. If the sum of the kinetic energies in the initial state is not equal to the final state, the reaction is considered to be inelastic. Inelastic reactions used for radiotherapy in the energy range generally lead to fragmentation of the interacting projectile and target nuclei [36], [37].

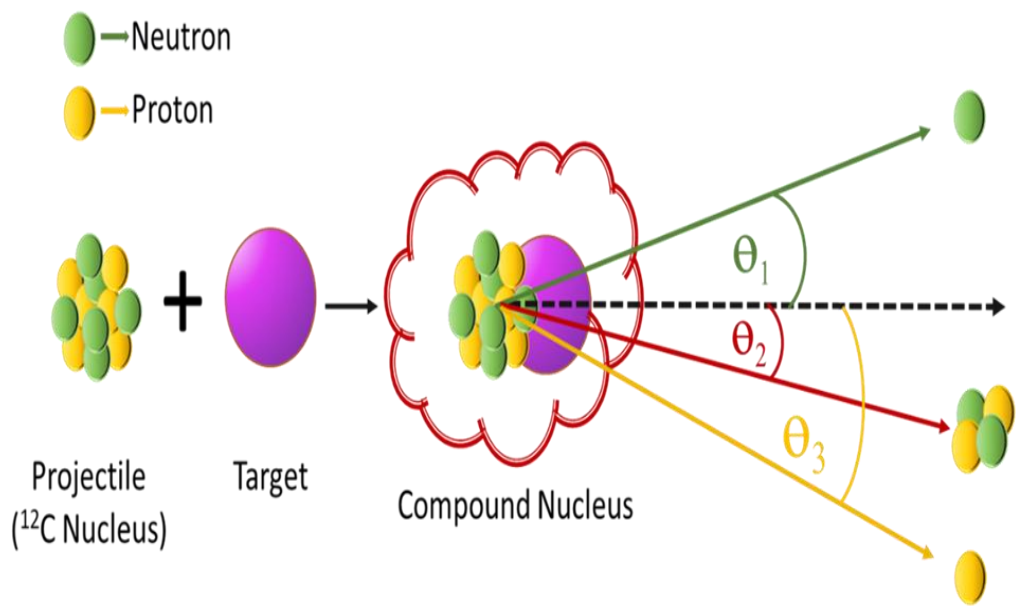

*Figure 2.6 Illustration of nuclear fragmentation reaction channel by incident $^{12}$C ion projectile colliding with target nuclei.*

Both experimental and theoretical research has been done on total reaction cross sections. The total reaction cross section σ(N, N)$_R$ is defined as

$$\sigma(N,N)_R = \sigma_{total} - \sigma_{elastic} \qquad (2.7)$$

with σ$_{elastic}$ the elastic cross section and σ$_{tot}$ the total cross section. Several empirical parameterizations as derived for nucleus–nucleus collisions are given below.

- ***The Sihver formula***

The simplest form of Sihver formula:

$$\sigma(N,N)_R = \pi r_0^2 [A_{projectile}^{1/3} + A_{target}^{1/3} - b_0[A_{projectile}^{-1/3} + A_{target}^{-1/3}]]^2 \quad (2.8)$$

Where $A_{target}$ and $A_{projectile}$ are the atomic mass numbers of the target nuclei and projectile. Here $r_0$ is 1.36 fm and $b_0$ is given by:

$$b_0 = 1.581 - 0.876(A_{projectile}^{-1/3} + A_{target}^{-1/3}) \quad (2.9)$$

The formula consists of a transparency parameter and a geometrical term ($A^{1/3}_{projectile}$ + $A^{1/3}_{target}$) for the nucleons in the nucleus. The cross section is considered to be independent of incident energy of less than 100 MeV/nucleon. For most of the collision systems, the formula agrees well with experimental evidence [30], [38].

In the case of nucleon–nucleus interactions, formula (2.9) is almost indistinguishable for proton–nucleus (with $Z_{target} \leq 26$) if $A_{projectile} = 1$ is a proton and the parameter $b_0$ is expressed as a polynomial function of the first order in $(1 + A^{-1/3}_{target})$. The formula then becomes

$$\sigma(p,N)_R = \pi \cdot r_0^2 [1 + A_{target}^{1/3} - b_0[1 + A_{target}^{-1/3}]]^2 \quad (2.10)$$

$$b_0 = 2.247 - 0.915(1 + A_{target}^{-1/3}) \quad (2.11)$$

For incident proton energies $E_{proton} \geq 200$ MeV and $r_0 = 1.36$ fm.

- ***The Kox formula***

The Kox formula is developed on the strong absorption model to describe low energy nuclear reactions. The formula is based on the analysis of total reaction cross sections for heavy ion collisions in the intermediate-energy range of about 10–300 MeV/nucleon. This formula determines the total reaction cross section in terms of the nucleus-nucleus interaction barrier, the interaction radius and the center of mass energy of the colliding system [30], [39]. In this framework, a reaction is defined as the occurrence of a substantial interaction between nuclear materials. The general form of this formula:

$$\sigma(N,N)_R = \pi(IR)^2 [1 - \frac{B_C}{E_{c.m.}}] \quad (2.12)$$

Here $B_C$ is a parameter which is the Coulomb barrier of the projectile–target system and is given by:

$$B_C = (Z_{projectile}.Z_{target}.e^2)/(r_C(A^{1/3}_{projectile} + A^{1/3}_{target})) \quad (2.13)$$

where $Z_{target}$ and $Z_{projectile}$ are the atomic numbers of the target nuclei and the projectile, e is the electron charge, $r_C$ = 1.3 fm, $A_{target}$ and $A_{projectile}$ are the atomic masses of the target nuclei and the projectile, respectively. The Kox formula divides the interaction radius IR into volume and surface terms, $R_{vol}$ and $R_{surf}$:

$$IR = R_{vol} + R_{surf} \quad (2.14)$$

The terms $R_{vol}$ and $R_{surf}$ correspond to energy-independent and energy-dependent components of the reactions, respectively. Collisions with smaller impact parameters are independent of energy and mass number and are characterized as a volume component of the interaction radius and therefore depends only on the volume of the projectile and the target nuclei. It is given by

$$R_{vol} = r_0(A^{1/3}_{projectile} + A^{1/3}_{target}) \quad (2.15)$$

The second term of the interaction radius IR in Eq. (2.14) is a nuclear surface contribution and is parameterized by

$$R_{surf} = r_0[a.\frac{A^{1/3}_{projectile}.A^{1/3}_{target}}{A^{1/3}_{projectile} + A^{1/3}_{target}} - c] + D \quad (2.16)$$

The first term in the brackets is the mass asymmetry term, which is related to the volume overlap of the projectile and the target. The second term c is an energy dependent parameter, which takes into account the increasing surface transparency as the projectile energy increases. The parameter D is the neutron excess which is important in collisions with heavy or neutron-rich targets. It is given by

$$D = \frac{5(A_{target}.Z_{target})Z_{projectile}}{A_{projectile}.A_{target}} \quad (2.17)$$

The parameters $r_0$ = 1.1 fm and a = 1.85 are fixed values where the parameter c is a function of the beam energy per nucleon which is a simple analytical function and used in GEANT4 to

$$c = -\frac{10}{x^5} + 2.0 \quad \text{for x > 1.5,}$$

$$c = (-\frac{10}{1.5^5} + 2.0).(\frac{x}{1.5})^3 \quad \text{for } x < 1.5,$$

$$x = \log(E_{kin}),$$

where $E_{kin}$ is the kinetic projectile energy in unit MeV/nucleon in the laboratory system.

- ***The Tripathi formula***

Tripathi developed different algorithm for computing interaction cross sections for nucleon–nucleon and nucleus–nucleus interactions. The formula provides a basic universal parameterization of the total reaction cross section for any system of colliding nuclei in the energy range of a few A MeV to a few A GeV [30], [40]. The general form of this formula is given by:

$$\sigma(N,N)_R = \pi r_0^2 . [A_{projectile}^{1/3} + A_{target}^{1/3} + \delta_E]^2 . [1 - \frac{B}{E_{c.m.}}] \quad (2.18)$$

Here, the value of $r_0$ is 1.1 fm and B is a parameter which is the energy-dependent Coulomb barrier given by:

$$B = 1.44 . \frac{Z_{projectile} . Z_{target}}{R} \quad (2.19)$$

$$R = r_{projectile} + r_{target} + \frac{1.2[A_{projectile}^{1/3} + A_{target}^{1/3}]}{E_{c.m.}^{1/3}} \quad (2.20)$$

where $r_i$ is the equivalent sphere radius and is related to the $r_{rms,i}$ radius by $r_i = 1.29 \cdot r_{rms,i}$ with i = projectile, target), and $E_{c.m.}$, the center of mass energy, is given in MeV. There is an energy dependence of the reaction cross section at intermediate and higher energies mainly due to two effects, the transparency and the Pauli blocking. This is represented by the energy-dependent term $\delta_E$ in formula (2.18) given as

$$\delta_E = 1.85 S_1 + 0.16 \frac{S_1}{E_{c.m.}^{1/3}} - C_E + 0.91 S_2 \quad (2.21)$$

$$S_1 = \frac{A_{projectile}^{1/3} . A_{target}^{1/3}}{A_{projectile}^{1/3} + A_{target}^{1/3}}$$

$$S_2 = \frac{(A_{target} - 2Z_{target})Z_{projectile}}{A_{projectile} \cdot A_{target}}$$

where $S_1$ is the mass asymmetry term and is related to the volume overlap of the collision system and the term $S_2$ in the formula is related to the isotope dependence of the reaction cross section.

The term $C_E$ in formula (2.21) is related to the transparency and the Pauli blocking and is given by

$$C_E = D_{Pauli} \cdot (1 - \exp(-\frac{E}{40})) - 0.292 \cdot \exp(-\frac{E}{792}) \cdot \cos(0.229 \cdot E^{0.453}) \qquad (2.22)$$

The parameter $D_{Pauli}$ in formula (2.22) takes into account the density dependence of the colliding system, scaled with respect to the density of the ($^{12}C + ^{12}C$) colliding system.

$$D_{Pauli} = 1.75 \cdot (\rho A_{projectile} + \rho A_{target}) / (\rho A_C + \rho A_C) \qquad (2.23)$$

where the density of a nucleus using the hard sphere model with a given nucleus of mass number $A_i$ is given by

$$\rho A_i = A_i / (\frac{3}{4} \pi r_i^3) \qquad (2.24)$$

where the radius of the nucleus $r_i$ is defined above with the root-mean-square radius, $(r_i)_{rms}$, obtained directly from experimental data. With the parameter $D_{Pauli}$ simulates the modifications of the reaction cross section caused by the Pauli blocking which was introduced by Tripathi in the parameterization formula for the first time. At lower energies where the overlap of interacting nuclei is small and Coulomb interaction modifies the cross sections significantly, the influence of the Pauli blocking is small. The modification of the reaction cross section due to Pauli blocking plays an important role at energies above 100 MeV/n in nucleus–nucleus collisions which lead to higher densities. For proton–nucleus collisions where the compression effect is low, a single constant value $D_{Pauli} = 2.05$ gives good results for all proton–nucleus collisions.

For alpha–nucleus, where there is also little compression, the following terms for the $D_{Pauli}$ parameter are useful:

$$D_{Pauli}(\alpha) = 2.77 - (8 \cdot A_{target} \cdot 10^{-3}) + (1.8 \cdot A_{target}^2 \cdot 10^{-5}) - 0.8 / [1 + \exp[(250 - E)/75]] \qquad (2.25)$$

## *2.3.2 Total charge changing cross sections*

The total reaction cross-section, $\sigma_R$ as a function of energy, E determines the probability of a nuclear reaction, is one of the most basic parameters for heavy-ion transport calculations. However, it is extremely difficult to assess all possible reaction channels during an experiment due to the various reaction channel [24],. Many experiments determine charge-changing cross-sections $\sigma_{cc}$, which show the probability that atomic number ('charge') of the projectile changes because an accurate charge identification is straightforward with simple particle detection systems. The charge-changing cross-section is a decent approximation to the reaction cross-section since most nuclear fragmentation channels result in the loss of at least one proton. Most heavy-ion projectiles, on the other hand, can undergo neutron-removal reactions (for example, fragmentation of $^{12}C$ into $^{10}C$ or $^{11}C$), where the charge is unchanged [35], [41].

total charge changing cross section ($\sigma_{TCC}$) that is dominated by the total cross section ($\sigma_T$) and is defined as

$$\sigma_{TCC} = \sigma_T - \sigma_{el} - \sigma_{nr} \quad (2.25)$$

where $\sigma_{el}$ is the elastic cross section and $\sigma_{nr}$ is the neutron removal cross section. From above these three cross sections (total cross section, elastic cross section, neutron removal cross section), we can get the formula of total charge changing cross section.

Cross section ($\sigma$) tell us how frequently particle interact with travelling medium. Mathematically it has the dimension of area but physically it describes the probability of interaction. If dN is the number of particles that scattered per unit time through per unit solid angle $d\Omega$ and $\Phi$ is the flux of the particle that travel through the medium, then the differential cross section can be defined as the

$$\frac{d\sigma}{d\Omega}(E,\Omega) = \frac{1}{\Phi}\frac{dN}{d\Omega}$$

The total cross section at certain energy can be obtained by integrating the above equation, i. e.

$$\sigma(E) = \int \frac{d\sigma}{d\Omega} d\Omega$$

The unit of the cross section is barn (b) with $1b = 10^{-24}$ cm$^2$. And the total charge changing cross-section (TCC) can determine the probability of production of fragment or number of projectiles that changes the charge to become fragment. The formula for TCC can be derived from the definition of mean free path ($\lambda_m$) of a particle inside the travelling medium. The mean free path is the average distance of the particle that travel through a medium between two collisions with the medium. In other word mean free path is the probability of the particle interaction [42]. So, it can be defined as

$$\lambda_m = \frac{1}{n\sigma_{tcc}} \qquad (2.26)$$

Where n is the number density of the travelling medium and $\sigma_{tcc}$ is the total charge changing cross section of the particle in that medium. The number density can be obtained from the following relation

$$n = \frac{N_A \rho}{M} \qquad (2.27)$$

$$\therefore \lambda_m = \frac{M}{N_A \rho \sigma_{tcc}} \qquad (2.28)$$

Where M is the atomic/molecular mass number (g/mole) and ρ (g/cm$^3$) is the density of the material and $N_A$ is the Avogadro`s number. Now the number of changes of particle in per unit thickness of the medium can be written as

$$-\frac{dN}{dx} = \frac{N}{\lambda_m}$$

$$\Rightarrow \frac{dN}{N} = n\sigma_{tcc}\, dx \qquad [Using\ equation\ (2.26)]$$

$$\Rightarrow \ln(N) = -n\sigma_{tcc}\, x + C \qquad [by\ integrating\ ] \qquad (2.29)$$

If the number of particles before travelling the thickness of medium is $N_{in}$ and after travelling the thickness x is $N_{out}$ then the attenuation of the particle with the thickness (x) can be expressed as

$$\Rightarrow N_{out} = N_{in}\, e^{-n\sigma_{tcc} x}$$

$$\Rightarrow \sigma_{tcc} = \frac{1}{nx} \ln\left(\frac{N_{out}}{N_{in}}\right)$$

$$\therefore \sigma_{tcc} = -\frac{M}{N_A \rho x} \ln\left(\frac{N_{out}}{N_{in}}\right) \times 10^{27} \ (mb) \qquad [\,by\ using\ (2.28)] \qquad (2.3)$$

The equation (2.3) is the formula with which we can calculate the total charge changing cross section by using CR-39. The number of incoming and outgoing particle can be determined experimentally. The following figure, Fig 2.1 illustrate the experimental determination of $N_{in}$ and $N_{out}$. In this figure the blue arrow shows the projectile that come to the target and after travelling the target of thickness x, the number reduced which gives the $N_{out}$. The number which reduced from the incoming particle they interact with the target and produced fragments. The green and red arrow shows those fragments in different direction.

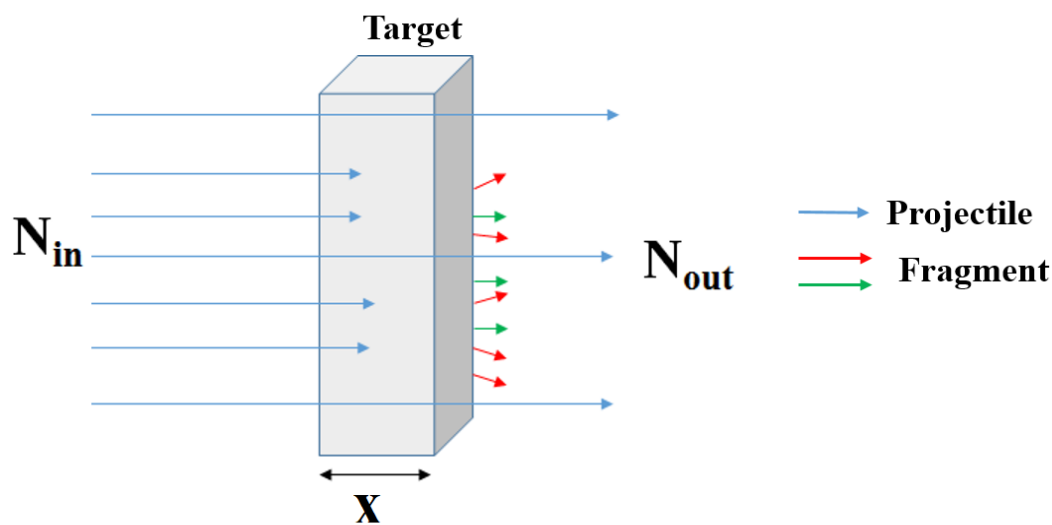

*Figure 2.7 Interaction of projectile with the target reduces the number of projectiles after the target. From these number of projectiles before and target charge changing cross section can be measured.*

## 2.4  Uncertainty estimation of $\sigma_{TCC}$

For calculation of TCC we will measure the x, $N_{in}$ and $N_{out}$ according to the equation (2.3). So, the source of uncertainty for the TCC are these three terms. The uncertainty measurement procedure depends on the analysis and getting the measurable value. The conventional error propagation method was used by the several groups like Westfall, Webber and C. Zeitlin [43]–[45]. Also S. Ota [46] in his study for precise measurement of total charge changing cross section ($\sigma_{tcc}$) used this method.

Let`s consider $N_{out} / N_{in} = \gamma$. Then according to the conventional error propagation method

$$\delta\sigma_{TCC} = \frac{M}{\rho N_A x} \sqrt{\frac{\delta\gamma}{\gamma} + \left\{\ln(\gamma) \times \frac{dx}{x}\right\}^2} \qquad (2.31)$$

According to the error in binomial distribution

$$\delta N_{out} = \sqrt{N_{in}\gamma(1-\gamma)} \qquad then$$

$$\delta\gamma = \frac{\delta N_{out}}{N_{in}} = \sqrt{\gamma(1-\gamma)/N_{in}} \qquad (2.32)$$

Therefore, the equation (2.32) become

$$\delta\sigma_{TCC} = \frac{M}{\rho N_A x} \sqrt{\frac{1-\gamma}{N_{in}} + \left\{\ln(\gamma) \times \frac{dx}{x}\right\}^2} \qquad (2.33)$$

which gives the uncertainty calculation for total charge changing cross section.

## 2.5 Energy and uncertainty of energy calculation

By using SRIM [47], the energy loss in each step in both CR-39 and Al has been calculated in this study. If $E_{in}$ is the energy of incident particle in some surface (Al/CR-39) and $E_{out}$ is the energy with which particle is come out from that material after travelling a specified thickness (Al/CR-39), then the energy for which total charge changing cross section will be calculated is

$$E\left(\frac{MeV}{n}\right) = \frac{E_{in} + E_{out}}{2} \qquad (2.34)$$

and the uncertainty of energy loss has been calculated as

$$\Delta E\left(\frac{MeV}{n}\right) = \frac{E_{in} - E_{out}}{2} \qquad (2.35)$$

## 2.6  Literature review

Carbon ions are a significant component of the GCR and of Solar Particle Events (SPEs), and they contribute to the dose delivered to astronauts on long duration space missions. Despite the fact that $^{12}$C ions are significantly less common in GCR than protons, they are crucial because of their high biological efficiency (RBE). As a result, $^{12}$C ions are being studied in a variety of experiments with various targets in the field of space shielding design as well as a cancer treatment. Aluminum is one of the important components of the materials surrounding many particle detectors in space and balloon experiments because of its medium mass. Furthermore, aluminum is a common structural material in spacecraft. As a result, a complete understanding of the interactions of various ions in the aluminum target will be critical for the corrections in the GCR fragmentation study in the material.

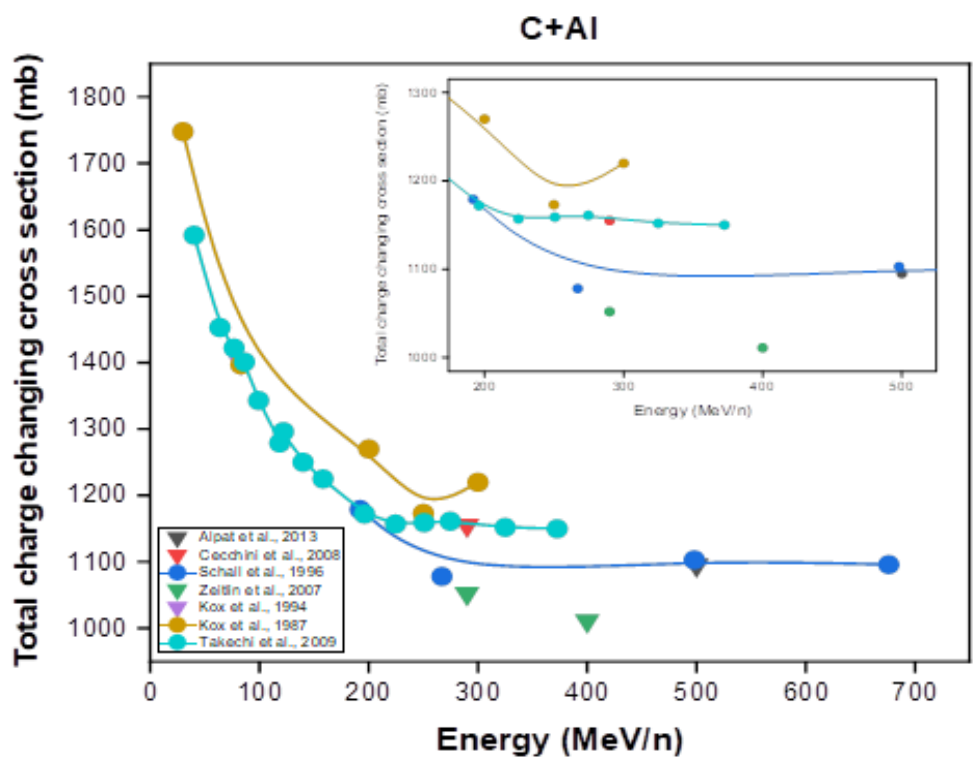

*Figure 2.8 Total charge changing cross section for energies (40-676) MeV/n $^{12}$C ion interacting with Al target. Different authors followed different theories of total charge changing cross section such as the parameterization of Sihver, Kox, Tripathi etc. So, there are discrepancies in the result of total charge changing cross section for different theories and calculation code.*

We know from the previous study that the total charge changing cross-section measurement for $^{12}$C ion interactions with Al has been conducted. But these studies are not still sufficient for the safety of astronauts or carbon therapy. Because of the limitation of detection efficiency and experiment methodology, we don't have enough accurate data for the total charge changing cross-section of $^{12}$C+Al. An accurate and precise database of the $^{12}$C+Al interaction cross-sections will be an important tool for both carbon therapy and space radiation protection. We collected total charge changing cross-sections of $^{12}$C+Al in energies ranging from 40 MeV/n to 676 MeV/n from previous studies. Different studies followed different theories of total charge changing cross-section, such as the parameterization of Sihver, Kox, Tripathi, etc. We have discussed these parameterizations in the above section (see section 2.3.1). There are discrepancies in the result of total charge changing cross-section for different theories and calculation codes. Also, there are lacking data specially in the energy above 400 MeV/n that need for further studies in particle therapy and space radiation shielding. The data from different authors (see: Alpat et al., 2013; Cecchini et al., 2008., Schall et al., 1996; Zeitlin et al., 2007; Kox et al., 1994; Kox et al., 1997; Takechi et al., 2009) are plotted for comparison their result [22], [48], [49], [50].

This study will measure the total charge changing cross-section for $^{12}$C ion interacting with Al in the energy below 100 MeV/n. These energy ranges of carbon ion beams are relevant to cancer therapy and the study of particle spectra in space radiation. We want to compare our results with those previous data that will show the effectiveness and reliability of this study.

# Chapter 3
# CR-39 Track Detector & Energy Loss Method

## 3.1 Introduction

When an ion encounters a nuclear track detector foil, it causes damage at the level of molecular bonds (a few tens of nanometers) along its path as a result of the excitation and ionization of atoms, creating the so-called "latent track." The latent tracks in crystals are atomic displacements, whereas the damage in plastic materials is caused by broken molecular chains that release free radicals (see Fig.3.1). Etch-pit cones are formed by an appropriate chemical etching of the detector [51].

In the following ways, a charged particle can transfer its energy to the insulating material:

- Electron displacements cause ionization and excitation of material atoms and molecules.
- Recoil tracks are caused by a nuclear interaction that is both elastic and inelastic with the target atoms.

Latent tracks can be viewed under a laboratory optical microscope once a proper chemical solution enlarges them. This is known as "detector etching" or "track visualization."

Aqueous solutions of NaOH or KOH with concentrations ranging from 1 to 12 N and temperatures ranging from 40 to 90 $^0$C are the most typical etchants for polymers. The majority of the etching was done using 47 % HF acid on glasses, mica, and minerals. For high REL particles, short etching periods are sufficient, while long etching times are necessary for low REL particles. The charge, mass, and velocity of the incident particle, as well as track detecting conditions, chemical solution type, normality, and temperature of the etchant, all influence the geometry of an etched track in a particular nuclear track detector. CR-39 NTDs have been used to examine nuclei with Z/β values ranging from 5 to 83. CR39 has the best sensitivity, resolution, and optical qualities of any other track recording detector. The surface of the detector stays smooth even after etching away 500 μm of thickness [52]. It is widely utilized in heavy ion studies, such as cosmic ray composition, heavy ion nuclear processes, radiation exposure due to heavy particles, extra-heavy element explorations, and the searching for magnetic monopoles, among other things. We will briefly discuss the CR-39 track detector in next section.

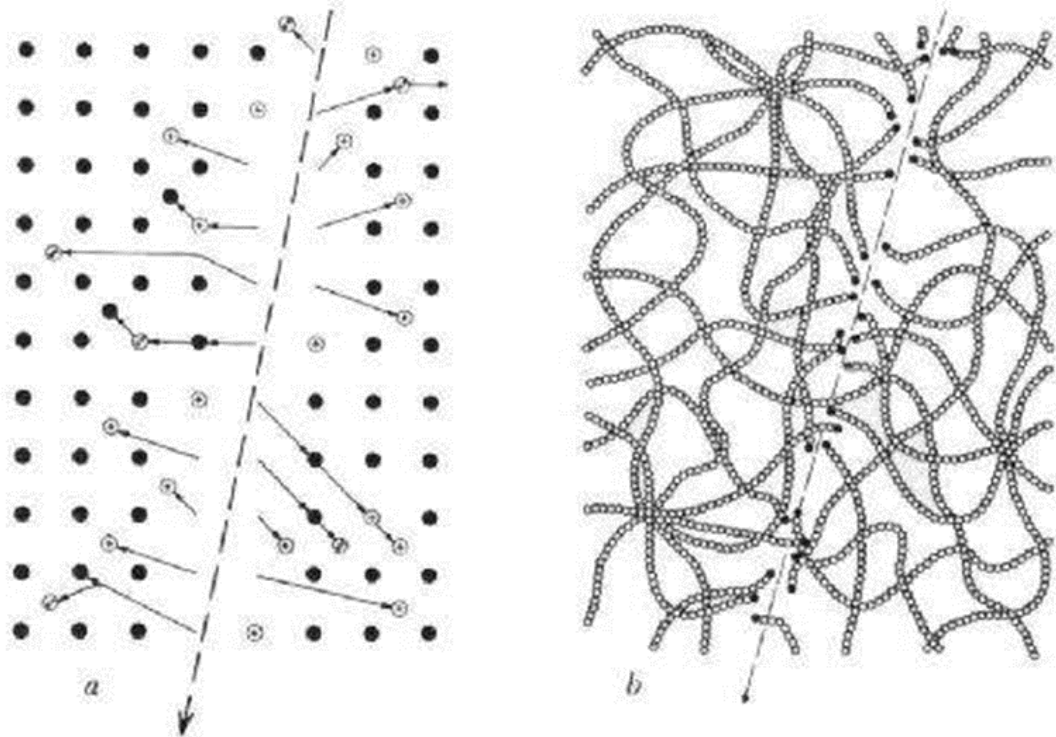

*Figure 3.1 A latent track is formed in (a) inorganic solids by the ionization and production of dense positive ions along the ion path and (b) the breakage of polymeric chains by the passage of charged particle [52].*

## 3.2 CR-39 track detector

The continuous development of ultrashort, high-intensity lasers contributes a perspective for an extensive range of applications in the field of intense-laser driven particle acceleration. Many of these applications necessitate well-defined ion beams with controlled parameters. As a result, it is necessary to develop instrumentation for accurate diagnosis of the spectral and spatial characteristics of these beams. Electrostatic, magnetic and magneto-electrostatic mass spectrometers have all been effectively used to measure the spectral distribution of various ion species over a wide energy range. The spatial characteristics of the beams, on the other hand, are often lost in the process. In order to estimate emittance, divergence, and source characteristics of ion beam, multiple film stacks, such as image plates, scintillators, radio chromic films, and track detectors are usually used. The energy-resolved, spatial profile of the beam is provided by these stacks, but they are unable to differentiate between different ion species. Furthermore, an absolute calibration is required to pursue quantitative measurements [53].

The size and shape of a track in ion track detectors depends on species, velocity($v_i$) and charge($Z$) of the ion a well as on the stopping power of the matter. In the current paradigm of solid state nuclear track detectors (SSNTDs), the CR-39 is a highly sensitive charged particle detector. CR-39 is Columbia Resin #39. It is allyl diglycol carbonate ($C_{12}H_{18}O_7$). The chemical form of CR-39 is illustrated in figure (1-1). Cartwright was the first to discover this detector, which is made up of short polyallyle chains connected together by carbonate and ethylene glycol groups to form a dense three-dimensional network. It's a high-grade plastic with excellent light transmission, a refractive index close to 1.50, and a density of 1.32 grams per cubic meter. It can withstand a sustained temperature of $100^0C$. In heating experiments, it was discovered that when CR-39 is heated above $100^0C$, it loses its transparency. When heated to $200^0C$, it becomes opaque, despite the fact that its melting temperature is higher than $200^0C$. Long-term storage of CR-39 detectors in refrigerators barely marginally degrades their properties. In radiation measurement laboratories, CR-39 is chemically stable against water and other common liquids.

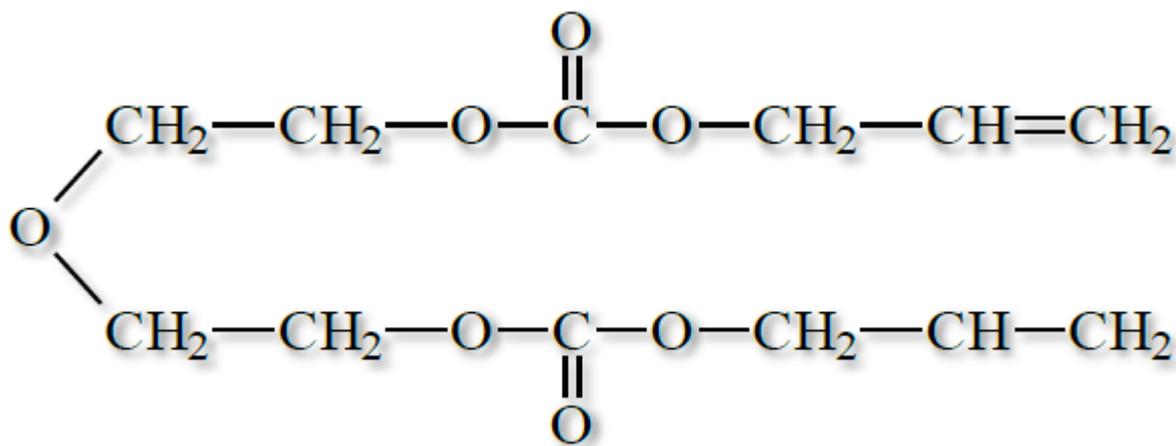

*Figure 3.2 Columbia Resin #39 is allyl diglycol carbonate (C12H18O7). This illustration shows chemical form of CR-39 detector [54].*

If the particles release enough energy as they interact with CR-39, it can detect the absolute number of particles with 100% detection accuracy. An etching process is needed to reveal the damaged zones in CR-39 which proceeds in certain "etching time" steps so as to analyze the formation of the tracks due to the passage of ions. Though, the etching mechanism depends on some parameters such as concentration, temperature, etching time and intrinsic purity of NaOH solution, all of which can vary from experiment to experiment, resulting in different properties of the tracks even for the same "etching time" [55]. A real picture of CR-39 track detector is given below:

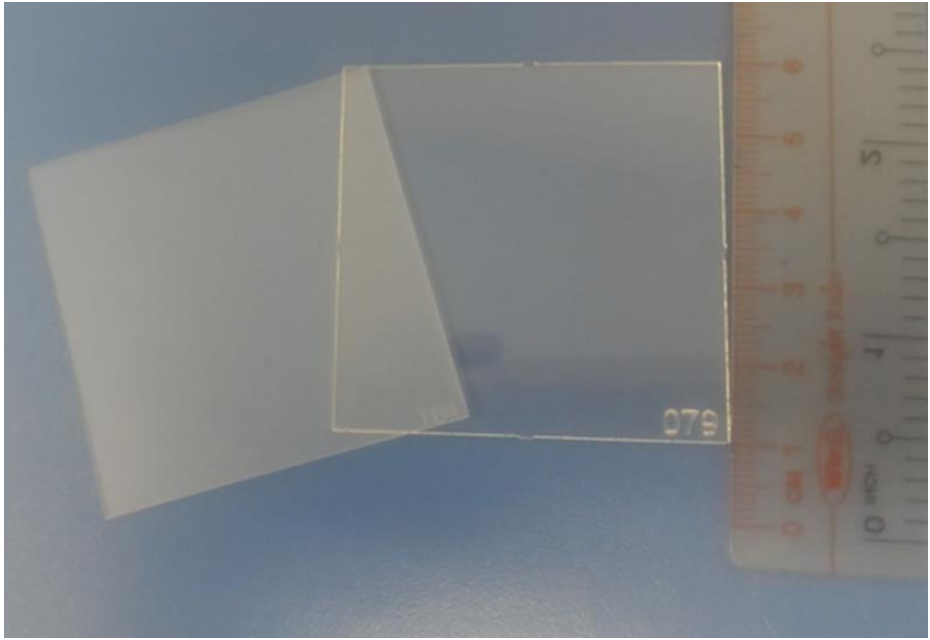

*Figure 3.3 Image of CR-39 detector of dimension 5 cm X 5 cm.*

The general characteristics of CR-39 can be summarized as:

1. Environmentally very stable.

2. Optically clear.

3. Amorphous polymer.

4. Having a closed packed and uniforms molecular structure.

5. Having non – solvent chemical etchant.

## 3.3 History of CR-39 track detector

D.A. Young at AERE Harwell reported the first observations in the field of Solid State Nuclear Track Detection (SSNTD) in 1958. After treatment with a chemical reagent (HF+CH3COOH) saturated with FeF3, he discovered that when lithium fluoride crystals were held 1 mm away from uranium oxide film and irradiated with thermal neutrons, the crystal surface exposed a number of shallows etch pits. In addition, the number of pits (as observed under an optical and electron microscope) corresponded to the theoretically estimated amount of fission fragments recoiling into the crystal from the uranium foil. As a result, a particle detector was born, but its full potential was not realized for another three years. Young had recognized the existence of tracks in his short publication in Nature (Young, 1958), proved that they could be etched and become optically visible, and explained their development as resulting from the damage trails left behind by the passage of fission fragments [56]. With the help of a transmission electron microscope, later Silk and Barnes were able to observe heavy charged particle tracks in mica. P. B. Price, R. M. Walker, and R. L. Fleischer, pioneers in the field of nuclear track detectors, developed the technology extensively by studying nuclear tracks in dielectric solid crystals, plastics, and glasses. These detectors have potential advantages over conventional detectors in some disciplines of physics because to qualities like as lightweight, large geometrical factor, simplicity, adaptability, and the capacity to discriminate among lightly ionizing particles. A sufficient chemical etching of the damaged trail reveals the trajectory of an individual heavily ionizing charged particle in NTDs.

The CR-39 polyallyldiglycol carbonate a NTD was first discovered by Cartwright et al.; CR stands for Columbia Resin. For detecting charged particles over a wide range of Z values and from low energy protons to relativistic U ions, CR-39 has the widest dynamic range. NTDs outperform conventional detectors when it comes to registering heavy ions over a wide-angle range. Even in the relativistic energy region, a single CR-39 NTD can detect nuclei with a wide range of charges down to $Z = 5e$. This requires a well-thought-out calibration technique based on high-energy heavy ions and fragments. Tarlè later showed that CR-39 had a charge resolution superior to scintillators, nuclear emulsions, and ionization chambers of comparable thickness for high Z nuclei. On a single sheet diameter measurement, CR-39 has an intrinsic charge resolution of $\sigma_Z \sim 0.2 - 0.3e$. CR-39 NTD has been used successfully for many years as charged particle detectors [57].

## 3.4  Track registration mechanisms

Many track registration criteria have been given for the formation of an etchable latent track in the detector material, which is primarily determined by the incoming particle's charge, mass, and velocity, as well as the detector parameters, such as density and chemical composition. Only three of them are mentioned briefly here.

## 3.4.1 Total energy loss (dE/dx)

According to this criterion, an etchabale track can occur if the total energy loss (dE/dx)$_{>critical}$ reaches a critical amount that varies depending on the detector material. Fleischer was the first to suggest it, and they employed mica and polycarbonate detectors with different projectiles with varied Z/β values. The criterion appears to explain track formation for low-energy ions, but it appears to fail to recreate some experimental data for high-energy ions. The key flaw in this criterion is that it ignores primary ionization for track formation [52]. The total energy loss is given by

$$\frac{dE}{dx} = 4\pi N_A m_e c^2 r_e^2 \frac{Z_2}{A_2} \frac{Z_1^2}{\beta^2} [\ln(\frac{2m_e c^2 \beta^2 \gamma^2}{I}) - \beta^2 - \frac{\delta}{2} + \Delta L] \qquad (3.1)$$

Where,

$N_A$ = Avogadro's number

$r_e$ = classical electron radius

$Z_2$ = Target atomic number

$A_2$ = target mass number

I = mean ionization potential of the target material

$Z_1$ = effective charge of the incident ions

δ = correction for the density effect

ΔL = sum of corrections needed for precision measurements

## 3.4.2 Primary ionization loss (J)

A new criterion was proposed by Fleischer: the δ-rays are produced directly by the incident ion and this is based on the ion explosion spike model. Only if the linear ion density produced by the initial particle along its trajectory is greater than a critical value for that material can it record an etchable track [52]. The primary ionization (dJ/dx) is calculated as follows:

$$\frac{dJ}{dx} = 2\pi N_A m_e c^2 r_e^2 \frac{Z_2}{A_2} \frac{Z_1^2}{\beta^2} \frac{f_{outer}}{I_{outer}} [\ln(\frac{2m_e c^2 \beta^2 \gamma^2}{I_{outer}}) - \beta^2 - \delta + K] \qquad (3.2)$$

Where,

$I_{outer}$ = binding energy of the outermost electron (~2 eV)

$f_{outer}$ = fraction of electrons in the outer state (0.05)

K = constant for target properties (3.04)

The other quantities have the same meaning as in equation (3.1). For heavy ions, primary ionization fits the experimental data, however it is subjected to the following criticisms:

- The initial ionization for the emission of very low energy δ-rays is not taken into account.
- The value $I_{outer}$ 2 eV might be enough to excite the atoms but not enough to ionize them. To ionize an atom, one requires about 10-15 eV.
- The δ-rays cause a higher order of ionization, which is ignored by the model (which is controversial).

### *3.4.3 Restricted Energy Loss (REL)*

In 1955, Benton suggested that only δ-rays with an energy less than $\omega_o$ contribute to the development of tracks. Only when the REL exceeds a certain value of the detecting material does an etchable track form. The REL (MeV cm$^2$ g$^{-1}$) is calculated as

$$REL = (\frac{dE}{dx})_{\omega<\omega_0} = 4\pi N_A m_e c^2 r_e^2 \frac{Z_2}{A_2} \frac{Z_1^2}{\beta^2}[\ln(\frac{2m_e c^2 \beta^2 \gamma^2 \omega_0}{I^2}) - \beta^2 - \frac{\delta}{2}] \quad (3.3)$$

Where, $\omega_o$ = the energy cut-off value; ~ 200 eV for CR-39 and ~ 350 eV for Makrofol or Lexan. The other quantities have the same meaning as in equation (3.1). For a wide range of incident ions, the REL model fits the experimental data better at low and relativistic energies [52].

## 3.5 Methodology of track detection and visualization

### *3.5.1 Track detection*

By using preferred chemical etching or electrochemical etching, the latent damage trail formed in NTDs can be expanded and detected. The simultaneous actions of the etchant along the latent track and on the bulk material determine the geometry of a charged particle track. Figure 3.1 illustrates a schematic diagram showing chemical etching of a charged particle in an NTD. The quality of the etchant should be carefully monitored for reproducibility of results: To avoid etch product deposits on the detector surface and build-up inside the solution, use fresh etchants and uniform stirring.

## 3.5.2 Bulk etch rate ($v_B$)

The bulk etch rate, $v_B$, is the rate at which the detector's undamaged material is etched out. It varies depending on the etching condition and the depth below the original surface [58]. The bulk etch rate of the detector is typically measured using the techniques listed below:

i. Thickness measurement method

ii. Track diameter method

ii. Change in detector mass method

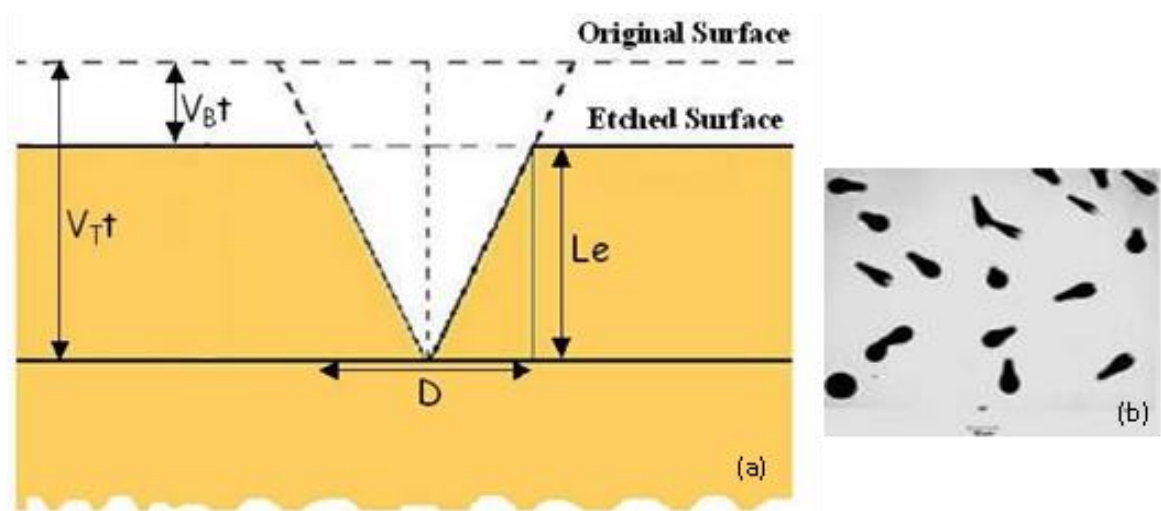

*Figure 3.4 This illustration shows (a) Sketch of a NTD track and (b) etched α tracks in a CR39 detector [52].*

## 3.5.3 Thickness measurement method

The thickness of the detector is measured at several points in order to determine $v_B$. After that, the detector is etched for a set amount of time $\Delta t$, and the thickness is measured after each etching step. The bulk etch rate is given by

$$v_B = \Delta x / 2\Delta t \quad (3.4)$$

where $\Delta x$ is the thickness variation after etching time $\Delta t$. On both sides of the detector, the bulk etching is considered to be the same. The bulk etch rate can be calculated by plotting $\Delta x$ as a function of etching time t and fitting the data points with a given fit [52].

## *3.5.4 Track diameter method*

The track diameter measuring technique can be used to determine $v_B$ if $v_T/v_B > 1$, in CR39.

$$D_{ff} = 2h\sqrt{\frac{p-1}{p+1}} \qquad (3.5)$$

where $D_{ff}$ is the diameter of fission fragments, $p = v_T/v_B$ and h is the thickness removed from both sides of the detector during an etching time t. If p >> 1 the above equation can be written as [52]

$$D_{ff} = 2h(h = v_b t)$$

$$D_{ff} = 2v_B t$$

$$v_B = D_{ff}/2t \qquad (3.6)$$

## *3.5.5 Mass Change method*

The difference in mass of the detector Δm before and after etching can be used to estimate the bulk etch rate. From the measurements of Δm and the density of the detector material, $v_B$ can be calculated as [52]

$$v_B = \frac{\Delta m}{2A\rho t} \qquad (3.7)$$

where A is the surface area, ρ is the density of the detector and t is the etching time. When determining Δm of the same relative humidity, care must be given.

## *3.5.6 Track etch rate ($v_T$)*

The track etch rate $v_T$ is the rate at which the detector material is chemically etched along the particle trajectory's damage trail, or the rate at which the etch cone's tip moves along the latent track during the etching process. The rate of track etching is determined by the incoming particle's energy loss, as well as the temperature and concentration of the etchant.

## *3.5.7 Critical angle of etching ($\theta_c$) and registration efficiency (η)*

The trajectory of a charged particle incident at an angle to the detector surface is shown in Fig. 3.5. The etched track length along the particle trajectory is $v_T t$, and the thickness of the bulk material removed is $v_B t$ after an etching time of t; the trajectory is seen as an etched cone if the vertical component of ($v_T$t) > $v_B$t. The critical angle c, measured with regard to the detector surface, is obtained when ($v_T$t)$_\perp$ = $v_B$t. Particles interacting with $\theta < \theta_c$ are undetectable. $\theta_c$ is a crucial metric for NTDs since it is connected to detector efficiency. The critical angle has a significant impact on the track registration efficiency [52].

NTDs have low critical angles for high REL nuclei, such as fission fragments, and large critical angles for low REL particles, such as protons and low Z nuclei ($\theta_c$ varies from $2^0$ to $5^0$ for these in CR-39). The track registration efficiency η of NTDs is defined as

$$\eta = 1 - \sin\theta_c \qquad (3.8)$$

It is obvious from this expression that for small critical angles, track registration efficiencies are high. Most plastic NTDs with critical angles of 2 - $5^0$ have high registration efficiency (85-99%).

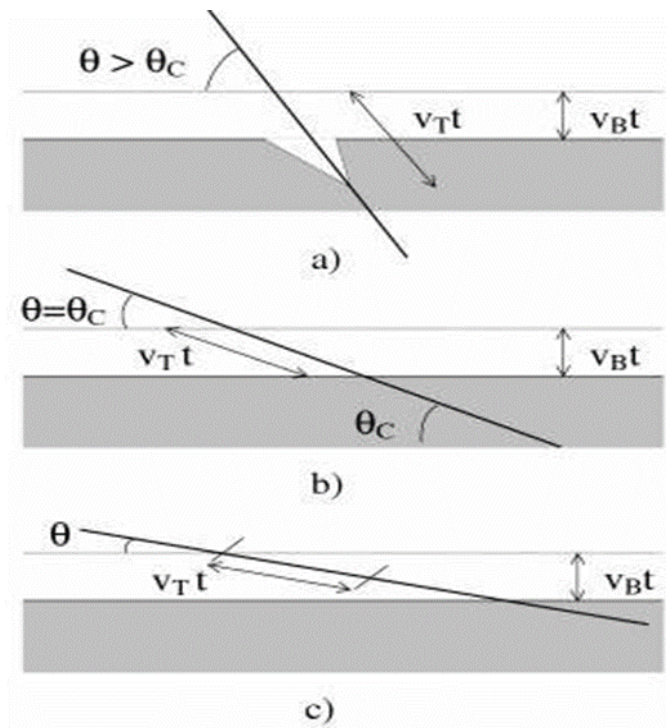

*Figure 3.5 Track geometry at different dip angle conditions: (a) formation of the post etched cone for a particle with an incident angle θ > θc ($v_T$t > $v_B$t), (b) limiting case when the incident angle θ = θc ($v_T$t ≥ $v_B$t) and (c) no track is detected if the incident angle θ < θc ($v_T$t ≤ $v_B$t) [52].*

## *3.5.8 Track geometry at normal incidence*

The essential parameters are the base cone diameter, major and minor axes, the etched track's cone height, the "bulk" etch rate "$v_B$," and the track-etch rate $v_T$, as shown in Fig. 3.6 for particles at normal incidence with constant REL. The "signal" or sensitivity "p" of the detector material is the ratio of track-etch rate to bulk-etch rate ($p = v_T/v_B$). In particular, the chemical solution etches the detector material's surface at a lower rate ($v_B$ on both sides) than the damaged region, where the etchant attacks at a higher rate ($v_T$). Because measuring the track etch rate during etching is challenging, its value is calculated using the geometry of an etch-pit [52].

The following relations hold for normal incidence geometry (Fig. 3.6):

$$L_e = (v_T - v_B)t \qquad (3.9)$$

$$v_B = \Delta x / 2\Delta t \qquad (3.10)$$

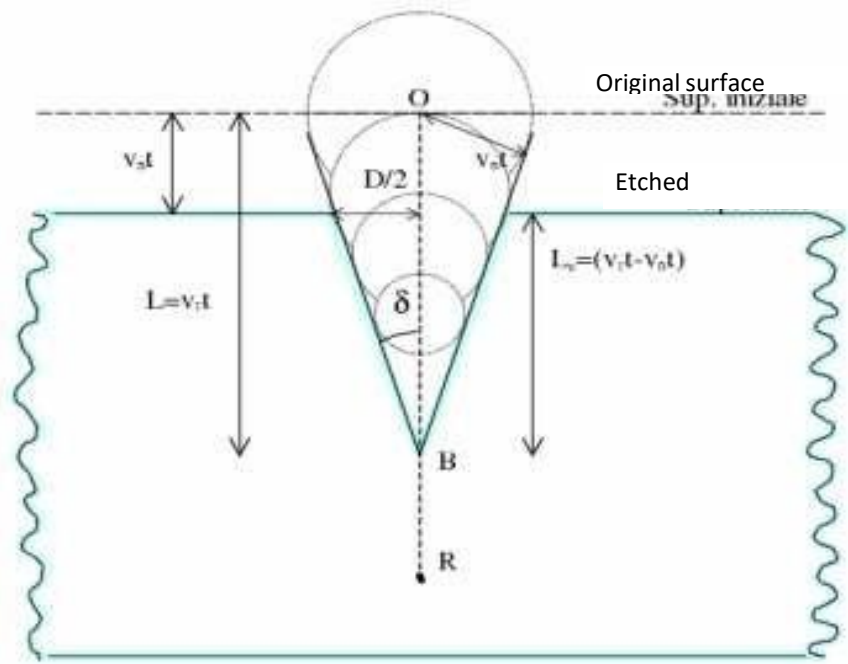

*Figure 3.6 Illustration shows track geometry for a charged particle impinging at normal incidence in a nuclear track detector* [52].

The semi cone angle $\delta$ of the etched cone is

$$\sin \delta = \frac{v_B}{v_T} = \frac{1}{p} \qquad (3.11)$$

From equation 3.9 the track etch rate $v_T$ is

$$v_T = v_B + \frac{L_e}{t} \qquad (3.12)$$

The relations between 'D' '$v_T$', '$v_B$' and 'p' are

$$D = 2v_B t \sqrt{\frac{(v_T - v_B)}{(v_T + v_B)}} \qquad (3.13)$$

$$p = \frac{1 + (D/2v_B t)^2}{1 - (D/2v_B t)^2} \qquad (3.14)$$

$$p = \frac{1 + A/(\pi v_B^2 t^2)}{1 - A/(\pi v_B^2 t^2)} \qquad (3.15)$$

The error on 'p' can be calculated from the following relations

$$\Delta p = \frac{\Delta L_e}{v_B t}$$

$$\Delta p = \frac{D.\Delta D}{v_B^2 t^2 (1 - \frac{D^2}{4v_B^2 t^2})^2}$$

$$\Delta p = \frac{2.\Delta A}{\pi v_B^2 t^2 (1 - \frac{A}{\pi v_B^2 t^2})^2} \qquad (3.16)$$

Equation 3.9 – 3.16 are apply to any particle of constant energy loss such as relativistic ions, magnetic monopoles etc.

### *3.5.9 Track Geometry at Oblique Incidence*

Fig. 3.7 shows the geometry of a particle incident at an angle θ with respect to the detector surface and constant REL. The following geometrical relations can be deduced from Fig. 3.7:

$$r_1 = \frac{\tan \delta [v_T - v_B / \sin \theta] t}{\sin \theta - \cos \theta \tan \delta} \qquad (3.17)$$

$$r_2 = \frac{\tan \delta [v_T - v_B / \sin \theta]}{\sin \theta + \cos \theta \tan \delta} \qquad (3.18)$$

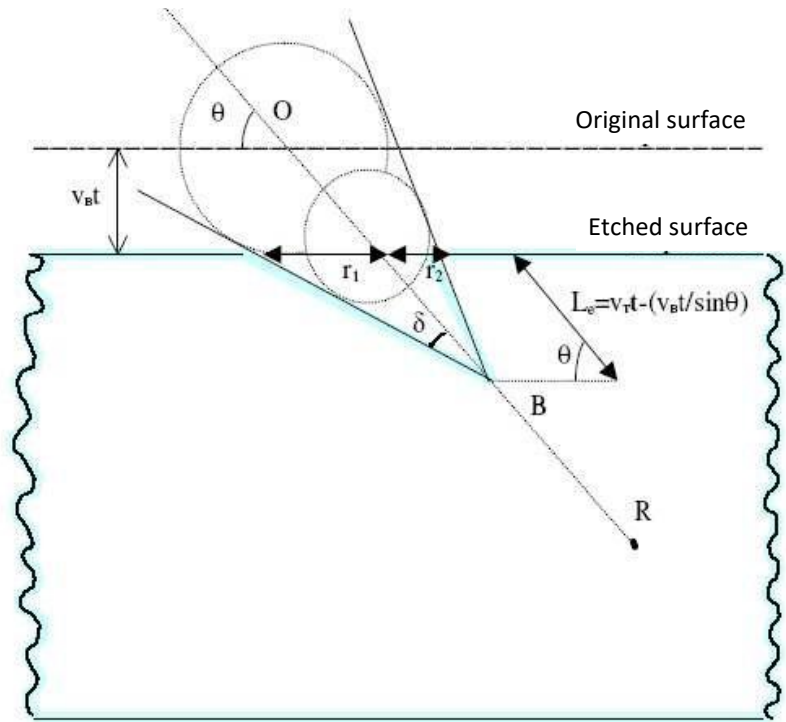

*Figure 3.7 Illustration shows geometry of a particle incident at an angle θ with respect to the detector surface* [52].

We deduce the relations for 'D', 'd' and 'p',

$$D = 2v_B t \frac{\sqrt{(p^2-1)}}{(p\sin\theta+1)} \qquad (3.19)$$

$$d = 2v_B t \frac{\sqrt{(p\sin\theta-1)}}{(p\sin\theta+1)} \qquad (3.20)$$

$$p = \sqrt{1+\frac{4A^2}{(1-B^2)^2}} \qquad (3.21)$$

where $A = (D/2v_B t)$ and $B = (d/2v_B t)$ and

$$\theta = \arcsin(\frac{1}{p}\cdot\frac{1+B^2}{1-B^2}) \qquad (3.22)$$

## 3.6 Calibration

The use of an NTD to identify a relativistic charged fragment relies on two types of information. To begin, we have a decent understanding of how a charged ion loses energy as it propagates through a stopping medium with well-known chemical and physical properties. Second, the ion trajectory leaves a physically detectable signal that can be accurately measured experimentally. The first of these inputs is derived from the expression of Relative Energy Loss (as discussed in section 3.4.3), which is a function of $Z/\beta$ and the second input is obtained through the measurement of the reduced etch-rate of the ion. The 'calibration curve' is a relationship between the two quantities. A purely geometrical measurement (such as the cone-length of a track or the area of its base) can be converted into a physical characteristic, namely the charge of the ion, using this curve (if all fragments are highly relativistic) [53], [57].

## 3.7 Etching

An important item is the determination of the optimal etching conditions to achieve the best surface quality and reduce the number of fake tracks in CR-39. For NTD experiments, chemical etching is a necessary step. Etching magnifies the nanometer-diameter particle trails in NTDs to microscopic dimensions suitable for optical microscopy observations [58]. The etching of nuclear tracks is complicated by a variety of factors that are difficult to manage precisely. Concentrations of active etching molecules or species in the etchant, stability, and temperature control are the most important factors. The etching behavior of the detector material varies, hence causing uncertainty in experimental etching results. Usually, aqueous solutions of NaOH and KOH have been used for etching the CR-39 track detector. Improvements in etching conditions have been studied extensively, including improved stirring and temperature control of the etching solution, as well as the addition of various percentages of ethyl alcohol [52]. The addition of ethyl alcohol in the etchant increases the etched surface quality while also reducing the number of surface defects and background tracks. Generally, two types of etching have been used in CR-39: Soft etching and Hard etching. The etchants of water solutions: 6N NaOH, and 6N KOH at 70 and 60ºC, respectively, with different fractions of ethyl alcohol. These are called "soft" (normal) etching conditions. Another type of "strong" etching condition is in order to fastly reduce the thickness of the detectors for the analysis of the CR39 NTDs. For strong etching, mainly 8N

NaOH, 7N KOH, 8N KOH water solution use at 75 and 77°C, with the addition of different fractions of ethyl alcohol in the solution.

The etchant solution is made in a volumetric flask using the equation:

$$W = W_{eq}.N.V \qquad (3.23)$$

Where,

W = Weight of NaOH/KOH needed to prepare the given normality.

$W_{eq}$ = Equivalent weight of NaOH/KOH.

N = Normality for CR-39.

V = Volume of distilled water.

## 3.8 Stopping of high-energy ions

When a heavy charged particle passes through matter it delivers energy to atomic electrons by the Coulomb interaction. The multiple electronic collisions that are undergone by the particle can in a good approximation be summarized to a continuous energy loss. The stopping power of material is defined as the ratio of the differential energy loss for the particle within the material to the corresponding differential path length:

$$S = -\frac{dE}{dx} \qquad (3.24)$$

Where, E is the kinetic energy of the charged particle, the value of –dE/dx along a particle track is also called its specific energy loss. For charged particles, S increases as the particle velocity decreases [31].

Ion therapy of deep-seated tumors requires ion beam ranges in tissue of up to 30 cm corresponding to specific energies up to 430 MeV/n for carbon ions with particle velocities β=v/c ≈ 0.7. At these velocities the energy-loss rate dE/dx in the slowing-down process is dominated by inelastic collisions with the target electrons (electronic stopping) and can be well described by the classical expression that describes the specific energy loss is known as Bethe-Bloch formula [59].

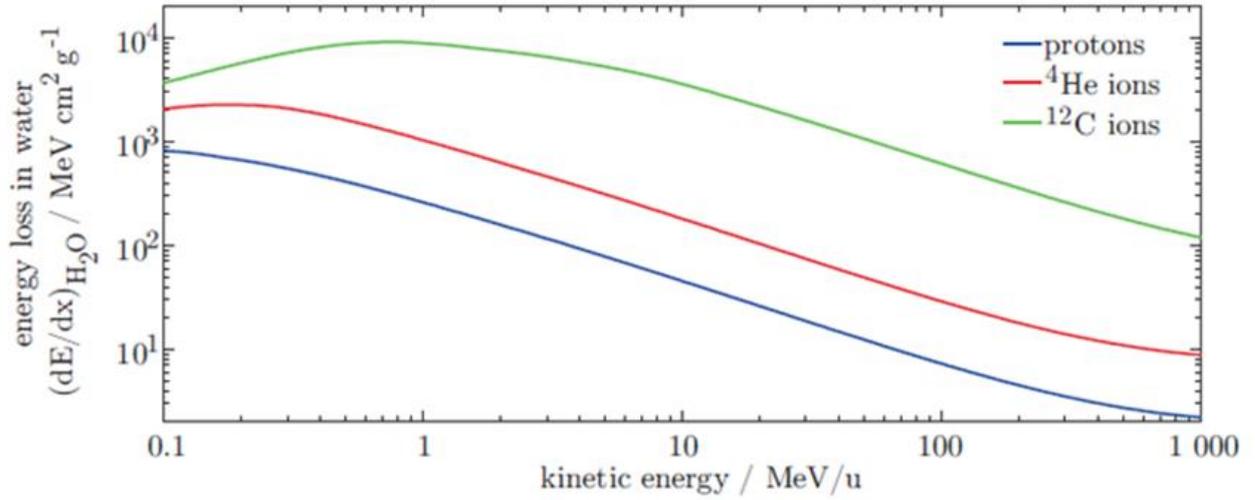

*Figure 3.8 Illustration shows energy loss of protons, $^4$He and $^{12}$C ions in water as a function of their kinetic energy* [36].

Relativistic version of this formula is:

$$\frac{dE}{dx} = \frac{4\pi e^4 Z_t Z_p^2}{m_e v^2}[\ln\frac{2m_e v^2}{<I>} - \ln(1-\beta^2) - \beta^2 - \frac{C}{Z_t} - \frac{\delta}{2}] \qquad (3.25)$$

$Z_p$ and $Z_t$ denote the nuclear charges of the projectile and target, $\delta/2$ is the density effect correction term, $m_e$ and e are the mass and charge of the electron, $C/Z_t$ is the shell correction term, <I> is the mean ionization energy of the target atom or molecule, β is the projectile velocity in units of c(=v/c). The energy loss increases as the particle energy decreases due to the 1/β dependence. This is the cause for Bragg peak at the end of the range of particle.

The atomic electrons are totally stripped off at high velocities and the projectile charge is equal to the atomic charge number, $Z_p$. The mean charge state decreases at lower velocities due to the interplay of recombination and ionization process. So, the effective charge $Z_{eff}$ must take the place of $Z_p$ in above equation. Corresponding to the Bragg peak, the maximum energy loss rate is reached at a projectile velocity of

$$v_p \approx Z_p^{2/3} v_0 \qquad (3.26)$$

This maximum occurs for $^{12}$C at a particular energy of ≈ 350 KeV/n. Elastic collisions with target nuclei at lower projectile energies, start to contribute significantly to the energy loss and dominate the stopping process at very end of the particle track. The corresponding dose contribution is, however, very small and can be neglected in radiotherapy applications [60].

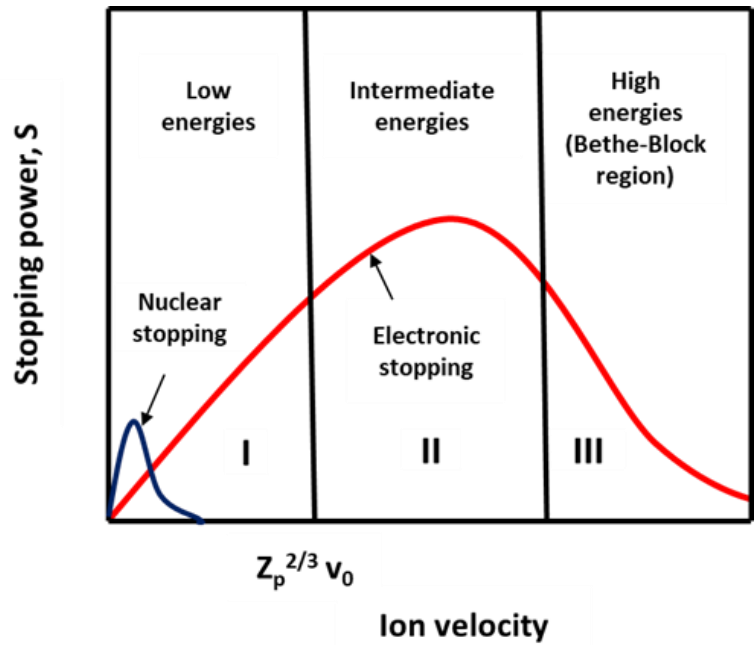

*Figure 3.9 Illustration shows nuclear and electronic components of the ion stopping power as a function of ion velocity. Nuclear stopping power is in the low energy range and electronic stopping power is in the intermediate energy range.*

The total path length of the particle's trajectory in the absorber is given by

$$R(E) = \int_0^E (\frac{dE'}{dx})^{-1} dE' \qquad (3.27)$$

Which for heavy charged projectiles is nearly the same as the mean Range. The range, R of a charged particle is defined as the distance it travels from its source to its target substance. It is difficult to calculate the path length of light particles like electron as they can be scattered widely in the path of the targets with large angles because of their low weight [61]. On the other hand, heavy particles such as helium, carbon, etc. are very little scattered and the path length is almost straight line. Ranges of protons, $^4$He and $^{12}$C ions in water are shown in Fig.3.9.

The range, R at the same velocity scales with $A/Z^2$, hence the protons and $^4$He ion curves are overlapping however a higher acceleration is required for $^{12}$C ion to reach the same depth. Proton and $^4$He ions require to be accelerated to an energy of 220 MeV/n for particle ranges of 30 cm in water while a specific energy of 430 MeV/n is required for $^{12}$C ions. These are the typical maximum energies therapy accelerators are designed for because with a range of 30 cm one can potentially reach every point in the human body with the Bragg peak [36].

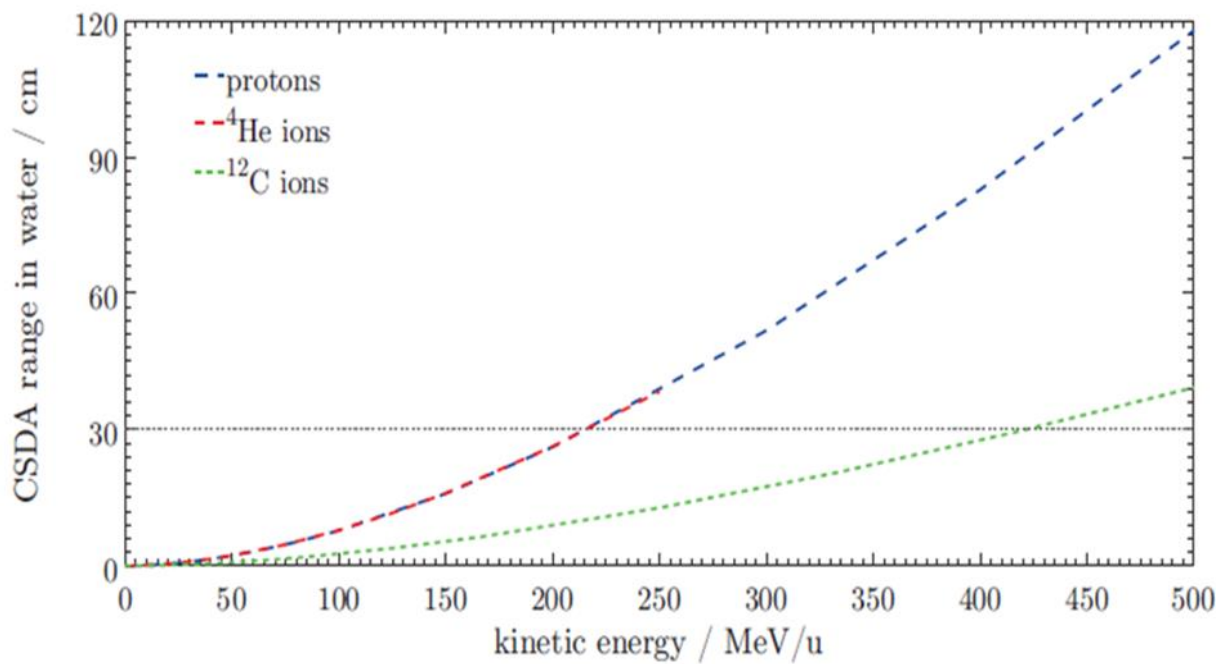

*Figure 3.10 Illustration shows CSDA ranges of protons, $^4$He and $^{12}$C ions in water as a function of kinetic [36].*

If an ion changes its $A/Z^2$-ratio in a nuclear fragmentation reaction then this affects also its range. As a result, the fragment dose is mainly deposited before and beyond the primary ion range, where the Bragg peak is formed (see Fig.1.2). Further details will be described on these nuclear fragmentation reactions which are the major phenomenon investigated within this study.

## 3.8.1 The Bragg curve

The Bragg curve, as shown in Fig. 3.11, is a representation of the specific energy loss along the track of a charged particle. The example is for an alpha particle with an initial energy of several MeV. The charge on the alpha particle is two electronic charges during the most of the track, and the specific energy loss increases nearly as 1/E. The charge is decreased near the end of the track due to electron pickup, and the curve begins to fall off. Plots for a single alpha particle track as well as the average behavior of a parallel beam of alpha particles with the same initial energy are presented. Due to the effects of straggling, which will be addressed further below, the two curves differ slightly.

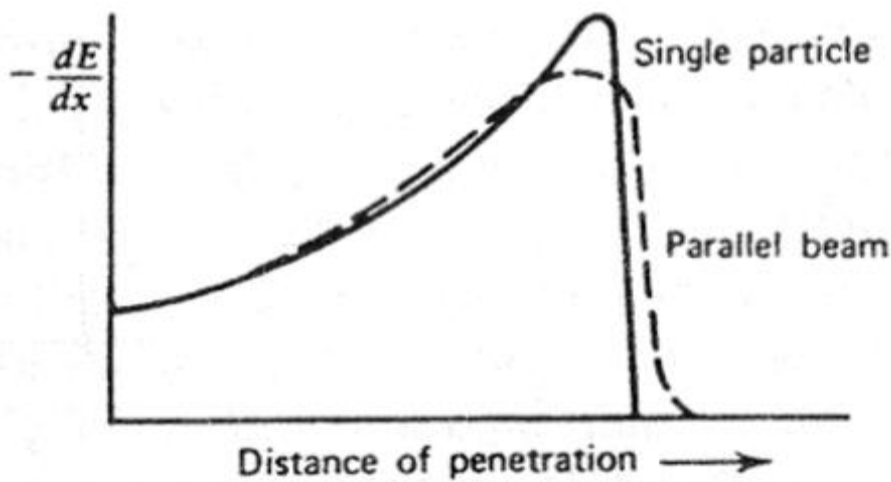

*Figure 3.11 Illustration shows the specific energy loss along a particle track* [31]. *The charge on the alpha particle is two electronic charges during the most of the track, and the specific energy loss increases nearly as 1/E. The charge is decreased near the end of the track due to electron pickup, and the curve begins to fall off.*

Because of the Bragg peak phenomenon, carbon therapy is distinctive. Carbons, unlike photons, deposit the majority of their energy in a single location, the Bragg peak. A carbon beam can be conformed to the shape and depth of the tumor, eliminating the exit dose and sparing surrounding organs and normal tissue from radiation exposure. As a result, patients can receive higher doses, improving therapy effectiveness. This allows radiation oncologists to give patients higher doses of radiation, boosting therapy effectiveness.

## *3.8.2 Energy loss straggling*

The mean energy loss of a heavy charged particle per unit path length is given in Equation 3.25, however the energy that is lost in a thinner absorber shows statistical fluctuations around this mean value, which is referred to as energy loss straggling. The shape of the resulting Bragg curve is significantly affected by these microscopic fluctuations, which add up to a macroscopic range straggling [62]. The energy loss straggling has a great impact on the peak widths and shapes in heavy charged particle energy loss spectra measured with thin detectors, such as plastic scintillators. Figure 3.12 illustrates the energy loss distribution of protons, $^{4}$He, and $^{12}$C ions in a 1 mm water slice with the same specific energy of 220 MeV/n. The Landau distribution can be used to approximate the shape of such energy loss distributions, which have a characteristic asymmetric shape. Since the velocity of the different ions in Fig. 3.12 is the same, according to equation 3.25 their mean energy loss scales with $Z^2$. In the case of heavier projectiles, the absolute peak width increases while the relative energy loss straggling decreases. It is noticeable that there is a minimum energy transfer, hence every charged particle has a finite maximum range, which is defined by the initial energy. While the energy loss fluctuates only very limited to the lower energy side, there is a long extension towards high energy transfers (the so-called Landau tail) which also reflects in the production of high-energy delta electrons that can travel up to millimeter distances away from their production point before stopping [63].

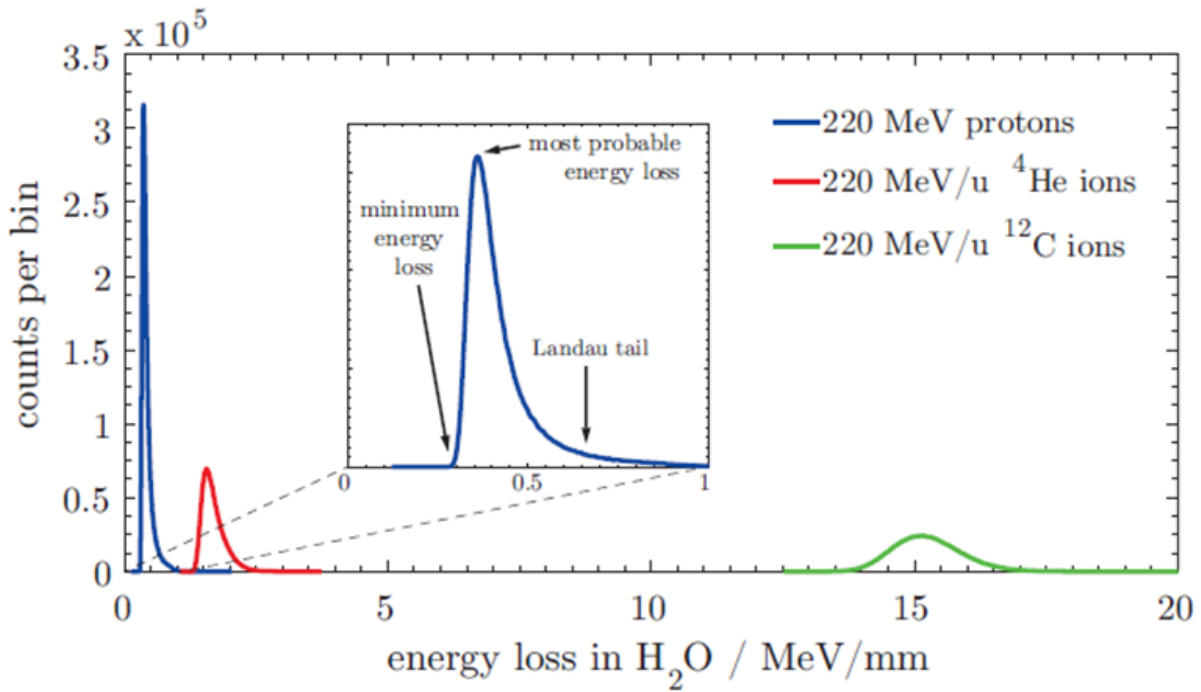

*Figure 3.12 Illustration shows the energy loss distribution in a water slice of 1 mm thickness by 220 MeV/n protons, $^{4}$He and $^{12}$C ions [36].*

### *3.8.3 Range and range straggling*

The range of a charged particle is the total length of its travel path before it is stopped. Range is defined as the sum of the distance covered across the crooked path (track), whereas penetration is specified as the net projection measured along the initial direction of motion. The scattering encountered by the particle during its path causes the difference between range and penetration distances. Large-angle scatterings are unusual for heavy charged particles with high initial velocities. For the most part, the associated trajectories are straight, and the difference between range and penetration distance is insignificant. Using a proper stopping-power formula, particle ranges can be calculated. Range is easier to measure experimentally than stopping power. Range is usually well-known for incident energies higher than this crucial value, and computation correlates with experiment within approximately 5%. The precision of aluminum, which is the most well-studied material, is within 0.5%. Range calculations are generally uncertain for incident energy smaller than the critical value, and agreement with experiment is unsatisfactory. The range–energy relationship is sometimes expressed as a power law, in which the range (R) is proportional to the energy (E) raised to a power (n); that is, R E$^n$. With the exponent n equal to 1.75, protons in the energy range of a few hundred MeV conform to this type of relation rather well. In particle identification, range measurements are extremely important. For a variety of heavy particle ranges, there is a wealth of experimental data and computations available [31].

Because energy loss is a statistical event, both stopping power and range should be understood as mean values over an ensemble of atoms or molecules. Fluctuations are to be expected, and these fluctuations are referred to as straggling. There are several types of straggling. The most important of them is range straggling, which implies that particles in the same medium have different path lengths between the same initial and final energy for statistical reasons. For long path lengths, Bohr demonstrated that the range distribution is roughly Gaussian. Emergent particles exhibit a kind of energy straggling known as Landau type for short path lengths, such as those encountered in thin film penetration [36].

## 3.9 SRIM simulation

SRIM [47] stands for Stopping and Range of Ions in Matter and is a software package. Since its introduction in 1985, it has been upgraded on a regular basis. The fundamental physics of the program is described in detail in the textbook "SRIM - The Stopping and Range of Ions in Matter." Since this time, modifications have been made based on further experimental data. It is a Monte Carlo simulation program that calculates the stopping and range of ions (up to 2 GeV/amu) into matter using a quantum mechanical treatment of ion-atom collisions. This calculation is made very efficient by the use of statistical algorithms, which allow the ion to make jumps between calculated collisions and then average the collision results over the intervening gap. During the collisions, the ion and atom have a screened Coulomb collision, including exchange and correlation interactions between the overlapping electron shells. The ion has long range interactions creating electron excitations and plasmons within the target. These are described by including a description of the target's collective electronic structure and interatomic bond structure when the calculation is setup (tables of nominal values are supplied). The charge state of the ion within the target is described using the concept of effective charge, which includes a velocity dependent charge state and long-range screening due to the collective electron sea of the target [64], [65].

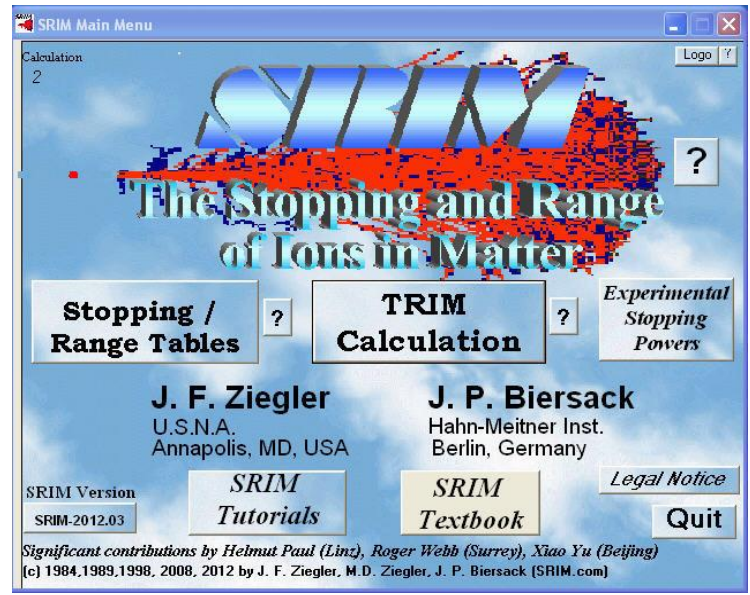

*Figure 3.13 Picture of SRIM-2012 program that process the data regarding the transport of ions in the matter. This is the main page of the SRIM calculation. In this page we have the option to run SRIM or TRIM.*

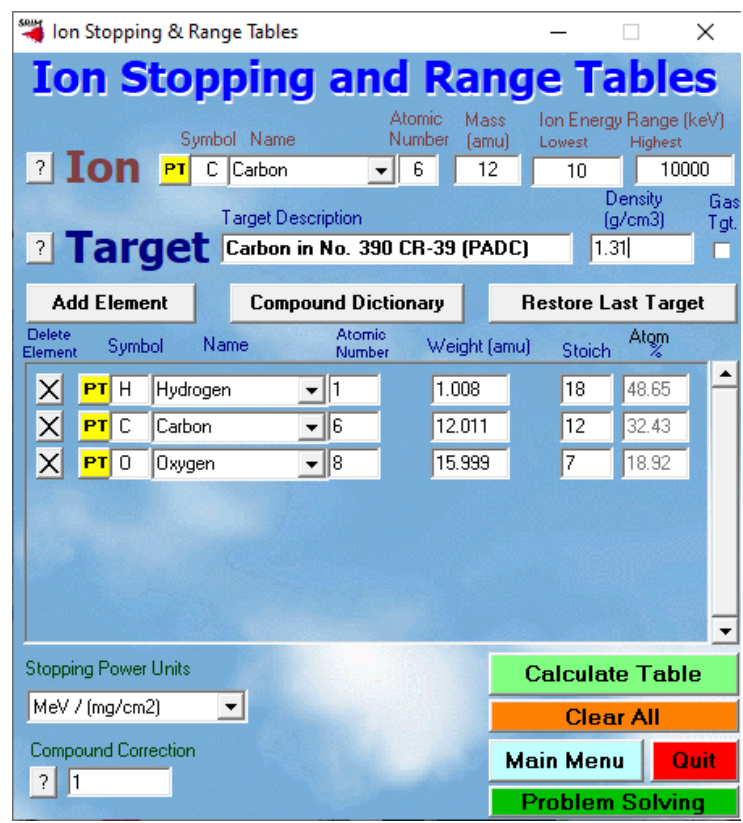

*Figure 3.14 After choosing SRIM we will have this window where we can choose projectile ion, energy range and target (element/compound form). We can choose the unit of energy loss and then order to start calculation.*

In previous radiation material research articles, SRIM has offered genuine and approximate experimental data, among other programs such as ASTAR, MSTAR, PSTAR, etc. We chose the SRIM program to generate data since experimental measurement of stopping power is a difficult process. The calculation approach used by the SRIM program is based on the modified Bethe formula:

$$S = \frac{4\pi r_0^2 m_e c^2 Z_2}{\beta^2} Z_1^2 [f(\beta) - \ln<I> - \frac{C}{Z_2} - \frac{\delta}{2}]$$

$$\text{Or, } S = \frac{kZ_2}{\beta^2} Z_1^2 [L_0(\beta) + Z_1 L_1(\beta) + Z_2^2 L_2(\beta)...] \quad (3.28)$$

$$\text{where } k \equiv 4\pi r_0^2 m_e c^2$$

And $L_0$, $L_1$, $L_2$ are called respectively correction factors of the Fano formula, Barkas correction, Bloch correction. When the stopping number $L(\beta)$ is introduced, the equation becomes:

$$S = \frac{kZ_2}{\beta^2} Z_1^2 L(\beta) \quad (3.29)$$

The Bethe formula has some limits, particularly when particles have a low velocity, because positively charged particles have a tendency to pick up electrons from target materials at this velocity. This method effectively reduces the charge of this charged particle, resulting in a linear energy loss. As a result, there is a significant chance that the charged particle will accumulate electrons and eventually become a neutral atom. The stopping power of any particle (low particle energy to high particle energy up to 2 GeV/amu) can be calculated simultaneously using the SRIM program, which is based on the above modified Bethe formula. We calculated the energy of $^{12}C$ ions in both CR-39 and aluminum target using this SRIM simulation program.

The mean excitation potential is an important parameter which characterizes the stopping of target materials. From above equation, mean excitation potential is defined as:

$$\ln<I> = \sum f_n \ln E_n \quad (3.30)$$

Bloch had shown that mean excitation potential is practically proportional to the atomic number (Z) of the target materials and close to Rydberg energy $I = 13.5 Z.eV$. The Rydberg energy equation is used to calculate the mean excitation potential of all materials in this program.

# Chapter 4
# Experimental Method

## 4.1 Accelerator

Particle accelerators are used to generate charged particle beams. Cyclotrons and synchrotrons are the most frequently applied accelerator technology in particle therapy centers. Particle accelerators are machines that increase the kinetic energy of charged particles by using electromagnetic fields. Their acceleration processes are established on the Lorentz force created by the presence of an electric field. An accelerator for medical applications should provide a stable and preferably continuous beam at maximum energies of about 450 MeV/n for carbon ions [66].

## 4.2 Overview of accelerator at WERC

The schematic structure of the accelerator complex at WERC is shown in Fig. 4.1. A 5 MV tandem accelerator, a synchrotron, the ion implanter, two ion sources, two experimental rooms for the experiment using tandem beams and two rooms for the irradiation of the beams from the synchrotron are all part of this system. The tandem accelerator is interfaced with the synchrotron and works as a synchrotron injector. The synchrotron can accelerate high energy proton and carbon beam up to 200 MeV at irradiation room 4. The tandem accelerator has also been used independently to accelerate medium energy protons and carbon beams up to 10 MeV at irradiation room 2. The ion implanter produces ions with energies ranging from 10 to 200 keV with a high beam current. This low energy and intense beam have been used in irradiation room 1. The next subsections will go over the detailed specification of each accelerator [67].

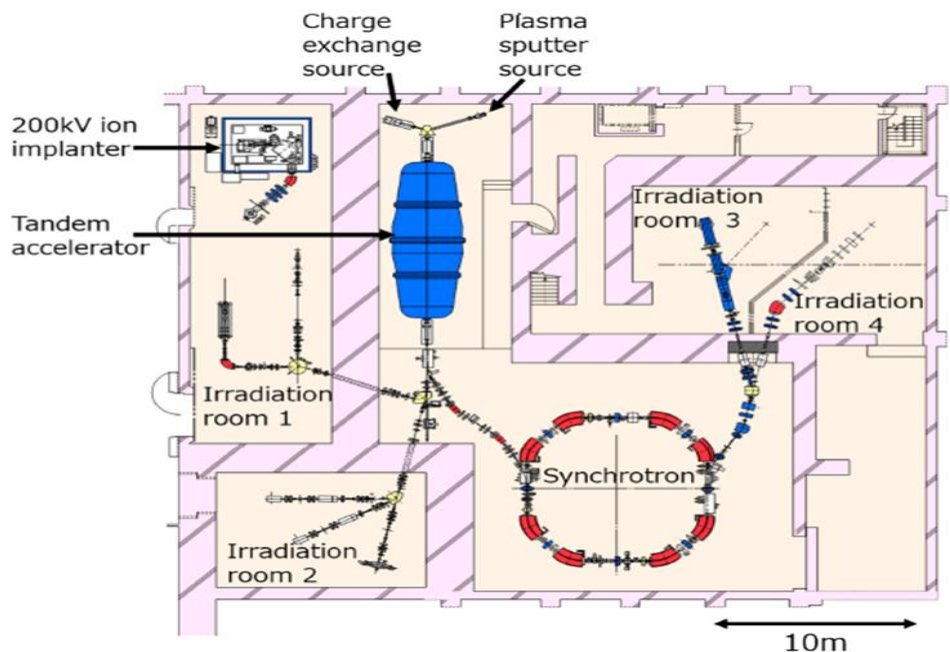

*Figure 4.1 Schematic layout of WERC accelerator system. A 5 MV tandem accelerator, a synchrotron, the ion implanter, two ion sources, two experimental rooms for the experiment using tandem beams and two rooms for the irradiation of the beams from the synchrotron are all part of this system.*

## *4.2.1 Tandem Accelerator at WERC*

The WERC tandem accelerator (see Fig. 4.2) accelerates negative ions that are alternately injected from two ion sources. A plasma sputter is the main ion source that produces protons and heavy ions such as carbon ions. For the ionization of gas elements, the charge exchange type is another ion source.

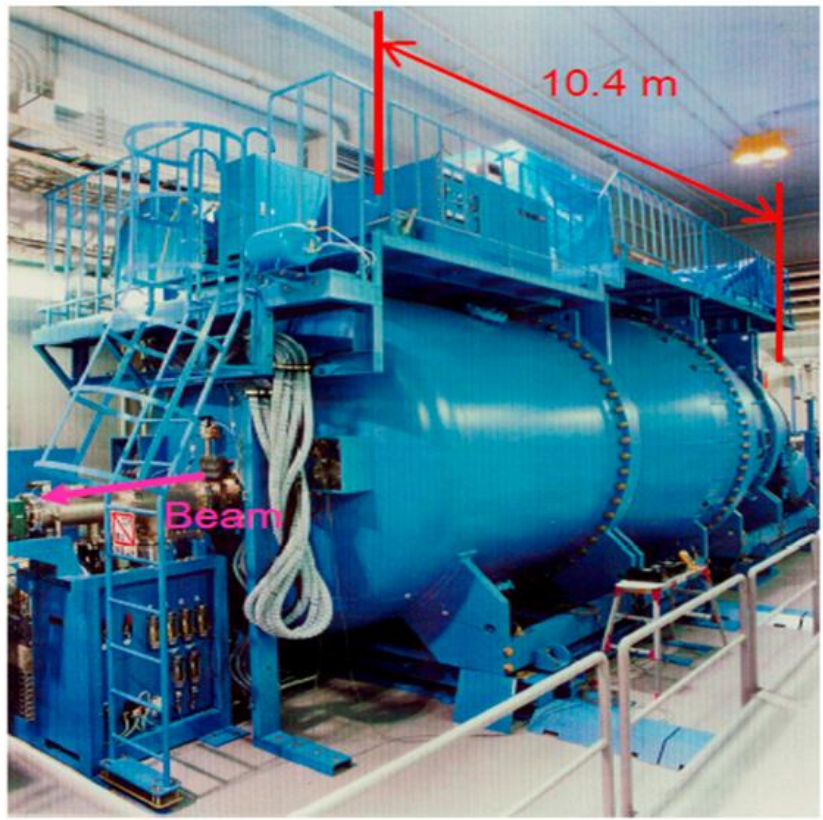

*Figure 4.2 Tandem accelerator at WERC. This tandem accelerator accelerates negative ions that are alternately injected from two ion sources.*

The tandem accelerator has a maximum acceleration voltage of 5 MV. The high tension is generated by the Schenkel circuit with a 58-step voltage doubler rectifier, which rectifies 40 kHz RF power by a resonant transformer and an oscillator. The insulation column support in the pressure tank holds the Schenkel rectifier, the resonant transformer, the high-tension terminal, and the acceleration tube. In order to enable the acceleration of the highly intense beam from the ion sources, the conveyor current through the Schenkel rectifier amounts to 1 mA at the terminal high tension of 5 MV. At a terminal high tension of 5 MV, the voltage ripple is 0.4 kVpp. The charge exchange canal in the terminal has a 15 mm inner diameter and a charge exchange of 1 m effective length. In the middle of the canal, argon gas is introduced for charge exchange. The concentration of charge exchange in the canal with a large inner diameter is made possible via recirculation by four turbo-molecular pumps and differential pumping. Negative ions are converted to positive ions at a rate of 99 percent in this charge exchange system. Furthermore, near-perfect transmission efficiency is attained along the following transport beam line to the synchrotron [67], [68].

## *4.2.2 Synchrotron*

A synchrotron needs a massive infrastructure dedicated to its construction and operation.

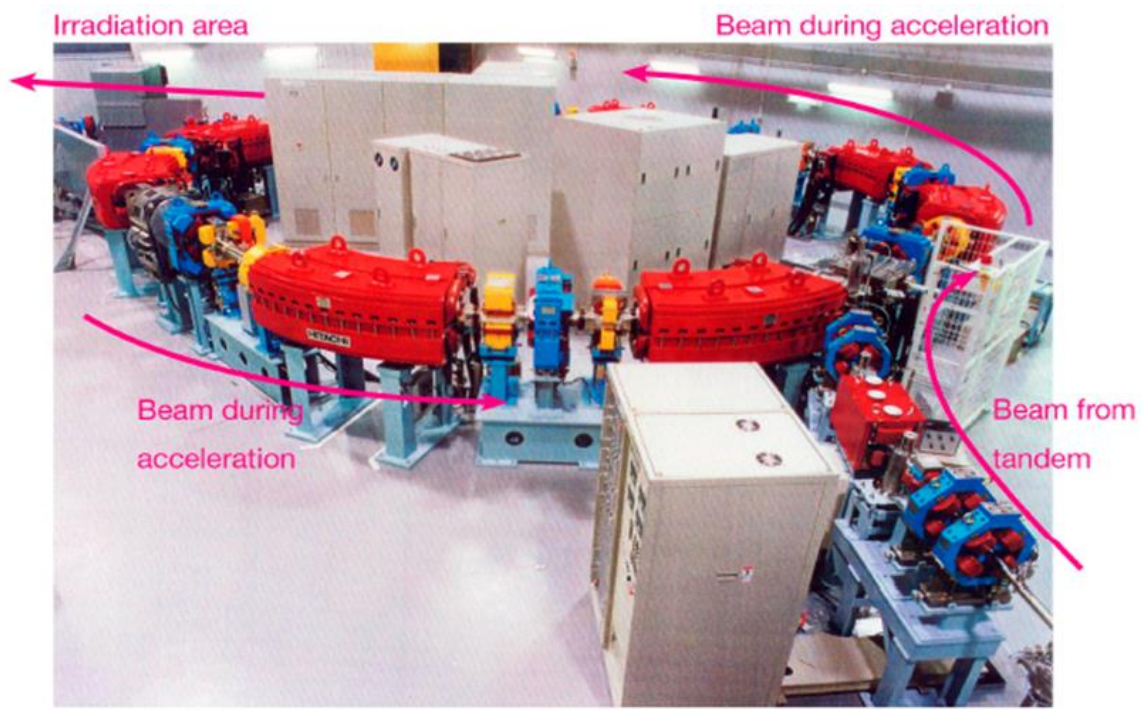

*Figure 4.3 Synchrotron is a circular machine with a fixed particle orbital radius. It is an incredibly powerful source of X-rays. This is a picture of an overview of the WERC synchrotron.*

Synchrotron is a circular machine with a fixed particle orbital radius. It is an incredibly powerful source of X-rays. The X-rays are generated by high energy electrons as they revolve around the synchrotron. The accelerating particle beam in a synchrotron travels around a fixed closed-loop path. During the accelerating process, the magnetic field that bends the particle beam into its closed path increases with time, synchronized with the increasing kinetic energy of the particles [69]. Fig. 4.3 depicts an overview of the WERC synchrotron. The synchrotron receives a beam of 2.08 MeV/n incident energy for carbon ions from the tandem accelerator, which is then accelerated to 55 MeV/n. The synchrotron has a diameter of 33.2 m and a super periodicity of 4. Each lattice has a permutation of QF-BM-QD-BM and is run as a separate function. In this case, QF and QD refer to quadrupole magnets that focus and defocus in the median plane, respectively, and BM refers to a bending magnet. Vertical and horizontal tunes are respectively 0.85 and 1.75. Injection, acceleration, extraction, and return to injection are the four modes of the synchrotron's period. The injection is carried out in a multi-turn technique. In horizontal phase space, the RF knock out mechanism slowly separates the accelerated beam from the separatrix. Excitation of sextupole magnets generates the separatrix.

## *4.2.3 Ion-Implanter*

A high-current ion beam irradiation system known as the 200 kV ion-implanter has been installed at WERC. Fig. 4.4 shows the schematic diagram of the 200 kV ion-implanter. A microwave ion source, a mass separation, a 50 kV first acceleration tube, and a 150 kV acceleration tube make up the 200 kV ion-implanter. Positive ions are produced by the microwave ion source from gas targets such as $H_2$, He, $N_2$, Ne, Ar, and others. The first acceleration tube accelerates positive ions generated in the ion source up to 50 keV. The mass separation magnet is then used to sort them by mass. The ions are again accelerated up to 200 keV by the second acceleration tube after mass separation. The energy of the ions can be changed from 10 to 200 keV. The first and second acceleration tubes provide maximum beam currents of 50 and 30 mA, respectively. In the actual irradiation experiments, however, the beam current is reduced to less than several tens of micro A/cm$^2$ to avoid beam heating of the target materials during the irradiation [68], [69].

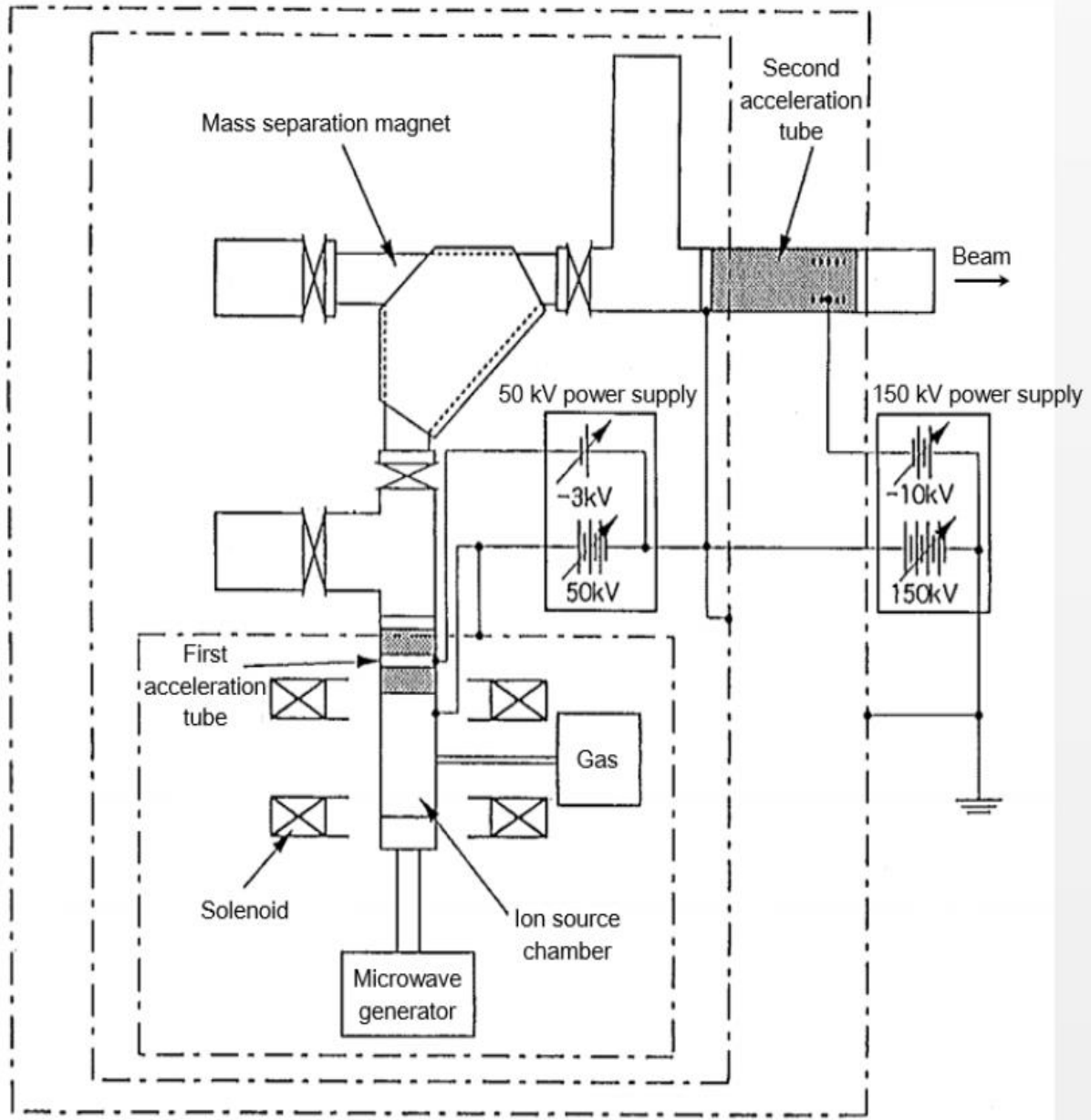

*Figure 4.4 A high-current ion beam irradiation system known as the 200 kV ion-implanter has been installed at WERC. Schematic diagram of 200 kV ion-implanter at WERC. A microwave ion source, a mass separation, a 50 kV first acceleration tube, and a 150 kV acceleration tube make up the 200 kV ion-implanter.*

## *4.2.4 Experimental beam delivery technique*

MeV-ion beam from the tandem accelerator is generally used for the irradiation for the improvement of the material and biological target and the ion-beam analyses such as "particle induced X ray Emission analysis (PIXE)", "Rutherford Back Scattering analysis (RBS)",

"channeling RBS", "Elastic Recoil Detection Analysis (ERDA)", "Nuclear Reaction Analysis (NRA)" and so on. A linear accelerator is often used to pre-accelerate the ion (commonly to a few MeV) before injecting it into the synchrotron ring. When the ion transits through the radiofrequency cavity, it is accelerated, acquiring energy and increasing velocity. To keep particles in a constant orbit, the magnetic field of the bending magnet must be increased to match this increase in velocity. As a result, the synchrotron accelerates the particle at a constant rate even when the particle energy increases. The particles are circulated around the storage ring by a series of magnets separated by straight sections. These dipole magnets or bending magnets are used to propel the particles around the ring (see Fig. 4.5). The particle loses energy in the form of light as it passes through each magnet. The light from the storage ring wall can then be directed into the experimental stations known as beamlines. Total beam line is 8 in 4 irradiation room. And the beam size enlarge in WERC is by Wobbler magnet (see Fig. 4.6). The lateral broadening of the beam completely relies on the magnetic deflection. Finally extracted beam passed through a beam monitor to irradiate the target [67], [70].

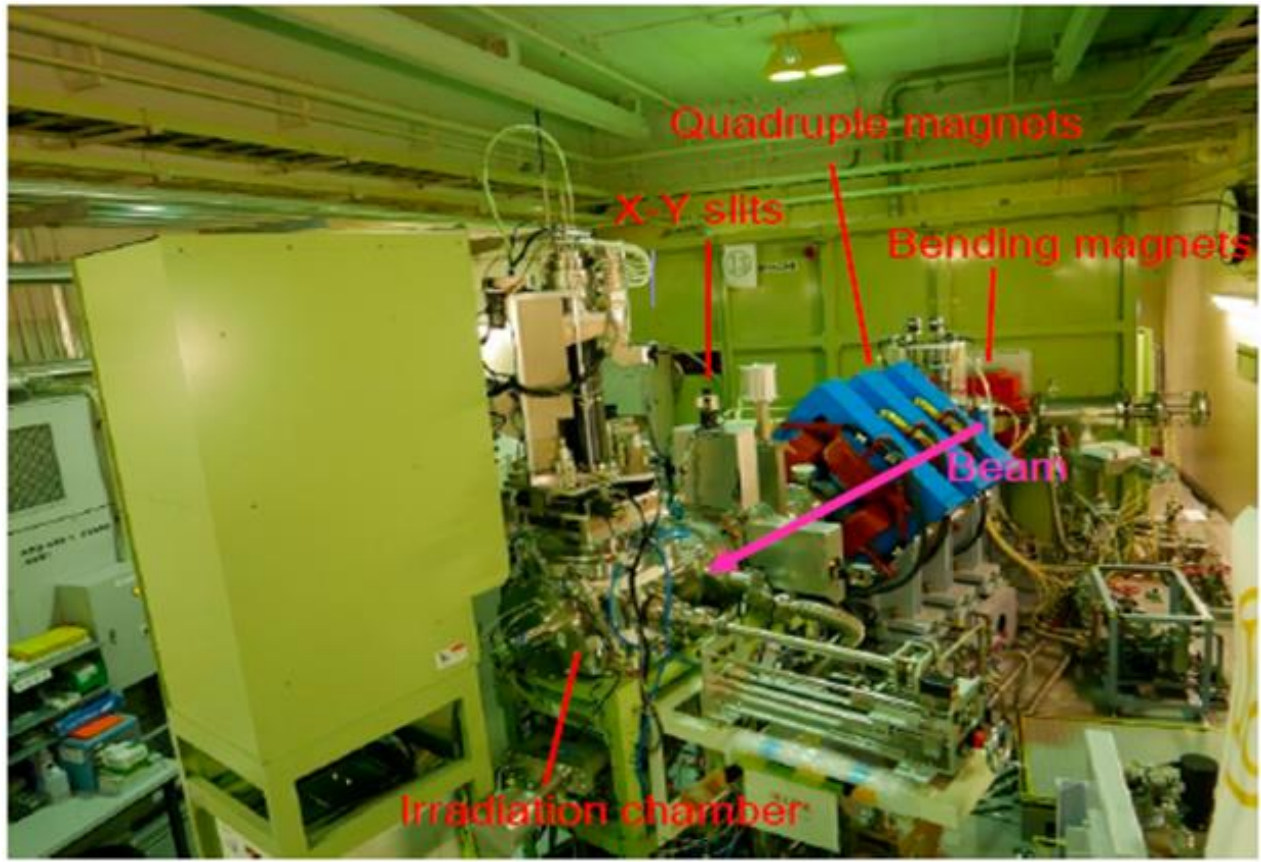

*Figure 4.5 The ion beam accelerated by the ion-implanter in irradiation room 1 is manipulated using a bending magnet and three quadrupole magnets. The beam size is defined by X-Y slits shown in the figure.*

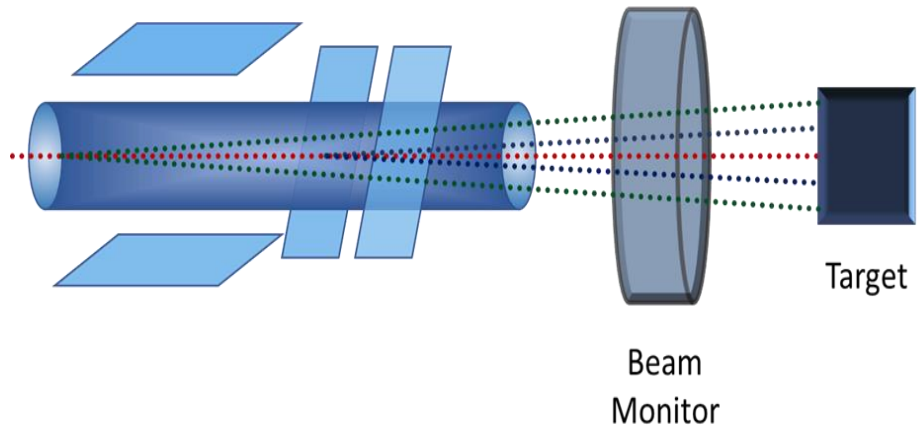

*Figure 4.6 Beam size enlarge in WERC is by Wobbler magnet shown by this illustration. The lateral broadening of the beam completely relies on the magnetic deflection. Finally extracted beam passed through a beam monitor to irradiate the target.*

## 4.3　HIMAC

At the end of February 1994, the heavy-ion medical accelerator in Chiba (HIMAC) was commissioned for preclinical trials. After 3.5 months of preclinical studies, a clinical trial using heavy-ion irradiation began on June 24, 1994. Over 300 patients were treated using fully stripped carbon-ion beams of 290, 350, and 400 MeV/n over the course of 3.5 years. Physical and biological tests for the preparation of heavy-ion radiotherapy have previously been carried out at the RIKEN ring cyclotron facility (The Institute of Physical and Chemical Research, Japan), employing 135-MeV/n carbon and neon beams. The physical properties of the beam-delivery system were measured during the past 3.5 years of heavy-ion treatments, and new methods for irradiating heavy-ion beams were developed. For irradiation synchronized with patient breathing and three-dimensional (3-D) irradiation, a synchronized irradiation technique was established [71]. We will discuss on irradiation system in a later section.

### *4.3.1 Overview of HIMAC accelerator*

The HIMAC's design specifications were determined by estimating radiological requirements. The energies of ion species ranging from He to Ar can be accelerated in this facility. For effective therapy, the beam energy was designed to range from 100 to 800 MeV/n. An injector linear accelerator (linac) cascade, dual synchrotron rings with independent vertical and horizontal beam lines, and three treatment rooms with passive beam delivery devices make up the HIMAC.

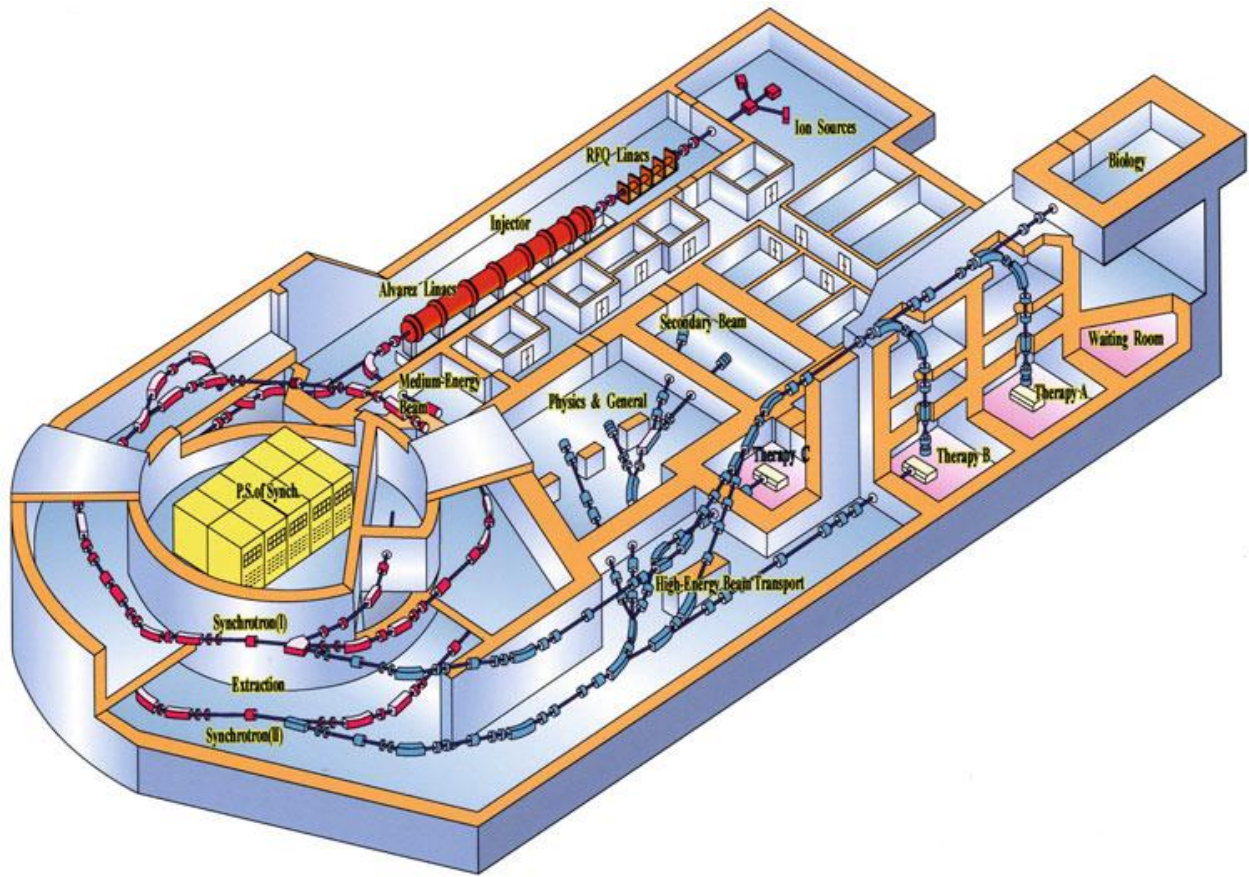

*Figure 4.7 A bird's-eye view of the HIMAC facility A unique doublesynchrotron ring heavy-ion accelerator system dedicated for medical use was designed and constructed. It consists of two ion sources, an RFQ (radio frequency quadrupole) linear accelerator (linac), an Alvarez linear accelerator (linac), two synchrotron rings, a high-energy beam transport system, and an irradiation system* [72].

$C^{2+}$ beam produced by the 10 GHz-ECR ion is injected into the linac cascade consisting of RFQ and Alvaretz linacs and accelerated up to 6 MeV/n for carbon ion radiotherapy. After entirely stripping the carbon beam with a carbon-foil stripper, the beam is injected into the synchrotron rings using a multiturn injection system and slowly extracted after reaching the necessary energy level. Finally, the beam delivery mechanism delivers the carbon beam collected from the synchrotron. Fig. 4.7 depicts a bird's-eye view of the HIMAC facility, and Table 4.1 lists the HIMAC's specifications. The HIMAC has a footprint of 120 by 65 meters and a construction cost of 32,600,000,000 JPY [73].

*Table 4.1 Specifications of the HIMAC* [72].

| Ion: H-Ar | |
|---|---|
| **Max energy** | 100-800 MeV/n |
| **Treatment room** | |
| Fixed vertical | Room A |
| Fixed horizontal | Room C |
| Fixed vertical and horizontal | Room B |
| **Accelerated energy** | |
| Vertical beam | 140 or 290 or 350 or 400 MeV/n |
| Horizontal beam | 140 or 290 or 400 or 430 MeV/n |
| **Range of carbon ion beam in water** | |
| 140 MeV/n | 5 cm |
| 290 MeV/n | 15 cm |
| 350 MeV/n | 20 cm |
| 400 MeV/n | 25 cm |
| 430 MeV/n | 30 cm |
| **Maximum field size** | 15 cm by 15 cm |

## *4.3.2 Irradiation system of HIMAC*

Two ion sources (ECR and PIG), an RFQ linac, an Arvaretz linac, and two synchrotrons make up the HIMAC accelerator complex (the upper ring and the lower ring). There are three clinical treatment rooms (Rooms A, B and C): one biological experiment room, two large general experimental halls, and a low-energy experimental room for the linac beams in the HIMAC facility. Treatment Room A has a vertical beam line, Room C has a horizontal beam line, and Room B has both vertical and horizontal beam lines that cross at the irradiation area. Each clinical treatment room and the biological experiment room have identical irradiation systems. Fig. 4.8 depicts the irradiation system of the treatment rooms at the HIMAC facility.

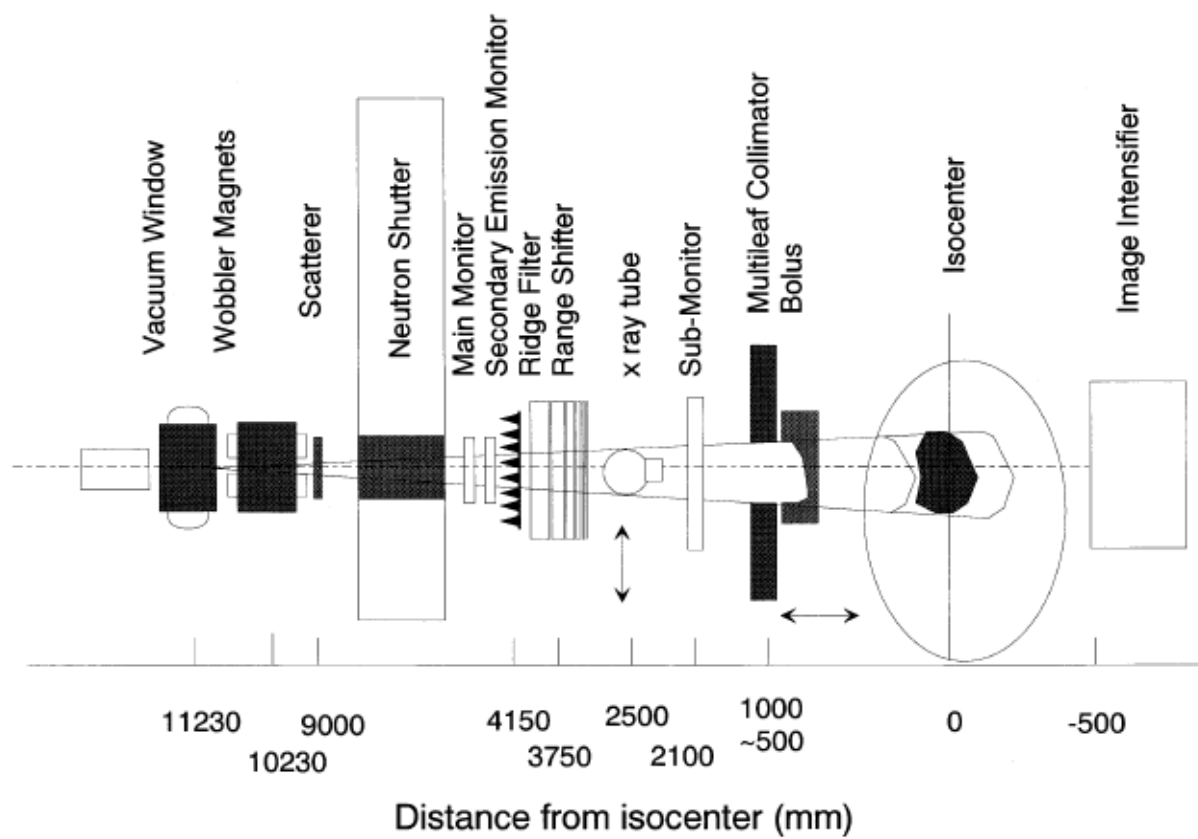

*Figure 4.8 Illustration of irradiation system of the treatment rooms at the HIMAC facility [71].*

### 4.3.3 Wobbler system

Figure 4.8 shows the treatment room's irradiation system at the HIMAC facility. The irradiation system includes wobbler magnets, beam monitors, a scatterer, a ridge filter, a range shifter, collimators, patient-positioning devices, and a patient couch. A pair of wobbler magnets and a scatterer are used to produce consistent irradiation fields. The range shifter is used to alter the heavy ion's residual range in the patient. In the depth-dose distribution of heavy ions, the ridge filter is employed to spread out the Bragg peak. The irradiation courses at Lawrence Berkeley National Laboratory, the NIRS cyclotron facility and the RIKEN ring cyclotron facility, all use the same conceptual method to create a homogeneous irradiation field. An accelerated beam is focused on the irradiation site of the irradiation course. With the wobbler magnets, the concentrated beam can draw a circular track at the irradiation site. The wobbled beam is then scattered at the irradiation point to create a uniform field. In the case of a 0.5 Hz repetition rate, the beam pulse width is approximately 200–250 ms [72], [73]. The beam pulse had a fine time structure of only 11–13 spikes with a 50-Hz recurrence before beam extraction utilizing an RF knock-out technique or enhancing the reliability of the synchrotron power supplies. The wobbler technique, in which the three spikes were spread in a circular trace and rotated approximately 1/3 of a turn during one beam pulse, was successful in achieving a homogeneous irradiation field even in this circumstance of spiked beam pulses. In such case, the frequency of the beam wobbling was 62.14 Hz. The beam pulses are now extended inside about 1 s with a repetition rate of 0.33 Hz while developing a new beam extraction technology or increasing power sources. The difference between the intensities at the central part and those at the peripheral highest part was less than 2% of the average intensity in large irradiation fields with a diameter of over 22 cm [71].

### *4.3.4 Beam-modulating system*

Inserting bar ridge filters in the beam course creates a SOBP. The ridge filter is made of aluminum. Each bar ridge has a 5 mm spacing and does not move during irradiation. Shades of the bar ridge are spread out at the irradiation point due to repeated scattering in the ridge filter and the angular dispersion of the wobbled beam. Aluminum ridge filters are lined up for SOBP in clinical trials from 2-cm to 12-cm width, at which point the height of the aluminum ridge becomes about 6 cm. A brass ridge filter is utilized for a SOBP with a width of 15 cm. We can choose a suitable ridge filter from an array of eight ridge filters installed on a large wheel. In the SOBP, the ridge filter is meant to ensure that the survival fraction of human salivary gland tumor cells (HSG cells) is uniform.

A range shifter is positioned immediately downstream of the ridge filter to modify the heavy-ion range in patients. The range of the heavy ion can be adjusted in 0.5 mm increments by changing the thickness of the range shifter's polymethyl metacrylate (PMMA) plates. The irradiation field for each patient is defined using a multileaf collimator. The leaf is 6.5 mm wide and 14 cm thick. A total of 23 leaves are stacked on one side. The gaps between the leaves are kept within 0.2 mm. Each leaf includes two 0.6-mm steps on both sides to prevent heavy-ion beams from leaking through the gaps and to move quickly enough to block the beam dynamically in the case of 3-D irradiation. For shaping the distal part of the irradiation field, a compensator can be installed on the housing of the multileaf collimator. The compensator is manufactured from a polyethylene block on an NC milling machine, according to the desired shape using a treatment-planning system. A patient collimator, which is manufactured by cutting a brass block, is used for fields smaller than 10 cm in diameter. The collimator system can be moved along the beam line. The collimator system is moved back from the location while the patient is being positioned on the couch. The collimator is brought as close to the patient as feasible when irradiating a patient in order to sharpen the irradiation field's penumbra. A target is fixed on the irradiation site using two or three sets of x-ray tubes and image intensifiers [71].

## *4.3.5 Monitor system*

In HIMAC's beam delivery system, five dose monitors are employed to track the irradiation dose to patients. Upstream monitors are situated directly above the ridge filter, whereas downstream monitors are positioned right above the multileaf collimator. Upstream monitors include parallel-plate ionization chambers and a secondary emission chamber (SEC). Two signal electrodes and three high-voltage electrodes make up the parallel-plate ionization chamber of the upstream monitors. One signal is utilized to monitor the time structure of the beam and is put into a high-speed amplifier. For every 1000 pC of the output signal, the other output signal is supplied into a charge-to-frequency (I/F) converter, which converts the collected charge to a pulse. The pulse train is delivered to a preset counter, which is connected to a beam shutter, that controls the patient's exposure dose. The downstream monitor is also a parallel-plate ionization chamber. One signal plane is sandwiched between two high-voltage electrodes in the ionization chamber. There are 29 signal electrodes, each 20 mm in diameter, distributed in the guard ground on one side of the signal plane. The homogeneity of the irradiation field is monitored using these ionization chambers. For example, it is used to check that the phase of the wobbler magnets is not zero degrees, to check field size, or to check unexpected materials inserted in the beam line. A signal electrode on the other side of the signal plane covers the whole uniform irradiation field. The dose supplied is monitored using an SEC. Because of recombination effects during the big spike, the ionization chamber cannot provide an accurate dose measurement. In the SEC, four signal electrode aluminum sheets are placed between five high-voltage electrode aluminum sheets. Aluminum sheets have a 7 mm thickness [71]. The SEC has 150-mm thick aluminum entrance and exit windows. In the SEC, the vacuum level is less than $10^{-5}$ Pa. The collected current of the secondary electron is proportional to the fluence rate of the heavy ions and the energy loss in the aluminum sheets.

## 4.4 Target-detector assemblies and exposures

With the use of proper target-stack designs, CR-39 nuclear track detectors can be used to measure projectile fragmentation cross sections. We have utilized CR-39 detectors for fragmentation studies of $^{12}$C projectiles on aluminum targets. The approach of these experiments is the detection of the $^{12}$C ion beam incident on the target Al and after the target Al where the surviving beams are present and the secondary particles or fragments produced in the interaction of the $^{12}$C ion beam with the Al target.

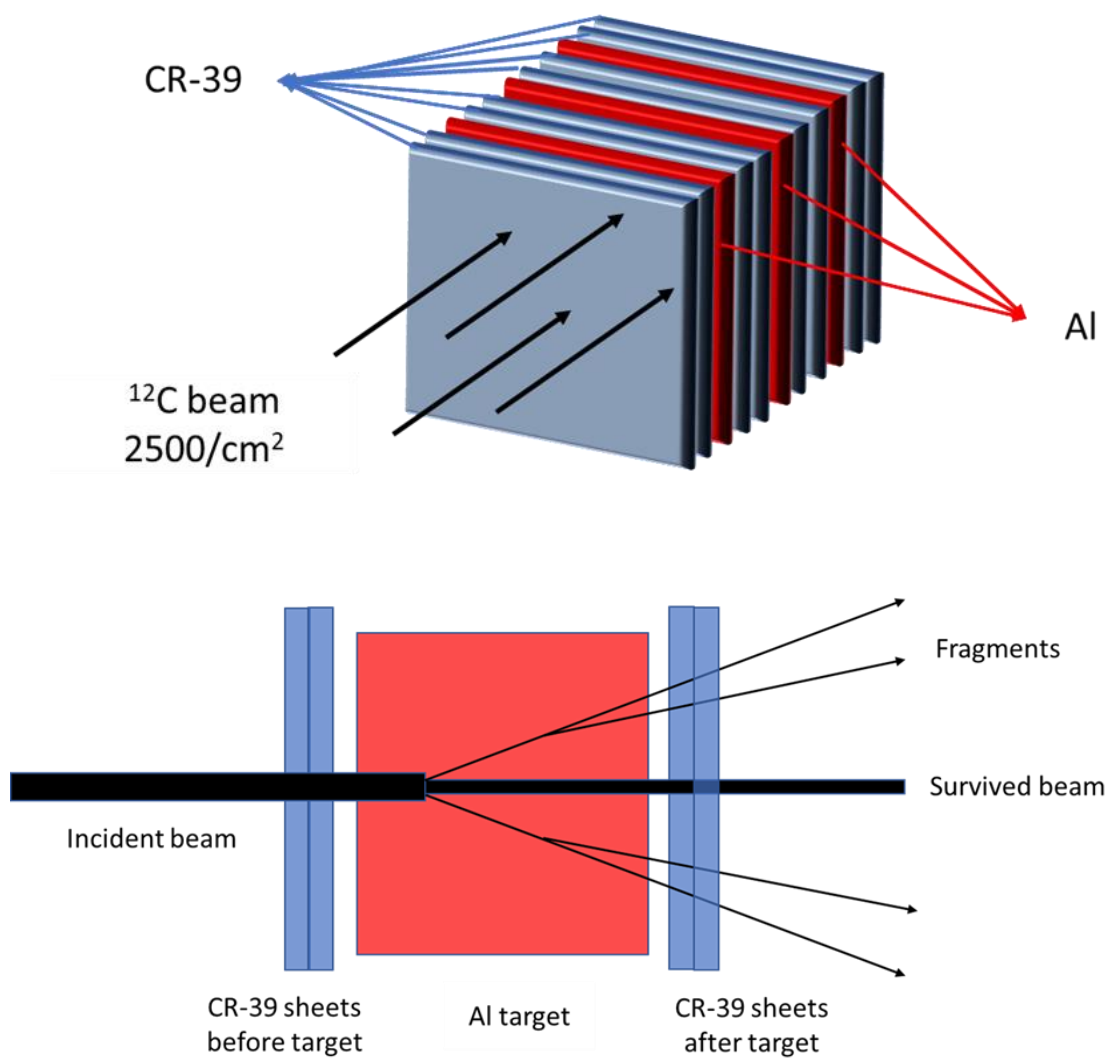

*Figure 4.9 Schematic showing (a) Stack of NTDs and (b) exposure geometry. The approach of these experiments is the detection of the $^{12}$C ion beam incident on the target Al and after the target Al where the surviving beams are present and the secondary particles or fragments produced in the interaction of the $^{12}$C ion beam with the Al target.*

The setup of CR-39 detector-Al target assemblies and exposure are shown in Fig. 4.9. Multiple such stacks of CR-39 detectors and Al targets were exposed to 55 MeV/n and 135

MeV/n $^{12}$C beams at the WERC and HIMAC facility, respectively. The target thicknesses were chosen in such a way that multiple interactions were negligible but still appropriate to produce sufficient fragments. The CR-39 detector plates were a dimension 5cm × 5 cm and 0.4 mm in thickness, and the thickness of target Al was 0.1 mm. The total number of incidents of $^{12}$C ions were determined using CR-39 detectors located upstream of the targets. The intensity of $^{12}$C beam ions was around 2500 and 2700 ions/cm$^2$ for WERC and HIMAC exposure, respectively. The beam then passed through the Al target of thickness typically half of the mean free path of the $^{12}$C ions in the given Al target. Both the survived $^{12}$C ions and their fragments produced in Al targets were recorded by CR-39 foils downstream of the Al target. NTD based experiments are feasible compared to those which utilize ionization chambers (ICs), multi wire proportional chambers (MWPCs), and Cherenkov detectors (CDs). The later experiments need online data acquisition, whereas NTDs allows for offline data acquisition and are appealing to research groups without direct access to high-energy accelerator facilities [53], [74].

## 4.5 Chemical etching

The change in etching temperature was less than ±1°C. During the etching of exposed CR-39 detectors, a standardized procedure was followed. An effective and balanced level of stirring was kept. The concentration of etchant was efficiently maintained under particular etching conditions by limiting the water evaporation from the etching solution. Exposed CR-39 detectors were etched in 7N NaOH water solution at a temperature of 70°C for 15 hours. Bulk etch rate, the rate of removal of layers of the undamaged surface of CR-39, was determined using the mass decrement method. This bulk etchs rate measurement was applied as a method to ensure that etching conditions in all experiments were the same as they were set [75], [76].

Additionally, the damaged zone of CR-39 from particle impact etches faster than the undamaged zone of the plate during the etching process. This results in the generation of nuclear tracks, usually referred to as pits. It is important to characterize the pits in order to determine the species and energy of the particles. The diameter and shape of the pits, on the other hand, are dependent not only on the linear energy transfer of the particle but also on the individual CR-39 plate and etching conditions. As a result, a set parameter is necessary to compare data analyzed under different etching conditions [53], [55], [76].

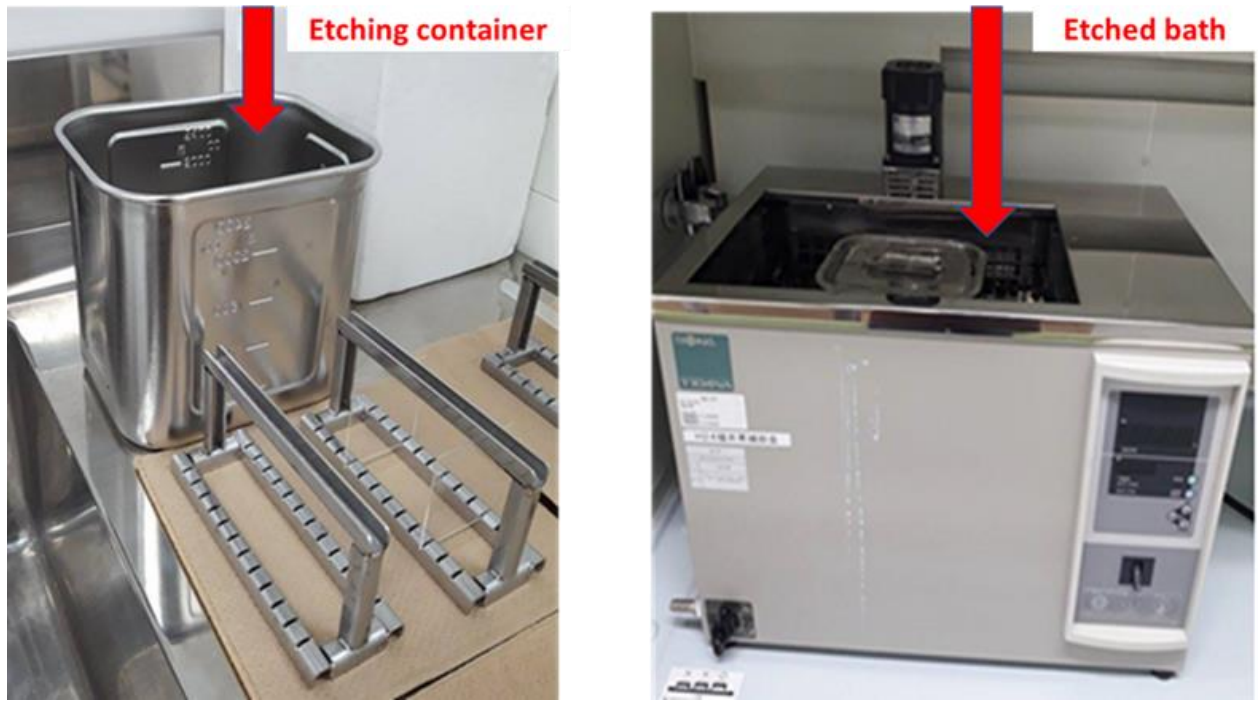

*Figure 4.10 Etching process of CR-39 nuclear track detector. The concentration of etchant was efficiently maintained under particular etching conditions by limiting the water evaporation from the etching solution. Exposed CR-39 detectors were etched in 7N NaOH water solution at a temperature of 70°C for 15 hours.*

## 4.6 Scanning of CR-39 detector

Track measurements are just as necessary as the exposure and chemical etching experiments mentioned earlier. During etching, nuclear track measurements are taken once or several times, depending on the experiment's purpose. Microscopic cracks and scratches develop differently from nuclear tracks; therefore, they may be differentiated from tracks even when they are still in the initial stages of growth [53], [51].

After chemical etching, the resulting track parameters (diameter, length, and density) were analyzed in a CR-39 detector using an FSP-1000 imaging microscope (see Fig. 4.11) automatically. All the track measurements at the WERC and HIMAC laboratory were done automatically. If track parameters are measured manually, the process is incredibly slow but provides visual insights of track forming particles. In manual measurements, discrimination between actual events and erroneous or background events is more reliable than discrimination done by automatic nuclear track scanning. The measurement of cone heights of tracks in two CR-39 foils for the case of each Al target used, two sets of foils before each Al target and two after, were measured with an optical microscope automatically. It was found the cone length increased as the ion charge grew. Observations of changes in cone heights were used to measure nuclear fragmentation. SEIKO FSP-1000 microscope with a magnification of 20X was used for these measurements, and the image resolution of this microscope is 0.25 x 0.25 μm$^2$/pixel. The depth measuring system paired with the microscope was used to measure the cone heights of $^{12}$C projectiles and fragments from the top view. Difference of readings on depth measuring instrument, one when track diameter at the detector surface is focused and other when endpoint of etchable range is focused in the micro- scope, gives the cone height of a track. Besides, a circle or a square was drawn in the exposed region of the detector whenever the growth of specific tracks was seen [76], [57].

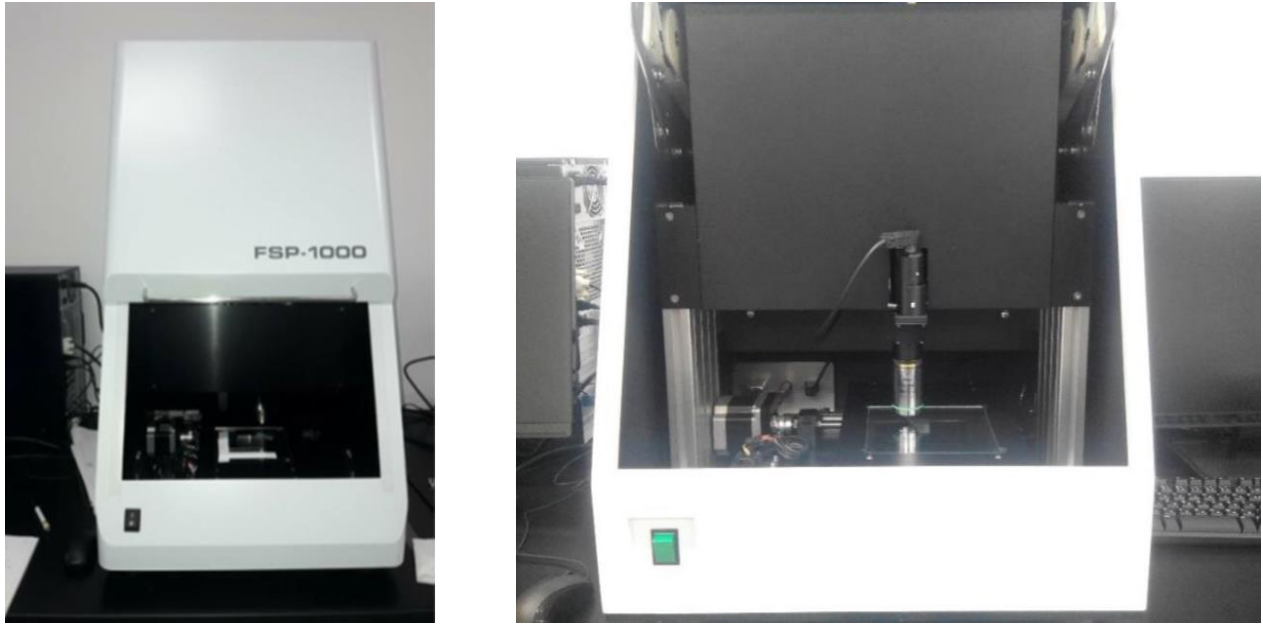

*Figure 4.11 The resulting track parameters (diameter, length and density) were analyzed in CR-39 detector using this FSP-1000 imaging microscope.*

# Chapter 5
# Results and Discussions

## 5.1 Imaging technique of CR-39 detector

We can take the front and back side images simultaneously by the automatic surface detection method of the microscope from the vertical movement of the objective lens, as shown in the Fig below:

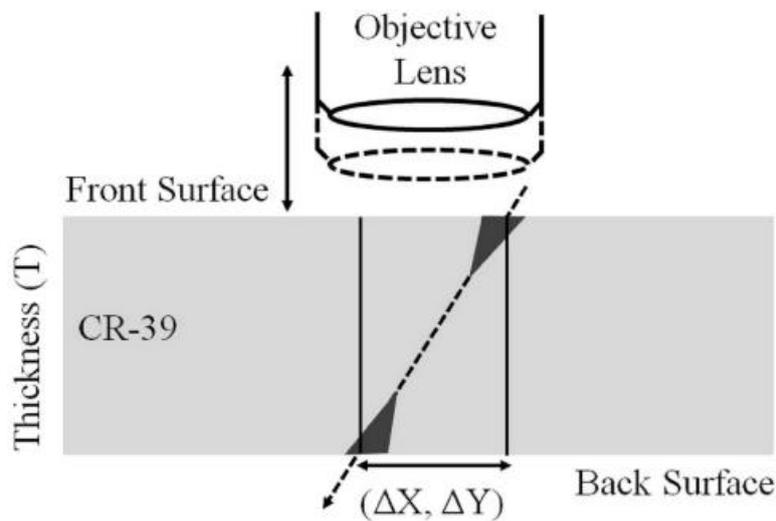

*Figure 5.1 Both surface imaging of CR-39 detector simultaneously [23].*

By this imaging technique we got the image for front and back surface as shown in the following figure:

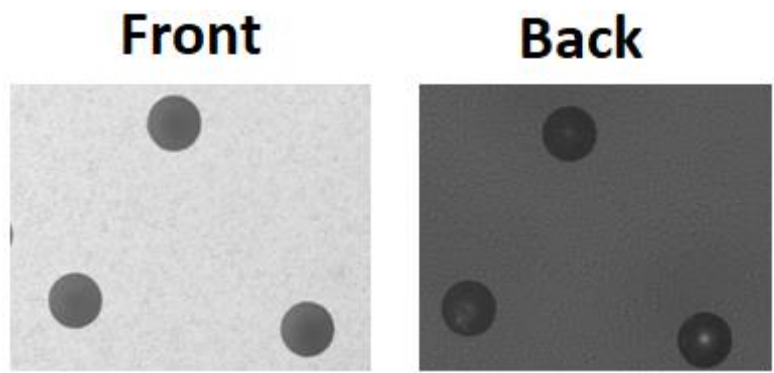

*Figure 5.2 The front shows the ion track, etch pit (solid black circles) in front surface of the CR-39 detector and back shows the corresponding etch pit in back surface. So, from both the figure we easily can track the projectile ion through the path. The view of the objective lens at 20x magnification is about (0.6 mm x 0.43 mm) 0.25 mm² area.*

In the figure, the front shows the ion track and etch pit (solid black circles) on the front surface of the CR-39 detector, and the back shows the corresponding etch pit on the back surface. So, from both the figures, we can easily track the projectile ion through the path. The view of the objective lens at 20x magnification is about (0.6 mm x 0.43 mm) 0.25 mm$^2$ area. Since the CR-39 was 50 mm x 50 mm in area, we divided the whole area into 16 divisions, each of about (12.6 mm x 12.5 mm) as shown in following Fig. 5.3.

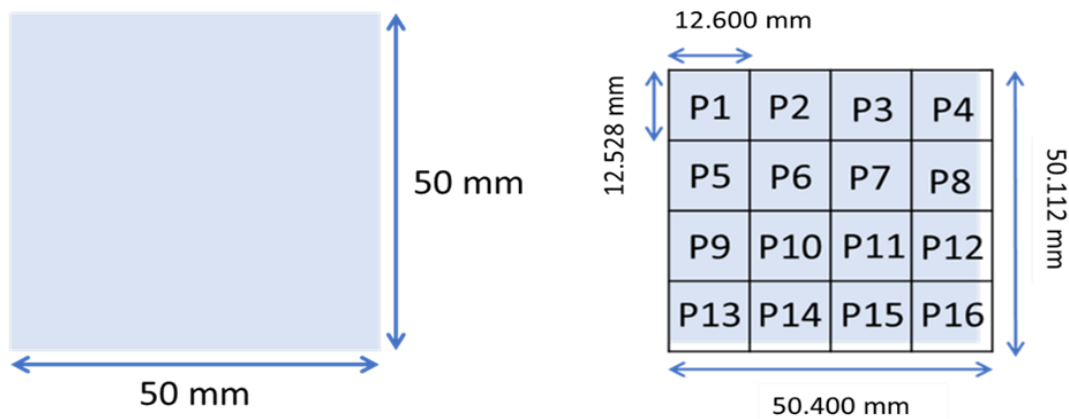

*Figure 5.3 Left side figure shows the whole CR-39 of 50 mm x 50 mm. Right side figure shows the 16 division of 12.6 mm x 12.5 mm dimension for the whole CR-39 detector.*

After dividing CR-39 into 16 divisions, each division scanned and stored the image of both the front and back surfaces in a directory. So, for each division (P1 to P16), we have 1218 images, 609 for the front and 609 for the back surface.

## 5.2 Synthesized method of images

Our target is to count the etch pit before and after the target. For that, we synthesized the etch pit images which are before the target on the corresponding etch pit images after the target, as described in Fig 5.4. In this figure, the top shows the irradiation system where an Al is sandwiched by two CR-39 detectors. The red line shows the back surface of the detector before the target, and the blue line shows the front surface of the detector after the target. In the lower portion, I show the etch pit images on the back surface, and II shows the corresponding images on the front surface after the detector. Then the etch pit on both sides before and after the targets were synthesized by the following formula:

205 - {(pixel value of I + 200) – pixel value of II}

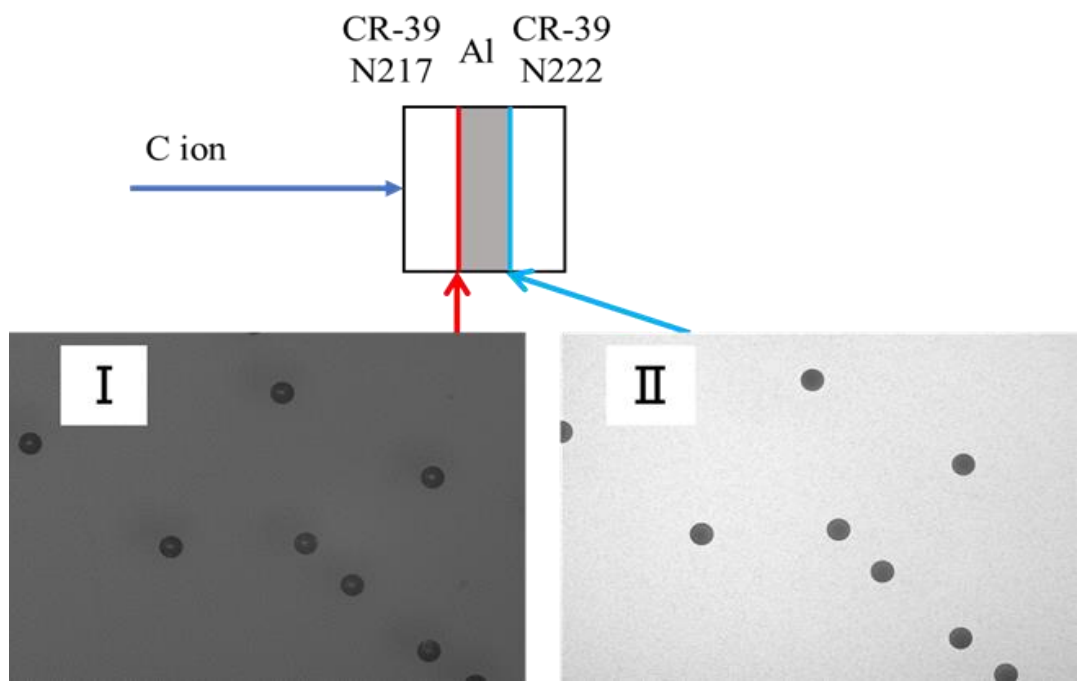

*Figure 5.4 Back and front surface of CR-39 image. In this fig, the top shows the irradiation system where an Al is sandwiched by two CR-39 detectors. Red line shows the back surface of the detector before the target and blue line shows the front surface of the detector after the target. In lower portion I shows the etch pit images in the back surface and II shows the corresponding images in the front surface after the detector.*

and obtained the image as below. The numerical value in the above formula is just to obtain a clear image as shown.

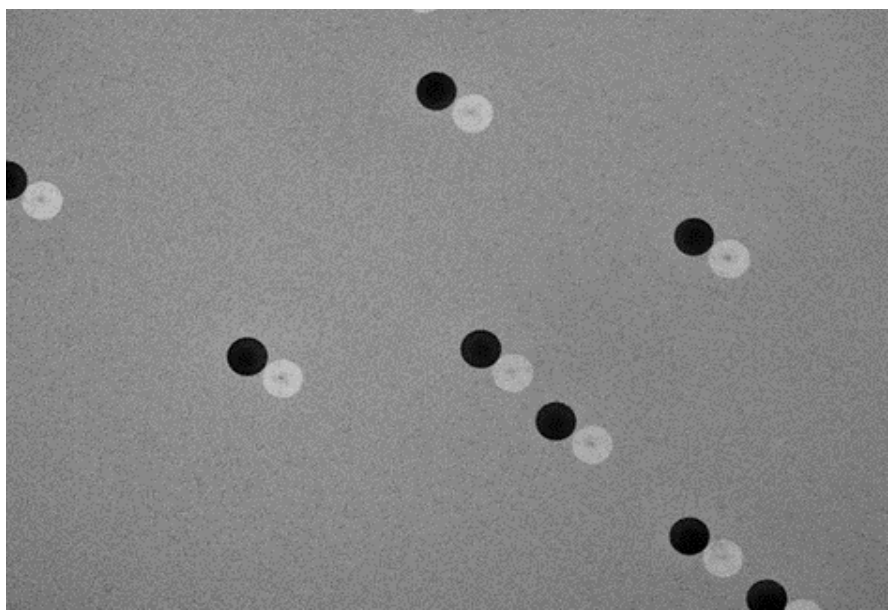

*Figure 5.5 Synthesized image and etch pits. White circles and black circles are almost the same size and are in the same position. The white circles are the number of incoming particles to the target, $N_{in}$ and the black circles are the number of survival particles after the target, $N_{out}$.*

In Fig. 5.5, white circles are the etch pits on the back surface (I) and black circles are the etch pits on the front surface (II). White circles and black circles are almost the same size and are in the same position. The white circles are the number of incoming particles to the target, $N_{in}$, and the black circles are the number of survival particles after the target, $N_{out}$. The total number of incoming carbon ions is referred to as $N_{in}$, and survival carbon ions, are referred to as $N_{out}$, which were counted manually.

## 5.3 Definition of Carbon

For measurement of charge changing cross-section, we need to know the number of carbons before and after the target. So, in this experiment, we will have the ion track, which is called etch pit in the CR- 39 detectors. Among the etch pits images, we need to define which are carbon so that we can ignore fragments (lighter and heavier) and noise etch pit images as shown in the following figure.

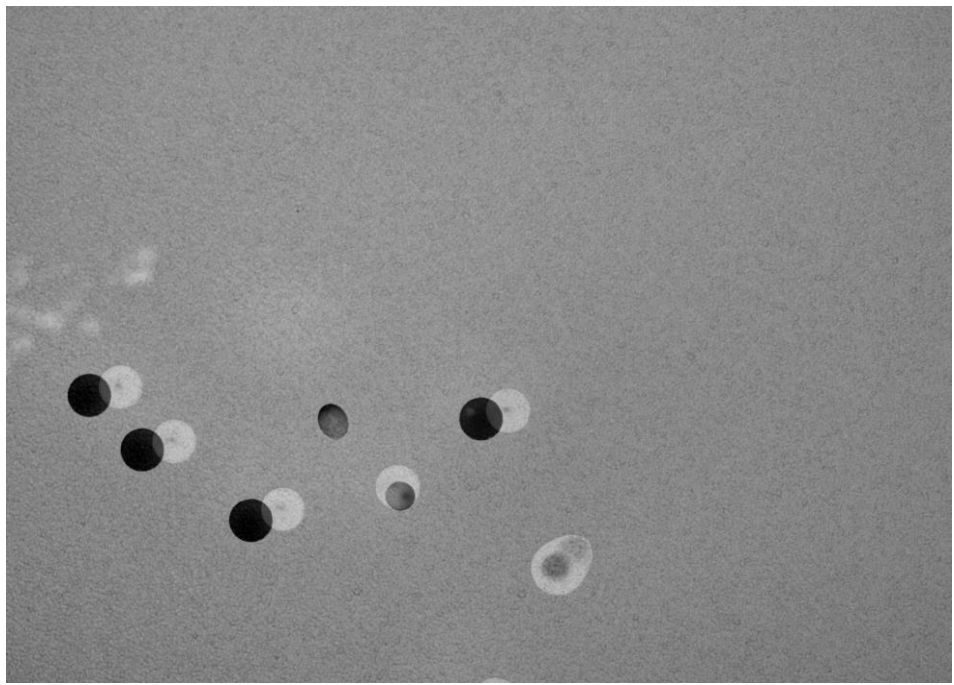

*Figure 5.6 Carbon tracks, lighter fragments and noises are present in this figure. We count the carbon manually which etch pits have the features of completely circular, same size and shape, ions have the same level of gray scale (white in front and black in back).*

In Fig. 5.6, there are carbon tracks, lighter fragments, and noises. These tracks are produced based on the Linear Energy Transfer (LET) of the ion. Since all carbon ion has almost the same energy, so there LET will be the same. Since we count the number manually, so based on the eye, we define the carbons whose etch pits have the following features.

i. Completely circular.

ii. Same size and shape.

iii. All most all ions have the same level of gray scale (white in front and black in back).

Based on those criteria, we can say that the above image has 4 carbon ions before and after the target.

## 5.4　Measurement method of $\sigma_{TCC}$

For measurement of $\sigma_{TCC}$ we need to know the accurate number of carbons before and after the target. For counting the number, a protocol has been developed in this study. The protocol has been described below:

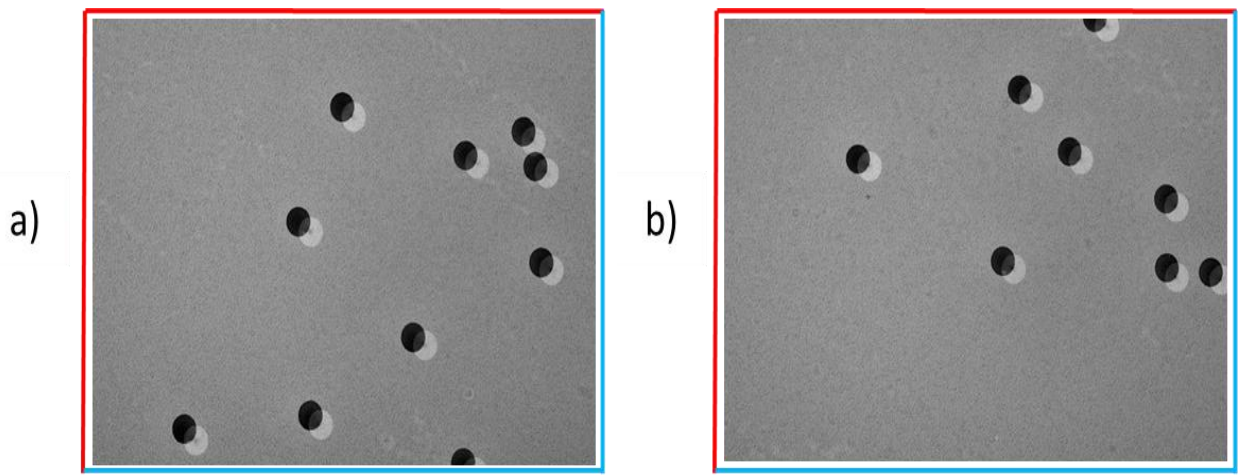

*Figure 5.7 Example of synthesized images. The frame of image has four side identified by the red and blue line. If the etch pit touch the red line they will be counted. And if the etch pits touch the blue line they will not be counted.*

Etch pits may appear in the four corners of the image.

The frame of the image has four sides identified by the red and blue lines. If the etch pit touches the red line, it will be counted. And if the etch pits touch the blue line, they will not be counted since they will appear in the next frame due to the overlap of the imaging area and will touch the red line, which will be counted there as shown in the following figure. To avoid the two ties counting some etch pits that overlapped with the next frame, these red and blue line boundary has been introduced. According to this protocol

If both etch pits (white and black) cover the upper left corner (indicated by a solid red line at the top and left), they are counted as $N_{in}$ and $N_{out}$.

- If both etch pits (white and black) cover the upper right corner, lower left corner, and lower right corner, they are not counted as $N_{in}$ and $N_{out}$.

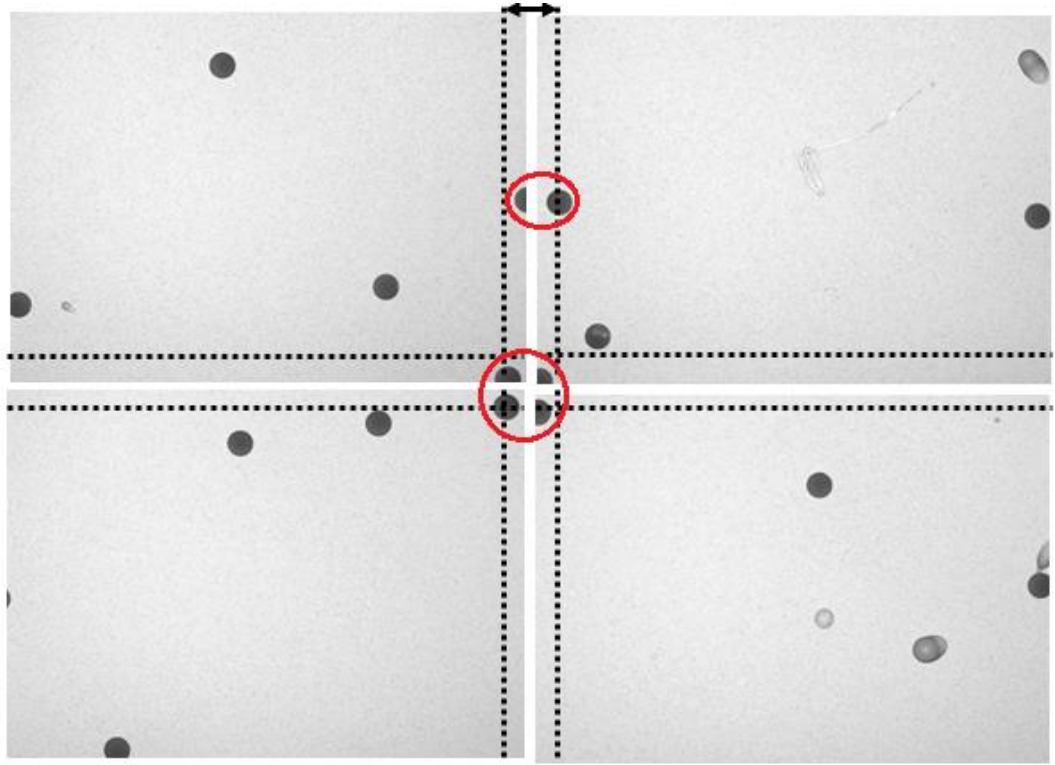

*Figure 5.8* In this figure it is shown that the etch pits enclosed by the red circles are appear twice in next frames.

- When etch pits overlap slightly (case-a) in following Fig. 5.9, and from the size of the etch pit, it can be judged that these are carbon, the number of $N_{in}$ and $N_{out}$ is counted as 2 carbon etch pit.
  - When etch pits overlap greatly (case-b) in the following Fig. 5.9, etch pits that are difficult to judge are ignored and recorded for further study.

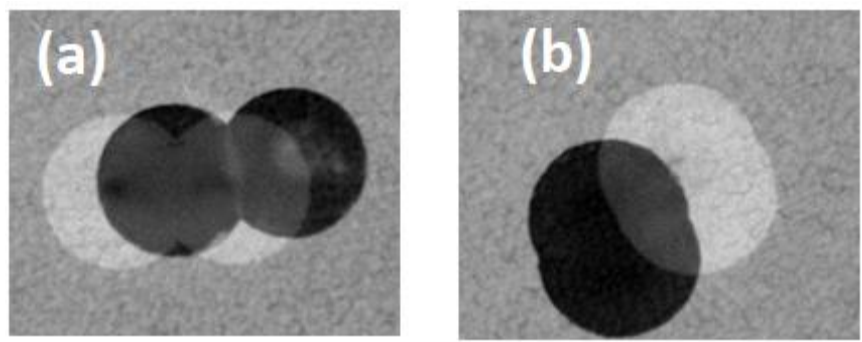

*Figure 5.9 Example of overlapping etch pits.*

- Etch pit size that is smaller than the surroundings are ignored as carbon and recorded for the charge identification of particles.

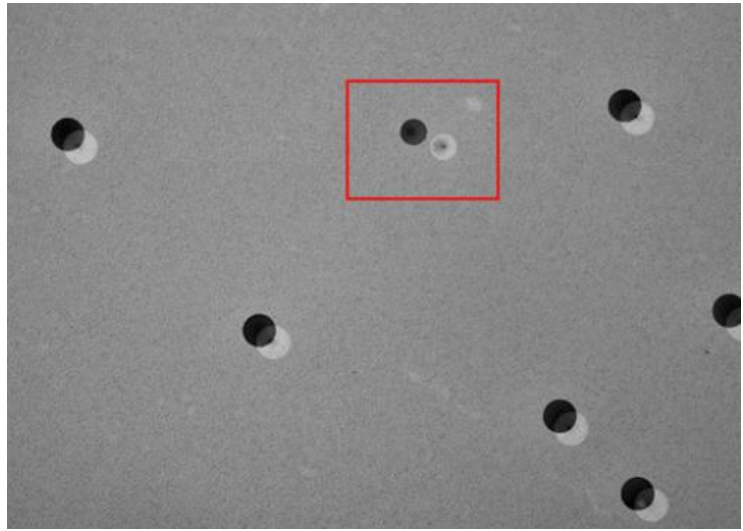

*Figure 5.10 Example of Synthesized images with light nucleus.*

- Other unidentified etch pits are also recorded for further study as shown on the following Fig. 5.11.

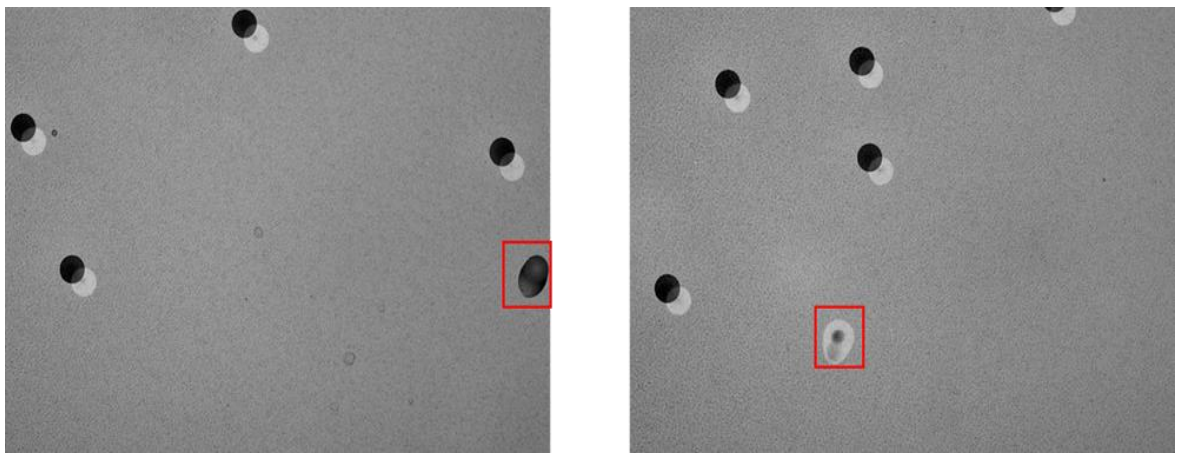

*Figure 5.11 Example of Synthesized images with unidentified etch pits.*

## *5.4.1 Fragment*

In the following figure, it is seen in the box surrounded by the red line that, instead of a black etch pit which is corresponds to a white etch pit, two smalls etch pits appeared. This means that a carbon ion has been fragmented into two lighter particles. In following Fig. 5.12, the etch pit surrounded by a red square shows that carbon has become two fragments. In this way, the etch pit in which the nuclear reaction occurred is recorded as a fragment. The etch pit surrounded by the blue square is not carbon. Etch pits other than carbon are also recorded for further study.

Chapter 5      Results and Discussions

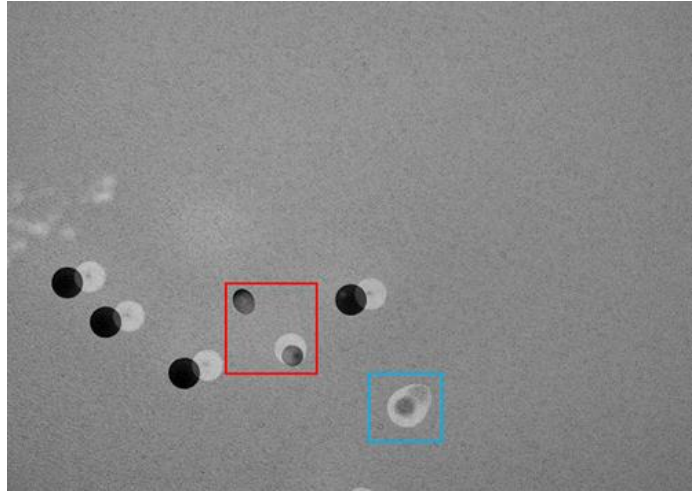

*Figure 5.12 Example of Synthesized images with fragments.*

## 5.5 Total charge changing cross section

The total reaction cross-section, $\sigma_R$ as a function of energy, E determines the probability of a nuclear reaction, is one of the most basic parameters for heavy-ion transport calculations. However, it is extremely difficult to assess all possible reaction channels during an experiment, so experimental cross-sections are frequently simply estimates of the total reaction cross-sections. Many experiments determine charge-changing cross-sections $\sigma_{cc}$, which show the probability that the atomic number ('charge') of the projectile changes because an accurate charge identification is straightforward with simple particle detection systems. The charge-changing cross-section is a decent approximation to the reaction cross-section since most nuclear fragmentation channels result in the loss of at least one proton. The total charge changing cross sections were calculated with the survival fraction of ions using the following formula

$$\sigma_{TCC} = -\frac{M}{\rho N_A X} \ln(\frac{N_{out}}{N_{in}}) \times 10^{27}$$

Where, $N_A$ = Avogadro number

$\rho$ = Density of the target

M = Atomic mass of the target ($M_{Al}$ = 27)

X = Thickness of the target

$N_{in}$ = Total number of incoming $^{12}$C ion

$N_{out}$ = Survival $^{12}$C ion after passing through the target

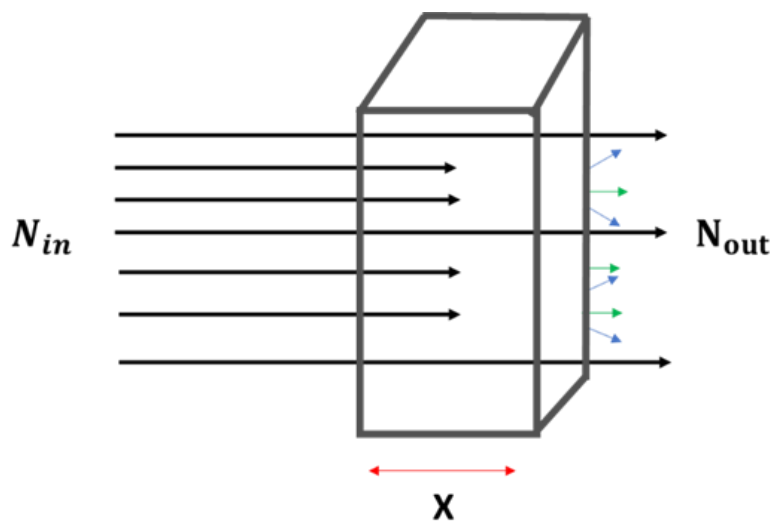

*Figure 5.13 $^{12}$C ion passing through the target. Interaction of projectile with the target reduces the number of projectiles after the target. From these number of projectiles before and target charge changing cross section can be measured.*

The following table shows the downstream detector ID, etching duration, and bulk etch, incident energy, outgoing energy of the projectile, energy at which TCC has been measured, and uncertainty in energy for the WERC stack. The grey row shows the target position and energy point at which cross-section has been measured.

*Table 5.1 This table shows the downstream detector ID, etching duration, Bulk etch, incident energy, outgoing energy of projectile, energy at which TCC has been measured and uncertainty in energy for WERC stack. The grey row shows the target position and energy point at which cross section have been measured.*

| Detector no. | Thickness (mm) | Etching duration (hr) | Bulk etch (μm) | $E_{in}$ (MeV/n) | $E_{out}$ (MeV/n) | E (MeV/n) | ΔE (MeV/n) |
|---|---|---|---|---|---|---|---|
| N210 | 0.412 | 20 hr | 25.5 | 37 | 34.5 | 35.7 | 1.2 |
| N217 | 0.397 | 20 hr | 25.7 | 34.5 | 32 | 33.6 | 1.2 |
| Al (0.1 mm) | | | | 32 | 30.9 | 31.5 | 0.5 |
| N222 | 0.413 | 20 hr | 25.7 | 30.9 | 28.1 | 29.5 | 1.4 |
| N223 | 0.422 | 20 hr | 25.9 | 28.1 | 24.9 | 26.5 | 1.5 |
| Al (0.1 mm) | | | | 24.9 | 23.1 | 24.1 | 0.9 |
| N224 | 0.441 | 20 hr | 26.6 | 23.1 | 19.4 | 21.3 | 1.8 |
| N212 | 0.397 | 20 hr | 26.6 | 19.4 | 15.4 | 17.4 | 1.9 |
| Al (0.1 mm) | | | | 15.4 | 13.5 | 14.5 | 0.95 |
| N213 | 0.406 | 20 hr | 26.4 | 13.5 | 7.9 | 10.7 | 2.7 |

*Table 5.2 This table shows the downstream detector ID, etching duration, Bulk etch, incident energy, outgoing energy of projectile, energy at which TCC has been measured and uncertainty in energy for HIMAC stack. The grey row shows the target position and energy point at which cross section have been measured.*

| Detector no. | thickness (mm) | Etching duration (hr) | Bulk etch (μm) | $E_{in}$ (MeV/n) | $E_{out}$ (MeV/n) | E (MeV/n) | ΔE (MeV/n) |
|---|---|---|---|---|---|---|---|
| **T051** | 0.467 | 25hr | 42.5 | 112.5 | 111.4 | 111.9 | 0.6 |
| **T052** | 0.469 | 25hr | 42.5 | 111.4 | 110.2 | 110.8 | 0.6 |
| **Al (0.1 mm)** | | | | 110.2 | 109.8 | 110.0 | 0.2 |
| **T053** | 0.471 | 30hr | 48.9 | 109.8 | 108.6 | 109.2 | 0.6 |
| **T054** | 0.473 | 30hr | 48.6 | 108.6 | 107.4 | 108.0 | 0.6 |
| **Al (0.1 mm)** | | | | 107.4 | 107.0 | 107.2 | 0.2 |
| **T055** | 0.476 | 25hr | 42.1 | 107.0 | 105.8 | 106.4 | 0.6 |
| **T056** | 0.463 | 25hr | 42 | 105.8 | 104.6 | 105.2 | 0.6 |
| **Al (0.1 mm)** | | | | 104.6 | 104.2 | 104.4 | 0.2 |
| **T057** | 0.462 | 25hr | 42.1 | 104.2 | 103.0 | 103.6 | 0.6 |
| **T058** | 0.463 | 25hr | 42.1 | 103.0 | 101.8 | 102.4 | 0.6 |
| **Al (0.1 mm)** | | | | 101.8 | 101.4 | 101.6 | 0.2 |
| **T059** | 0.467 | 25hr | 42.5 | 101.4 | 100.1 | 100.7 | 0.6 |
| **T060** | 0.471 | 25hr | 42.6 | 100.1 | 98.9 | 99.5 | 0.6 |
| **Al (0.1 mm)** | | | | 98.9 | 98.4 | 98.6 | 0.2 |
| **T061** | 0.459 | 25hr | 38.9 | 98.4 | 97.2 | 97.8 | 0.6 |
| **T062** | 0.458 | 25hr | 37.9 | 97.2 | 95.9 | 96.6 | 0.6 |
| **Al (0.1 mm)** | | | | 95.9 | 95.5 | 95.7 | 0.2 |
| **T063** | 0.462 | 30hr | 46.2 | 95.5 | 94.2 | 94.8 | 0.6 |
| **T064** | 0.466 | 30hr | 45.8 | 94.2 | 92.9 | 93.5 | 0.6 |
| **Al (0.1 mm)** | | | | 92.9 | 92.4 | 92.7 | 0.2 |
| **T065** | 0.469 | 25hr | 37.5 | 92.4 | 91.0 | 91.7 | 0.7 |
| **T066** | 0.455 | 25hr | 37.1 | 91.0 | 89.8 | 90.4 | 0.6 |
| **Al (0.1 mm)** | | | | 89.8 | 89.3 | 89.5 | 0.2 |
| **T067** | 0.458 | 25hr | 38.2 | 89.3 | 88.0 | 88.6 | 0.7 |

*Table 5.3 Following tables shows the number of incoming and outgoing particle in each surface of the downstream detectors in WERC. From which we can calculate the number of interactions. Since we can analyze the event by event interaction so we found how many carbons has fragmented into 1, 2, and 3 fragments among all ion.*

| Back surface of detector | Front surface of detector | No. of total $^{12}$C etch pit before C+Al interaction | Fragments | | | Missing | No. of total interactions | No. of total $^{12}$C etch pit after C+Al interaction |
|---|---|---|---|---|---|---|---|---|
| | | | 1 | 2 | 3 | | | |
| N217 | N222 | 22501 | 12 | 10 | 3 | 8 | 33 | 22468 |
| N223 | N224 | 22374 | 16 | 9 | 4 | 8 | 37 | 22337 |
| N212 | N213 | 22234 | 21 | 7 | 3 | 9 | 40 | 22194 |

*Table 5.4 Following tables shows the number of incoming and outgoing particle in each surface of the downstream detectors in HIMAC. From which we can calculate the number of interactions. Since we can analyze the event by event interaction so we found how many carbons has fragmented into 1, 2, and 3 fragments among all ion.*

| Back surface of detector | Front surface of detector | No. of total $^{12}$C etch pit before C+Al interaction | Fragments | | | Missing | No. of total interactions | No. of total $^{12}$C etch pit after C+Al interaction |
|---|---|---|---|---|---|---|---|---|
| | | | 1 | 2 | 3 | | | |
| T052 | T053 | 18589 | 14 | 2 | 0 | 6 | 22 | 18567 |
| T054 | T055 | 18513 | 13 | 0 | 0 | 6 | 19 | 18494 |
| T056 | T057 | 18447 | 10 | 0 | 0 | 9 | 19 | 18428 |
| T058 | T059 | 18385 | 6 | 0 | 0 | 10 | 16 | 18369 |
| T060 | T061 | 18321 | 2 | 1 | 0 | 3 | 6 | 18315 |
| T062 | T063 | 18263 | 3 | 1 | 0 | 2 | 6 | 18257 |
| T064 | T065 | 18190 | 4 | 1 | 0 | 8 | 13 | 18177 |
| T066 | T067 | 18117 | 4 | 2 | 0 | 5 | 11 | 18106 |

Then from the above counted number of projectiles before and after the target, the total charge changing cross-section has been measured. The measured cross section has been compared with the theoretical result calculated by Particle and Heavy Ion Transport code System (PHITS) [77] and the Glauber model[1] [78]. It is found that the measured result is in good agreement with the calculated result except for two points at energies 95.7 and 98.6 MEV/n. These two points need more analysis in the future.

*Table 5.5 Total charge changing cross-sections, with statistical standard deviations, for $^{12}C$ ions of different energies on aluminum target.*

| E (MeV/n) | ΔE (MeV/n) | This Experiment (mb) | | PHITS Calculation (mb) | | Glauber Model Calculation $\sigma_{TCC}$ (mb) |
|---|---|---|---|---|---|---|
| | | $\sigma_{TCC}$ | $\delta\sigma_{TCC}$ | $\sigma_{TCC}$ | $\delta\sigma_{TCC}$ | |
| 14.5 | 1.0 | 2990.0 | 472.0 | 2313.9 | 43.9 | 1627.9 |
| 24.1 | 0.9 | 2748.4 | 451.0 | 1762.9 | 38.3 | 1626.9 |
| 31.5 | 0.5 | 2437.2 | 424.0 | 1699.7 | 35.6 | 1590.7 |
| 89.5 | 0.2 | 1008.5 | 304.0 | 1350.3 | 33.3 | 1248.0 |
| 92.6 | 0.2 | 1187.2 | 329.0 | 1341.2 | 33.4 | 1234.2 |
| 95.7 | 0.2 | 545.6 | 222.0 | 1331.2 | 33.1 | 1221.1 |
| 98.6 | 0.2 | 545.0 | 222.0 | 1319.5 | 33.1 | 1209.3 |
| 101.5 | 0.2 | 1445.7 | 361.2 | 1312.9 | 33.0 | 1197.6 |
| 104.4 | 0.2 | 1711.2 | 391.0 | 1302.1 | 32.9 | 1187.1 |
| 107.2 | 0.2 | 1705.1 | 391.2 | 1294.5 | 32.0 | 1177.1 |
| 110.0 | 0.2 | 1966.5 | 419.0 | 1287.9 | 32.2 | 1167.5 |

For comparison with the previous result, we collected data from different authors like Takechi et al., 2009; Alpat et al., 2013; Cecchini et al., 2008; Schall et al., 1996; Kox et al., 1987. The measured data in this study have been compared with previous data in the following figure. Fig (a) shows the measured experimental data in this experiment with other previous measurements since there is no previous data point in the energy below 100 MeV/n, so we cannot see the comparison. But the PHITS and Glauber model calculated result has good agreement with the previous data in high energy ~200 MeV/n. These data are in good agreement with the PHITS and Glauber model data. So (b) shows a closer look at the data below 100 MeV/n and it is seen that these data have good agreement with the calculation.

---

[1] https://www.isotopea.com/nurex/#!/result

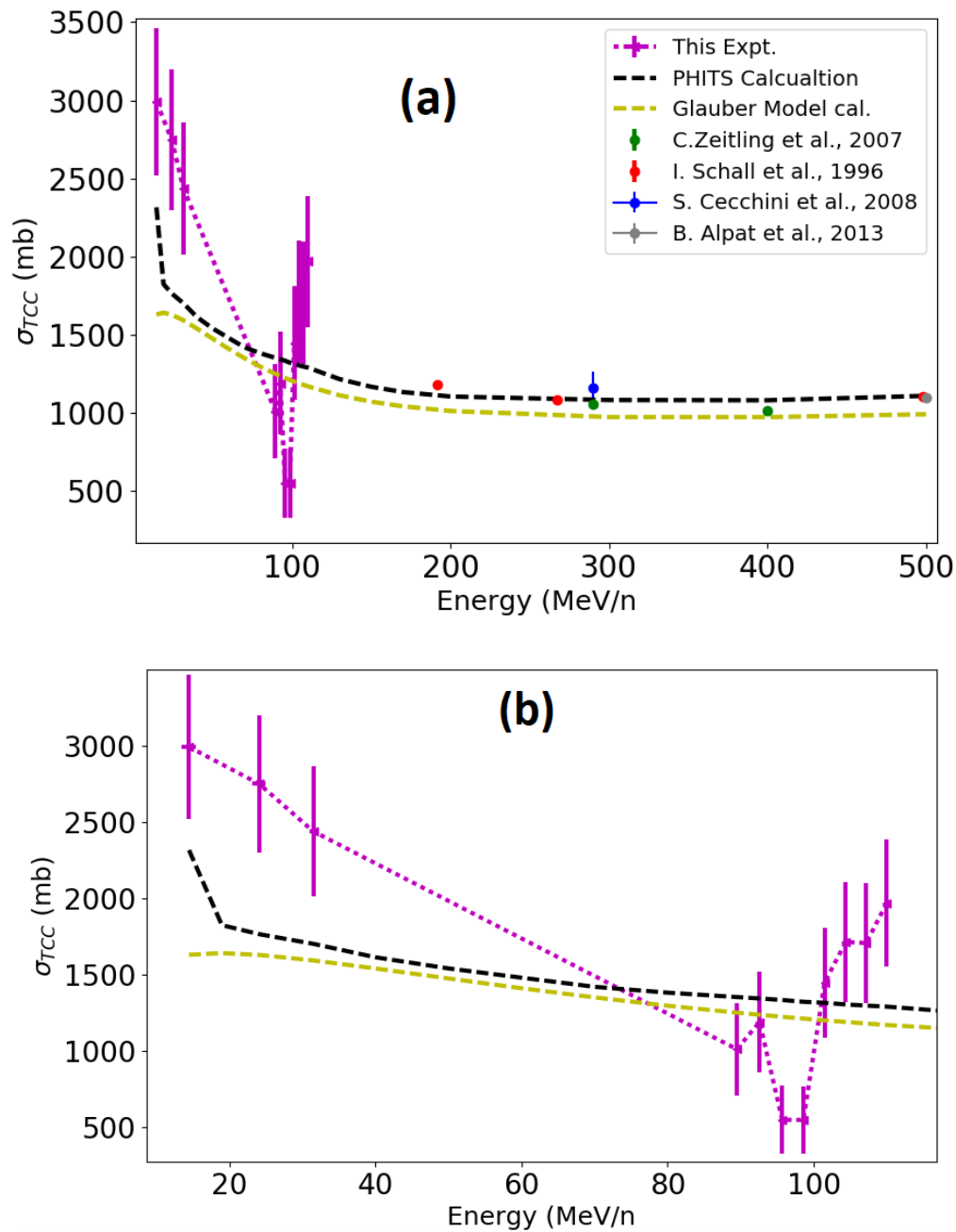

*Figure 5.14 (a) shows the measured experimental data in this experiment with other previous measurement. Since there are no previous data point in the energy below 100 MeV/n so we cannot see on comparison. But the PHITS and Glauber model calculated result has good agreement with the previous data in high energy ~200 MeV/n. These data are in good agreement with the PHITS and Glauber model data. So (b) shows the closer look at the data below 100 MeV/n and it is seen that these data have good agreement with the calculation.*

# Chapter 6
# Conclusions and Future Works

## 6.1 Conclusions

The aim of the present work was to measure total charge changing cross sections at an energy level of around 100 MeV/n for the interaction of $^{12}$C+Al, which can be useful for heavy ion cancer therapy and space radiation shielding. For that purpose, the target stack, Al interleaved with CR-39 detector, has been irradiated in HIMAC and WERC accelerator facilities in Japan. After irradiation, the etching and imaging with SEIKO Inc. FSP-1000 imaging microscope of CR-39 detector have been done at the Research Institute of Nuclear Engineering, University of Fukui, Japan. Then the image of etch pits, which are the carbon ion track, is analyzed. For that, a carbon etch pit counting protocol has been developed. Among different etch pit, based on the shape, size, and grey level (highest for incoming particle and lowest for outgoing particle), carbon eth pit has been identified. Manual counting has been performed to count the total number of incoming and outgoing particles before and after the target. From that total measurement, the total charge changing cross-section has been measured for the energy in between the incident and the outgoing particle's energy on the target. The measured data for the energies 14.5±1.0, 24.1±0.9, 31.5±0.5, 89.5±0.2, 92.6±0.2, 95.7±0.2, 98.6±0.2, 101.5±0.2, 104.4±0.2, 107.2±0.2, and 110.0±0.2 MeV/n were 2990.0±472.0, 2748.4±451.0, 2437.2±424.0, 1008.5±304.0, 1187.2±329.0, 545.6±222.0, 545.0±222.0, 1445.7±361.2, 1711.2±391.0, 1705.1±391.2, 1966.5±419.0 mb. Since there are no previous experimental data to compare, PHITS and Glauber model calculated theoretical values were compared with the previous and this experimental result. From that, the measured data were found in good agreement with the previous and calculated value except for two points at energy 95.7 and 98.6 MeV/n. These two points need more analysis which will be done in the next phase. Measured data in this study will be the first experimental TCC data for $^{12}$C+Al interactions in the energy ~100 MeV/n.

## 6.2  Future works

### *6.2.1 Measurement of the emission angle of fragments*

The straggling of light ion projectiles is low, and the beam's lateral size is precisely defined. Even so, this is not valid for the emitted projectile like fragments. The produced projectile fragments are emitted at larger angles after the primary ion collides with the target atom, which can result in a significant increase in the irradiation volume. When utilizing ions for tumor therapy, determining the fragment emission angles is very crucial since the produced projectile fragments can cause damage to vital tissues near the tumor. When so-called pencil beams for precise local irradiation are put into operation in the near future, precise information about fragment emission angular distributions will be very crucial. Since we can analyze event by event interaction and can see the fragments in each interaction, precise measurements of angle are possible, which is our next target.

### *6.2.2 Measurement of target fragmentation*

When a projectile interacts with a target material, then two types of fragment we can consider, either production of projectile fragments or target fragments. A considerable effort has been devoted to the measurement of the cross-section of projectile fragments. Nowadays, the production cross sections of the target fragments are a topic of great interest, but poor data are available yet. The direct measurement of target fragments is extremely difficult due to the short range of produced secondaries that have a low probability of escaping the target. The FOOT (Fragmentation Of Target) experiment will explore target fragments using a complicated experimental setup and an inverse kinematic technique. With CR-39, it will be possible to measure the production cross-section of target fragmentation.

# References

[1] A. Zucker, "Nuclear interactions of heavy ion," *Annu. Rev. Nucl. Sci.*, vol. 10, no. March, pp. 27–62, 1960.

[2] T. D. Malouff, A. Mahajan, S. Krishnan, C. Beltran, D. S. Seneviratne, and D. M. Trifiletti, "Carbon Ion Therapy: A Modern Review of an Emerging Technology," *Front. Oncol.*, vol. 10, no. February, pp. 1–13, 2020, doi: 10.3389/fonc.2020.00082.

[3] W. Tinganelli and M. Durante, *Carbon ion radiobiology*, vol. 12, no. 10. 2020.

[4] E. Takada, "Carbon Ion Radiotherapy at NIRS-HIMAC," *Nucl. Phys. A*, vol. 834, no. 1–4, pp. 730c-735c, 2010, doi: 10.1016/j.nuclphysa.2010.01.132.

[5] T. Kamada *et al.*, "Carbon ion radiotherapy in Japan: An assessment of 20 years of clinical experience," *Lancet Oncol.*, vol. 16, no. 2, pp. e93–e100, 2015, doi: 10.1016/S1470-2045(14)70412-7.

[6] H. Do Huh and S. Kim, "History of radiation therapy.," *Hist. Radiat. Ther. Technol.*, vol. 31, no. September,30, pp. 124–134, 2020, doi: 10.3769/radioisotopes.60.385.

[7] A. Kitagawa, T. Fujita, M. Muramatsu, S. Biri, and A. G. Drentje, "Review on heavy ion radiotherapy facilities and related ion sources invited," *Rev. Sci. Instrum.*, vol. 81, no. 2, 2010, doi: 10.1063/1.3268510.

[8] D. K. Ebner and T. Kamada, "The emerging role of carbon-ion radiotherapy," *Front. Oncol.*, vol. 6, no. JUN, pp. 6–11, 2016, doi: 10.3389/fonc.2016.00140.

[9] T. Ohno *et al.*, "Carbon Ion Radiotherapy at the Gunma University Heavy Ion Medical Center: New Facility Set-up," no. July, pp. 4046–4060, 2011, doi: 10.3390/cancers3044046.

[10] M. I. Dobynde, Y. Y. Shprits, A. Y. Drozdov, J. Hoffman, and J. Li, "Beating 1 Sievert: Optimal Radiation Shielding of Astronauts on a Mission to Mars," *Sp. Weather*, vol. 19, no. 9, pp. 1–13, 2021, doi: 10.1029/2021SW002749.

[11] M. Maalouf, M. Durante, and N. Foray, "Biological effects of space radiation on human cells:History, advances and outcomes," *J. Radiat. Res.*, vol. 52, no. 2, pp. 126–146, 2011, doi: 10.1269/jrr.10128.

[12] G. Onorato, E. Di Schiavi, and F. Di Cunto, "Understanding the Effects of Deep Space Radiation on Nervous System: The Role of Genetically Tractable Experimental Models," *Front. Phys.*, vol. 8, no. October, pp. 1–11, 2020, doi: 10.3389/fphy.2020.00362.

[13] C. Zeitlin and C. La Tessa, "The role of nuclear fragmentation in particle therapy and space radiation protection," *Front. Oncol.*, vol. 6, no. MAR, pp. 1–13, 2016, doi: 10.3389/fonc.2016.00065.

[14] J. C. Chancellor, G. B. I. Scott, and J. P. Sutton, "Space radiation: The number one risk to astronaut health beyond low earth orbit," *Life*, vol. 4, no. 3, pp. 491–510, 2014, doi: 10.3390/life4030491.

[15] P. Tsai, "Study of Secondary Particles Produced From Heavy-Ion," University of Tennessee, Knoxville, 2015.

[16] A. Ferrari, M. Ferrarini, and M. Pelliccioni, "Secondary particle yields from 400 MeV/u carbon ion and 250 MeV proton beams incident on thick targets," *Nucl. Instruments Methods Phys.*

*Res. Sect. B Beam Interact. with Mater. Atoms*, vol. 269, no. 13, pp. 1474–1481, 2011, doi: 10.1016/j.nimb.2011.04.003.

[17]   D. Schulz-Ertner and H. Tsujii, "Particle radiation therapy using proton and heavier ion beams," *J. Clin. Oncol.*, vol. 25, no. 8, pp. 953–964, 2007, doi: 10.1200/JCO.2006.09.7816.

[18]   J. K. Xu *et al.*, "Evaluation of neutron radiation field in carbon ion therapy," *Chinese Phys. C*, vol. 40, no. 1, pp. 1–5, 2016, doi: 10.1088/1674-1137/40/1/018201.

[19]   S. Agosteo, "Radiation protection at medical accelerators," *Radiat. Prot. Dosimetry*, vol. 96, no. 4, pp. 393–406, 2001, doi: 10.1093/oxfordjournals.rpd.a006627.

[20]   S. Yonai, N. Matsufuji, and T. Kanai, "Monte Carlo study on secondary neutrons in passive carbon-ion radiotherapy: Identification of the main source and reduction in the secondary neutron dose," *Med. Phys.*, vol. 36, no. 10, pp. 4830–4839, 2009, doi: 10.1118/1.3220624.

[21]   L. Sihver *et al.*, "Projectile fragment emission angles in fragmentation reactions of light heavy ions in the energy region <200 MeV/nucleon: Experimental study," *Radiat. Meas.*, vol. 48, no. 1, pp. 73–81, 2013, doi: 10.1016/j.radmeas.2012.08.006.

[22]   B. Alpat *et al.*, "Total and partial fragmentation cross-section of 500 MeV/nucleon carbon ions on different target materials," *IEEE Trans. Nucl. Sci.*, vol. 60, no. 6, pp. 4673–4682, 2013, doi: 10.1109/TNS.2013.2284855.

[23]   Q. M. Rashed Nizam, K. Yoshida, T. Sakamoto, E. Benton, L. Sihver, and N. Yasuda, "High-precision angular measurement of 12C ion interaction using a new imaging method with a CR-39 detector in the energy range below 100 MeV/nucleon," *Radiat. Meas.*, vol. 131, no. June 2019, p. 106225, 2020, doi: 10.1016/j.radmeas.2019.106225.

[24]   F. Luoni *et al.*, "Total nuclear reaction cross-section database for radiation protection in space and heavy-ion therapy applications Total nuclear reaction cross-section database for radiation protection in space and heavy-ion therapy applications," *New J. Phys.*, vol. 23, 2021.

[25]   M. Giorgini, "Fragmentation cross-sections of Fe26+, Si14+ and C6+ ions of 0.3 ÷ 10 A GeV on polyethylene, CR-39 and aluminum targets," *Radiat. Meas.*, vol. 44, no. 9–10, pp. 853–856, 2009, doi: 10.1016/j.radmeas.2009.10.032.

[26]   R. Gupta, A. Kumar, G. Giacomelli, L. Patrizii, and V. Togo, "Calibration of CR39 detectors with new system for Fe 26+ ion beam and measurement of total charge changing cross-section in Al target," *Radiat. Meas.*, vol. 47, no. 10, pp. 1023–1029, 2012, doi: 10.1016/j.radmeas.2012.07.007.

[27]   R. Gupta and A. Kumar, "Response of CR39 detector to 5A GeV Si14+ ions and measurement of total charge changing cross-section," *Radiat. Phys. Chem.*, vol. 92, pp. 8–13, 2013, doi: 10.1016/j.radphyschem.2013.07.012.

[28]   L. Huo, L. Wang, J. Zhu, H. Li, and J. Li, "The total charge-changing cross sections and the partial cross sections of 56 Fe fragmentation on Al , C and CH 2 targets," *Chinese J. Phys.*, vol. 60, no. January, pp. 88–97, 2019, doi: 10.1016/j.cjph.2019.04.022.

[29]   A. N. Golovchenko *et al.*, "Total charge-changing and partial cross-section measurements in the reactions of [Formula Presented] in carbon, paraffin, and water," *Phys. Rev. C - Nucl. Phys.*, vol. 66, no. 1, p. 8, 2002, doi: 10.1103/PhysRevC.66.014609.

[30]   D. Filges and F. Goldenbaum, *Handbook of Spallation Research*. 2009.

[31]   G. Knoll, "Radiation Detection and Measurement, 3rd ed - Glenn F." pp. 1–796, 2009.

[32]   P. Grafström, "Lifetime, cross-sections and activation," *CAS 2006 - Cern Accel. Sch. Vac. Accel. Proc.*, pp. 213–225, 2007.

[33]   C. Sengupta, "Precision Elastic Scattering and Reaction Cross-section Measurements of," The Australian National University, 2020.

[34]   L. F. Canto, V. Guimarães, J. Lubián, and M. S. Hussein, *The total reaction cross section of heavy-ion reactions induced by stable and unstable exotic beams: the low-energy regime*, vol. 56, no. 11. Springer Berlin Heidelberg, 2020.

[35]   F. Luoni *et al.*, "Total nuclear reaction cross-section database for radiation protection in space and heavy-ion therapy applications," *New J. Phys.*, vol. 23, no. 10, 2021, doi: 10.1088/1367-2630/ac27e1.

[36]   F. Ernst, "Measurement of Nuclear Reaction Cross Sections for Applications in Radiotherapy with Protons , Helium and Carbon Ions," 2019.

[37]   E. J. Burge, "The total proton reaction cross section of carbon from 10-68 MeV by a new method.," *Nucl. Phys.*, vol. 13, no. 511, 1959.

[38]   L. Sihver, C. H. Tsao, R. Silberberg, T. Kanai, and A. F. Barghouty, "Total reaction and partial cross section calculations in proton-nucleus and nucleus-nucleus reactions.," *Phys. Rev. C*, vol. 47, no. 3, pp. 1225–1236, 1993.

[39]   P. Eudes, Z. Basrak, G. Royer, and M. Novak, "Fusion excitation function revisited," no. September, 2012, doi: 10.1088/1742-6596/420/1/012133.

[40]   R. K. Tripathi, A. Cucinotta, J. W. Wilson, and N. Aeronautics, "Universal Parameterization Cross Sections of Absorption Systems," no. December, 1999.

[41]   P. Ferrando *et al.*, "Measurement of 12C, 16O, and 56Fe charge changing cross sections in helium at high energy, comparison with cross sections in hydrogen, and application to cosmic-ray propagation," *Phys. Rev. C*, vol. 37, no. 4, 1988.

[42]   S. N. Ahmed, *Physics and Engineering of Radiation Detection*, 1st ed. Queen's University, Kingston, Ontario: Academic Press is an imprint of Elsevie, 2007.

[43]   G. D. Westfall, L. W. Wilson, P. J. Lindstrom, H. J. Crawford, D. E. G. And, and H. H. Heckman, "Fragmentation of relativistic 56Fe," *Phys. Rev. C*, vol. 19, no. 4, p. 1309, 1979.

[44]   W. R. Webber, J. C. Kish, and D. A. Schrier, "Total charge and mass changing cross sections of relativistic nuclei in hydrogen, helium, and carbon targets," *Phys. Rev. C*, vol. 41, no. 2, pp. 520–532, 1990, doi: 10.1103/PhysRevC.41.520.

[45]   C. Zeitlin *et al.*, "Heavy fragment production cross sections from 1 . 05 GeV / nucleon 56 Fe in C , Al , Cu , Pb , and CH 2 targets," *Phys. Rev. C*, vol. 56, no. 1, pp. 388–397, 1997.

[46]   S. Ota, "Precise measurements of projectile charge changing cross sections for intermediate energy heavy ions using CR-39 track detectors," Waseda Universiyt, 2011.

[47]   J. F. Ziegler, "Stopping of energetic light ions in elemental matter," *J. Appl. Phys.*, vol. 85, no. 3, pp. 1249–1272, 1999, doi: 10.1063/1.369844.

[48]   S. Cecchini *et al.*, "Fragmentation cross sections of Fe26+, Si14+ and C6+ ions of 0.3-10 A GeV on polyethylene, CR39 and aluminum targets," *Nucl. Phys. A*, vol. 807, no. 3–4, pp. 206–213,

2008, doi: 10.1016/j.nuclphysa.2008.03.017.

[49]   I. Schall *et al.*, "Charge-changing nuclear reactions of relativistic light-ion beams (5 ≤ Z ≤ 10) passing through thick absorbers," *Nucl. Instruments Methods Phys. Res. Sect. B Beam Interact. with Mater. Atoms*, vol. 117, no. 3, pp. 221–234, 1996, doi: 10.1016/0168-583X(96)00325-4.

[50]   C. Zeitlin *et al.*, "Fragmentation cross sections of 290 and 400 MeV/nucleon C12 beams on elemental targets," *Phys. Rev. C - Nucl. Phys.*, vol. 76, no. 1, pp. 1–21, 2007, doi: 10.1103/PhysRevC.76.014911.

[51]   S. Manzoor *et al.*, "Charge identification in CR-39 Nuclear Track Detector using relativistic lead ion fragmentation," *Nucl. Instruments Methods Phys. Res. Sect. A Accel. Spectrometers, Detect. Assoc. Equip.*, vol. 453, no. 3, pp. 525–529, 2000, doi: 10.1016/S0168-9002(00)00470-8.

[52]   S. Manzoor, "Improvements and calibrations of nuclear track detectors for rare particle searches and fragmentation studies," 2007.

[53]   M. A. Rana, "CR-39 nuclear track detector : An experimental guide," *Nucl. Inst. Methods Phys. Res. A*, 2018, doi: 10.1016/j.nima.2018.08.077.

[54]   B. Mustafa Yousuf Rajab and H. Ali Al-Jobouri, "Digital Processing and Analysis for the Tracks Produced From the Irradiation with Neutrons Source 241 Am-9 Be on Some of Solid State Nuclear Track Detectors," no. January, 2016.

[55]   T. W. Jeong *et al.*, "CR-39 track detector for multi-MeV ion spectroscopy," pp. 2–9, 2017, doi: 10.1038/s41598-017-02331-w.

[56]   A. M. Bhagwat, "Solid State Nuclear Track Detection: Theory and Applications Indian Society for Radiation Physics Kalpakkam Chapter 1993," p. 34, 1993, [Online]. Available: http://www.iaea.org/inis/collection/NCLCollectionStore/_Public/25/019/25019093.pdf.

[57]   S. Kandaiya, *Characterization of CR 39 nuclear track detector for use as a radon/thoron dosemeter*. 1988.

[58]   R. Przybocki, "Investigation of non-normal incidence charged-particle response of CR39 nuclear track detectors , with applications in nuclear diagnostics for inertial confinement fusion by," 2020.

[59]   D. Schardt, T. Elsässer, and D. Schulz-Ertner, "Heavy-ion tumor therapy: Physical and radiobiological benefits," *Rev. Mod. Phys.*, vol. 82, no. 1, pp. 383–425, 2010, doi: 10.1103/RevModPhys.82.383.

[60]   T. Elsässer, A. Gemmel, M. Scholz, D. Schardt, and M. Krämer, "The relevance of very low energy ions for heavy-ion therapy," *Phys. Med. Biol.*, vol. 54, no. 7, 2009, doi: 10.1088/0031-9155/54/7/N03.

[61]   I. Ahmed, H. Nowrin, and H. Dhar, "Stopping power and range calculations of protons in human tissues," vol. 17, no. 4, pp. 1223–1233, 2020.

[62]   U. Weber and G. Kraft, "Design and construction of a ripple filter for a smoothed depth dose distribution in conformal particle therapy," vol. 44, pp. 2765–2775, 1999.

[63]   G. K. M. S. U. Bechthold, "Tumor therapy and track structure," pp. 229–237, 1999.

[64]   J. F. Ziegler, M. D. Ziegler, and J. P. Biersack, "SRIM - The stopping and range of ions in matter (2010)," *Nucl. Instruments Methods Phys. Res. Sect. B Beam Interact. with Mater. Atoms*, vol.

268, no. 11–12, pp. 1818–1823, 2010, doi: 10.1016/j.nimb.2010.02.091.

[65] R. E. Stoller, M. B. Toloczko, G. S. Was, A. G. Certain, S. Dwaraknath, and F. A. Garner, "On the use of SRIM for computing radiation damage exposure," *Nucl. Instruments Methods Phys. Res. Sect. B Beam Interact. with Mater. Atoms*, vol. 310, pp. 75–80, 2013, doi: 10.1016/j.nimb.2013.05.008.

[66] X. Zhang and S. Sheehy, "Current and future synchrotron designs for carbon ion therapy," *AIP Conf. Proc.*, vol. 2346, no. March, 2021, doi: 10.1063/5.0047815.

[67] S. Hatori *et al.*, "Applications of accelerators for industries and medical uses at the Wakasa Wan Energy Research Center," *AIP Conf. Proc.*, vol. 680, pp. 981–985, 2003, doi: 10.1063/1.1619873.

[68] S. Hatori, R. Ishigami, K. Kume, and K. Suzuki, "Ion accelerator facility of the wakasa wan energy research center for the study of irradiation effects on space electronics," *Quantum Beam Sci.*, vol. 5, no. 2, 2021, doi: 10.3390/qubs5020014.

[69] M. De Simoni and T. Advisor, "Development of tools for quality control on thera- peutic carbon beams with a fast-MC code ( FRED )," Universita Di Roma, 2020.

[70] S. Hatori *et al.*, "Tandem accelerator as an injecter for the medical-use synchrotron at the Wakasa Wan Energy Research Center," *IPAC 2010 - 1st Int. Part. Accel. Conf.*, pp. 714–716, 2010.

[71] T. Kanai *et al.*, "Biophysical characteristics of HIMAC clinical irradiation system for heavy-ion radiation therapy," *Int. J. Radiat. Oncol. Biol. Phys.*, vol. 44, no. 1, pp. 201–210, 1999, doi: 10.1016/S0360-3016(98)00544-6.

[72] H. Tsujii, T. Kamada, S. Toshiyuki, N. Koji, H. Tsuji, and K. Karasawa, *Carbon-Ion Radiotherapy*. Springer, 2013.

[73] H. Tsujii, T. Kamada, T. Shirai, K. Noda, H. Tsuji, and K. Karasawa, "Carbon-ion radiotherapy: Principles, practices, and treatment planning," *Carbon-Ion Radiother. Princ. Pract. Treat. Plan.*, pp. 1–312, 2014, doi: 10.1007/978-4-431-54457-9.

[74] H. A. Al-Jobouri and M. Y. Rajab, "Image processing analysis of nuclear track parameters for CR-39 detector irradiated by thermal neutron," *AIP Conf. Proc.*, vol. 1722, 2016, doi: 10.1063/1.4944209.

[75] M. El Ghazaly and N. M. Hassan, "Characterization of saturation of CR-39 detector at high alpha-particle fluence," *Nucl. Eng. Technol.*, vol. 50, no. 3, pp. 432–438, 2018, doi: 10.1016/j.net.2017.11.010.

[76] M. A. Rana, G. Sher, S. Manzoor, F. Malik, and K. Naz, "Nuclear Track Detectors for Relativistic Nuclear Fragmentation Studies: Comparison with Other Competitive Techniques," *Mod. Instrum.*, vol. 02, no. 03, pp. 49–59, 2013, doi: 10.4236/mi.2013.23008.

[77] T. Sato *et al.*, "Features of Particle and Heavy Ion Transport code System (PHITS) version 3.02," *J. Nucl. Sci. Technol.*, vol. 55, no. 6, pp. 684–690, 2018, doi: 10.1080/00223131.2017.1419890.

[78] M. Takechi *et al.*, "Reaction cross sections at intermediate energies and Fermi-motion effect," *Phys. Rev. C - Nucl. Phys.*, vol. 79, no. 6, pp. 1–5, 2009, doi: 10.1103/PhysRevC.79.061601.

# List of Publications

- Presented the paper titled as "**Total Charge Changing Cross Sections Measurements for $^{12}$C ion Interactions with Al using CR-39 Track Detector**" in the 4$^{th}$ International Conference on "Physics For Sustainable Development & Technology (ICPSDT-2022)" in the Oral Session (Nuclear and Health Physics-II) held during 22-23 January 2022 on the virtual platform organized by Department of Physics, Chittagong University of Engineering & Technology, Chattogram-4349, Bangladesh.

- Presented the paper titled as "**Measurement of total charge changing cross sections for $^{12}$C+Al interactions using CR-39 nuclear track detector**" in poster presentation at the 1$^{st}$ research symposium organized by the Institutional Quality Assurance Cell (IQAC) of the University of Chittagong.